\documentclass[a4paper,11pt]{article}
\usepackage{amsmath}
\usepackage{graphicx}
\usepackage[utf8x]{inputenc}
\usepackage{wasysym}
\usepackage{amsthm}
\usepackage{algorithm}
\usepackage{algorithmic}
\usepackage{fontenc}
\usepackage{textcomp}
\usepackage{hhline,latexsym}
\usepackage{color}
\usepackage{epsfig}
\usepackage{array}
\usepackage{multicol}
\usepackage{supertabular}
\usepackage{amssymb}
\usepackage{pstricks}
\usepackage[english]{babel}
\usepackage{fancyhdr}
\usepackage[FIGTOPCAP]{subfigure}
\usepackage{authblk}
\usepackage{appendix}
\usepackage{xfrac}
\usepackage{multicol,multirow}

\newcommand{\ds}{\displaystyle}

\newcommand*\dbbrac[1]{[\![ #1 ]\!]}

\def\u{{\bf u}}
\def\v{{\bf v}}
\def\p{{\bf p}}
\def\X{{\bf X}}
\def\F{{\bf F}}
\def\f{{\bf f}}
\def\X{{\bf X}}
\def\W{{\bf W}}
\def\up{{\bf u}^+}
\def\une{{\bf u}^-}
\def\upd{{\bf u}_d^+}
\def\uned{{\bf u}_d^-}
\def\upG{{\bf u}_G^+}
\def\uneG{{\bf u}_G^-}
\newcommand{\dd}{\,\mathrm{d}}
\newcommand{\Rb}{\,\mathbb{R}}

\newcommand{\Gc}{\Gamma_{\text{C}}}
\newcommand{\Gd}{\Gamma_{\text{D}}}
\newcommand{\Gn}{\Gamma_{\text{N}}}
\newcommand{\Gf}{\Gamma_{\text{F}}}
\newcommand{\Gz}{\Gamma_{0}}

\newcommand{\bu}{\mathbf{u}}

\newcommand{\br}{\mathbf{r}}
\newcommand{\bR}{\mathbf{R}}
\newcommand{\bP}{\mathbf{P}}

\newcommand{\bw}{\mathbf{w}}
\newcommand{\CC}{\mathbf{C}}

\def\tra{\mathsf{T}}

\def\W{{\mathbf{W}}}
\def\n{{\mathbf{n}}}
\def\t{{\mathbf{t}}}
\def\g{{\mathbf{g}}}
\def\d{{\mathbf{d}}}
\def\s{{\mathbf{s}}}
\def\y{{\mathbf{y}}}
\def\la{{\mathcal{L}}}
\newcommand{\ee}{\mathrm{E}}

\title{A new method to invert InSAR data to resolve stress changes on a fracture embedded in a 3D heterogeneous crust}
\author[]{Oliver ~Bodart\thanks{The Lyon University, Universit\'e Jean Monnet Saint-\'Etienne, CNRS UMR 5208, Institut Camille Jordan, F-42023 Saint-Etienne, France, 
email: olivier.bodart@univ-st-etienne.fr } }
\author[]{Val\'erie ~Cayol \thanks{Laboratoire Magmas et Volcans, . Universit\'e Clermont Auvergne-CNRS-IRD, OPGC, Clermont-Ferrand F-63038, France, email: v.cayol@uca.fr.} }
\author[]{Farshid ~Dabaghi\thanks{The Lyon University, Universit\'e Jean Monnet, Institut Camille Jordan, 
F-42023 Saint-Etienne, France, email: farshid.dabaghi@univ-st-etienne.fr}}

\affil[]{ }
\begin{document}
\maketitle
\begin{abstract}
We present a new method to invert variable stress changes of fractures from InSAR ground displacements. Fractures can be either faults or magma intrusions, embeded in a 3D heterogeneous crust with prominent topographies. The method is based on a fictituous domain approach using a finite element discretization of XFEM type.  A cost function involves the misfit between the solution of the physical problem and the observed data together with the smoothing terms. Regularization parameters are determined by using L-curves. The method is then reformulated to be applied to InSAR data (masked and noisy), projected in Earth-Satellite directions. Synthetic tests confirm the efficiency and effectiveness of our method.

\end{abstract}


\section{Physical and computational model}
\subsection{Mathematical modelling of the solid}

\begin{figure}[htp!]
\centering
 \includegraphics[scale=1]{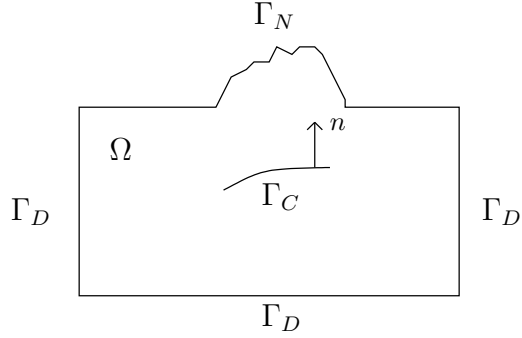}
 \caption{Volcanic domain $\Omega$ and its boundaries}\label{volcano}
\end{figure}

From a mathematical point of view, a volcano is a bounded open set $\Omega \subset \Rb^3$, occupied by an elastic solid. This set is assumed to have a smooth boundary $\partial \Omega$ which we separate in two (non empty) parts $\partial \Omega :=
 {\Gd} \cup\overline{\Gn}$, with ${\Gd} \cap
{\Gn} = \emptyset$. As depicted on Figure \ref{volcano}, the subset $\Gn$ is actually the ground of the volcano and free to move. The subset $\Gd$ is an artificial boundary defined in order to work on a finite size object and is assumed to to satisfy a 0 displacement condition. 
We assume that the elastic solid occupying $\Omega$ is subject to a body force field $\mathbf{f}$. 
In the sequel, boldface letters will be used to denote (usually $3$-dimensional) vector fields. Plain letters will represent scalar quantities.

We denote by $ \mathbf{x} = (x,y,z)$ the generic point in $\Rb^3$, and by $\mathbf{u} = \mathbf{u}(\mathbf{x})$ the displacement field of the solid $\Omega$. The Cauchy stress  $\sigma(\mathbf{u})$ and strain $\varepsilon(\mathbf{u})$ are given by
\begin{equation*}
\sigma(\mathbf{u}) = \lambda \textrm{Tr}  \bigl(\varepsilon (\mathbf{u})\bigr) \mathbf{I}_{\mathbb{R}^3} + 2 \mu \varepsilon(\mathbf{u})
  \quad \text{and}
 \quad \varepsilon(\mathbf{u}) = \frac{1}{2}(\nabla \mathbf{u} + \nabla \mathbf{u}^{\tra}),
\end{equation*}
where $(\lambda ,\mu)$ are the Lam\'e coefficients of the material, $\mathbf{I}_{\mathbb{R}^3}$ the
identity matrix, and $\textrm{Tr}(\cdot)$  the matrix trace.

\begin{figure}[htp!]
\centering
 \includegraphics[scale=1]{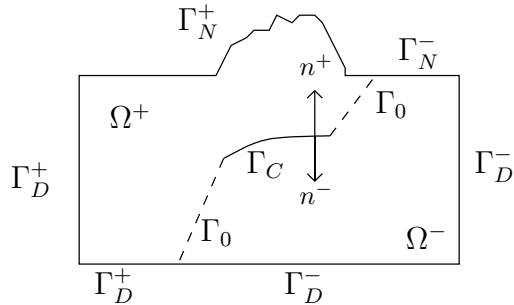}
 \caption{Splitting a volcanic cracked domain by domain decomposition method}\label{volcanic_dom}
\end{figure}

We also assume the presence of a crack in the volcano, which is represented mathematically by a 2-D surface $\Gc\subset \Omega$ on which  a traction or pressure force will be exerted (see also Figure \ref{volcano}). In a volcanic context,  $\Gc$ might represent a magma--filled crack on a fault. As already said in the introduction, the shape and position of the crack are supposed to be known in this work. 
The deformation field $\mathbf{u}$ of the solid is supposed to satisfy the following elastostatic system:

\begin{eqnarray}
   - \mathrm{div}~ \sigma(\u)  &= \mathbf{f}&\textrm{in}~\Omega, \label{problem1-1}\\
    \u&=  0 &\textrm{in}~ \Gd, \label{problem1-2}\\
    {\sigma}(\u)\cdot{\n}&= 0&\textrm{on}~\Gn, \label{problem1-3}\\
    {\sigma}(\u)\cdot{\n^\pm}&  = \t^\pm&\textrm{on}~\Gc.  \label{problem1-4}
\end{eqnarray}
The first equation is the equilibrium law describing the (linear) elastic behaviour of the material. Condition \eqref{problem1-2} describes the fact that the displacement vanishes on the underground boundary of the solid. In equation \eqref{problem1-3}, and in the sequel, $\n$ denotes the unit normal vector on the boundary of $\Omega$, oriented externally. This condition describes the free movement of the ground part of the volcano.
Finally, equation \eqref{problem1-4} describes the force acting on the crack $\Gc$. The vectors $\n^+$ and $\n^- = -\n^+$ are unit normal vectors on the crack $\Gc$ (see Figure \ref{volcanic_dom}). The vector functions $\t^\pm(\mathbf{x})$ denote the force exerted on the crack. They are such that $\t^+ = -\t^-$. When this force acts normally to the crack, that is $\t^\pm(\mathbf{x}) = p(\mathbf{x})\n^\pm$, the force is called a pressure, otherwise in general it is called a traction force.

This traction force is the actual unknown of our problem. From the mathematical point of view it is supposed to be quadratically integrable as well as its gradient.

In order to derive a finite element approximation of the state equations \eqref{problem1-1}--\eqref{problem1-4}, we need a weak formulation of the system. This is done by multiplying the equation \eqref{problem1-1} by a regular vector field $\v$ such that $\v =0$, and then integrating by parts over $\Omega$. Using the conditions \eqref{problem1-2}--\eqref{problem1-4} then yields:
\begin{equation}
\ds\int_{\Omega} \sigma(\u) : \varepsilon(\v) \dd  {\Omega}=\ds\int_{\Omega} \mathbf{f} \cdot \v  \dd {\Omega}+ \int_{\Gc}  (\t^\pm)\cdot\v \dd \Gc,\label{forml} 
\end{equation}
where $:$ is the double inner product for matrices, i.e. for $A=(a_{ij})_{i,j=1,2,3}$ and $B=(b_{ij})_{i,j=1,2,3}$,
$$
A:B = \sum_{i=1}^3 \sum_{j=1}^3 a_{ij}b_{ij}.
$$

Notice that this approach allow us to deal with situations where the elastic material is nonhomogeneous and anisotropic, that is when the Lamé coefficients are not constant.

\subsection{Finite element model}\label{discrete-model}
The direct problem \eqref{problem1-1}--\eqref{problem1-4} ef\/f\/iciently will have to be solved iteratively during the inversion procedure. Therefore we need an efficient algorithm to do so. To this aim, we use a domain decomposition method. 
More precisely, following \cite{BCCK16}, the domain $\Omega$ is split into two
sub-domains such that each point of the domain lies on one side of the crack or on the crack. For this purpose, we use  an artificial extension $\Gamma_0$ of the crack $\Gc$.
Assuming that $\Gf=\Gc \cup \Gz$ splits the domain into two subdomains $\Omega^{+}$ and  $\Omega^{-}$, we have 
$ \Omega=\Omega^{+}\cup \Gf \cup\Omega^{-}$, $\Gn=\Gn^{+}\cup\Gn^{-}$ and $\Gd=\Gd^{+}\cup\Gd^{-}$. We define on $\Gc$ two opposite unit outward
normal vectors $\n^{+}$ (from $\Omega^{+}$) and $\n^{-}$ (from  $\Omega^{-}$). The global unknown $\u$ is split into 
$\up=\u_{\mid\Omega^{+}}$ and $\une=\u_{\mid\Omega^{-}}$.

Therefore, the equations \eqref{problem1-1}--\eqref{problem1-4} can be rewritten with $\up$ and $\une$ as unknowns:
\begin{eqnarray}
    -\mathrm{div}~ \sigma(\u^{\pm})  &= \mathbf{f}^\pm&\textrm{in}~\Omega^{\pm}, \label{problem2-1}\\
    \u^{\pm} &=  0 &\textrm{on}~ \Gd\cap \partial \Omega^{\pm},\label{problem2-2}\\
    ({\sigma}(\u)\cdot{\n})^{\pm} &= 0&\textrm{on}~\Gn\cap \partial \Omega^{\pm}, \label{problem2-3}\\
    ({\sigma}(\u)\cdot{\n})^{\pm} &= \t^{\pm}&\textrm{on}~\Gc,\label{problem2-4}\\
    \dbbrac{\u} &=0 &\textrm{on}~\Gamma_{0}, \label{problem2-5}\\
   \dbbrac{{\sigma}(\u)}\cdot{\n}^{+} &=0&\textrm{on}~\Gamma_{0}, \label{problem2-6}
\end{eqnarray}
where $\dbbrac{\v}=\v^+-\v^-$ denotes the jump of a function across $\Gamma_0$.
The equations and conditions \eqref{problem2-1}--\eqref{problem2-4} are a straightforward rewriting of \eqref{problem1-1}--\eqref{problem1-4}. The two last conditions \eqref{problem2-5} and \eqref{problem2-6} are imposed to  enforce the continuity of displacement and stress across $\Gamma_0$ to ensure that $\u=(\u^-,\u^+)$ solves the original problem
\eqref{problem1-1}--\eqref{problem1-4}. 

Let us now describe the discretization of this problem via the finite element method.
A lagrange multiplier $\lambda$ defined on $\Gamma_0$ is introduced to enforce the continuity of the displacement across the $\Gamma_0$. Defining the associated lagrangian functional and expressing the associated saddle point conditions give a weak formulation for Problem \eqref{problem2-1}--\eqref{problem2-6} which
unknown is in the form $\X =(\u^+,\u^-,\boldsymbol{\lambda})^\top$ (see \cite{BCCK16} for details).

In order to numerically approximate the system, consider a tetrahedral mesh of $\Omega$ (as e.g. in Figure \ref{topo_flat}). Let $\boldsymbol\varphi^{\pm}_i$ be a finite element basis defined on this mesh. Also defining a mesh of  the extended crack, we build finite elements bases $\boldsymbol\psi_i$ and $\boldsymbol\theta_j$ on  $\Gamma_0$ and $\Gc$ respectively. 

Identifying the unknowns $\up$, $\une$ and $\boldsymbol{\lambda}$ with there value on the nodes of the mesh, the discretized form of Problem \eqref{problem2-1}--\eqref{problem2-6}, then reads
\begin{equation}\label{Discrete-Problem}
\mathrm{K}\X = \F.
\end{equation}
The matrix $\mathrm{K}$ of this system is built from the stiffness matrices $A^{\pm}$, and the constraint coupling matrices on $\Gamma_0$ denoted by $B^\pm$. More precisely, we have
$$
\mathrm{K}=\left( \begin{array} {ccc}
A^+ & 0 & {B^+}^T \\
0 & A^- & -{B^-}^T \\
B^+& -B^- & 0
\end{array}
\right), 
$$
where
$$
A^{\pm}  :=\left[\int_{\Omega^{\pm}} \sigma(\boldsymbol\varphi_i^{\pm}):\varepsilon(\boldsymbol\varphi_j^{\pm})\dd\Omega^{\pm}\right]_{ij},\qquad
B^{\pm} :=  \left[ \int_{\Gz} \boldsymbol\varphi_i^{\pm}\cdot \boldsymbol\psi_j\dd \Gz \right]_{ij}
$$

The right hand side vector $\F^\pm$ is given by
$$
\F=\left( \begin{array} {l}
\mathbf{F}^+ \\
\mathbf{F}^- \\
0
\end{array}
\right), \qquad\mathbf{F}^\pm:=\left[\int_{\Omega^{\pm}}\mathbf{f}^\pm\cdot \boldsymbol\varphi_i^\pm \dd\Omega^{\pm}+\int_{\Gc} \t^\pm\cdot\boldsymbol\varphi_i^\pm \dd\Gc
\right]_{i},
$$
which boils down to the algebraic formulation
$$
\F=\left( \begin{array} {l}
\mathbf{F}^+ \\
\mathbf{F}^- \\
0
\end{array}
\right)=\left( \begin{array} {c}
M_{\Omega}^+ \f^+\\
M_{\Omega}^- \f^-\\ 
~~0
\end{array}
\right)
+\left( \begin{array} {c}
+M_C\t \\
-M_C\t \\ 
~~0
\end{array}
\right),
$$
where the mass matrices $M_{\Omega}^\pm$ and the coupling matrix on $\Gc$, denoted by $M_C$, are given by
\begin{equation*}
\ds \left[M_{\Omega}^\pm\right]_{ij}=\int_{\Omega^{\pm}}\boldsymbol\varphi_i^\pm\cdot \boldsymbol\varphi_j^\pm\dd\Omega^{\pm}, \qquad
\ds \left[M_{c}\right]_{ij}=\int_{\Gc}\boldsymbol\varphi_i^{\pm}\cdot \boldsymbol\theta_j\dd\Gc.
\end{equation*}
Eventually, we can naturally define two matrices $L_\Omega$ and $L_C$ such that we have
$$
\F = L_{\small\Omega}\f+L_C\t.
$$

The matrix $K$ is symmetric and positive definite, which allows the use of powerful classical solvers.  
The implementation of this model uses the finite element library GetFem++ \cite{REGET++}. The library provides all the necessary tools to handle various types of PDEs,
and links to classical powerful solvers (conjugate gradient, SuperLU, MUMPS) as well as other routines for mesh management, definition of cracks and boundaries via level set functions)
which make it quite versatile and powerful.
We refer the reader to \cite{BCCK16} for a detailed mathematical and computational analysis of this method. See also \cite{BCDK2020, BCDK2021} for a mathematical analysis of similar optimal control problems aking use of the domains decomposition technique.

\section{Inversion from full ground measurements}

\subsection{Mathematical formulation of the problem}\label{math-form}

Let us assume the displacement field of the volcano to be fully measured on the ground $\Gn$, and call it $\bu_d$. The inversion problem consists in finding a traction vector $\t$ such that the solution of Problem \eqref{problem1-1}--\eqref{problem1-4} satisfies $\bu = \bu_d$ on $\Gn$. However this is known to be an ill-posed problem (e.g. the solution is not unique), and moreover the measurements are perturbed by random uncertainties. Therefore we adopt a least squares approach, combined with a regularization (i.e. smoothing) technique to build the inversion process.

Consider the following cost function:
\begin{equation}\label{min_prob}
J(\t)=\frac{1}{2}\int_{\Gn} (\u-\u_{d})^\top{\CC^{-1}} (\u-\u_{d}) \dd \Gn 
+\frac{\alpha_0}{2}\ds\int_{\Gc}\lvert \t\rvert^2\dd \Gc+\frac{\alpha_1}{2}\ds\int_{\Gc} \lvert\nabla \t\rvert^2\dd \Gc,
\end{equation} 
where ${\CC}$ denotes the covariance operator of the measurements uncertainties (see e.g. \cite{Tarantola}), and is assumed to be positive definite, $\top$ denotes the transpose operator, 
and finally, the non--negative constants $\alpha_0$ and $\alpha_1$ are the Tikhonov regularization (or smoothing) parameters.

We aim at minimizing this cost function over the space of feasible traction vectors. This space consists in quadratically integrable vectors, which gradient is also quadratically integrable.

In \cite{BCDK2020, BCDK2021}, a simplified version of this problem is mathematically studied. The cost function only features the first smoothing term. As it will be shown later on, our tests have proved the necessity of smoothing the gradient, in a practical framework, to obtain a physically admissible solution $\t$.
The optimal choices for the values of $\alpha_0$ and $\alpha_1$ will be discussed further. 

To derive the optimality conditions for the Problem \eqref{min_prob}, we introduce the following Lagrangian:
\begin{align*}
 \la(\u,\t,\boldsymbol\phi)=J(\t)-\displaystyle \int_{\Omega}(  \sigma(\u) : \varepsilon(\boldsymbol\phi)- f )\cdot \boldsymbol\phi \dd {\Omega} +
 \int_{\Gc}{\t^\pm} \cdot\boldsymbol\phi \dd \Gc,
\end{align*}
where $\boldsymbol\phi$ is the Lagrange multiplier for the constraint \eqref{problem1-1}--\eqref{problem1-4} in the weak form \eqref{forml}. 
When $\nabla_{\u}\la=0$ and $\nabla_{\boldsymbol\phi}\la=0$, we have $\nabla_{\t}\la= \nabla J$. Cancelling $\nabla_{\boldsymbol\phi}\la$ gives problem \eqref{problem1-1}--\eqref{problem1-4}. Cancelling $\nabla_{\u}\la$ and performing integration by parts yields that $\boldsymbol\phi$ is the solution of 
\begin{equation}\label{problem_adjoint}
  \begin{cases}
   - \mathrm{div}~ \sigma(\boldsymbol\phi)  = 0&\textrm{in}~\Omega, \\
    \boldsymbol\phi=  0 &\textrm{in}~ \Gd,\\
    {\sigma}(\boldsymbol\phi)\cdot{\n}=\mathrm{C}^{-1}(\u-\u_{d}) &\textrm{on}~\Gn,\\
  \end{cases}
\end{equation}
called the  \textit{adjoint state} system. Finally, computing $\nabla_{\u}\la$, for $\u$ and $\boldsymbol\phi$ satisfying \eqref{problem1-1}--\eqref{problem1-4} and \eqref{problem_adjoint}, gives: 
\begin{equation}\label{gradient}
 \nabla J(\t)= \alpha_0 \t +\alpha_1 \nabla \t + (\boldsymbol\phi\cdot \n^\pm).
\end{equation}
We refer the reader to \cite{BCDK2020, BCDK2021}, for further details.

After splitting the computational domain $\Omega$, the cost function $J$ in  the optimization problem \eqref{min_prob} becomes:  
\begin{equation}\label{objective_fun_split}
\begin{split}
 J(\t)&=\frac{1}{2}\int_{\Gn^+} (\u^+-\u_{d}^+)^\top{\CC^{-1}} (\u^+-\u_{d}^+) \dd \Gn^+ +\frac{1}{2}\int_{\Gn^-} (\u^--\u_{d}^-)^\top{\CC^{-1}} (\u^--\u_{d}^-) \dd \Gn^-\\
 &+\frac{\alpha_0}{2}\ds\int_{\Gc}\lvert \t\rvert^2\dd \Gc+\frac{\alpha_1}{2}\ds\int_{\Gc} \lvert\nabla \t\rvert^2\dd \Gc.
\end{split}
\end{equation}
The previous optimality conditions rewrite naturally from this new formulation.

\subsection{The Discrete Problem}\label{discrete-problem}
The next step is to adapt the cost function, the adjoint state and the gradient of the cost function to the discrete system \eqref{Discrete-Problem}. To this aim, it has to be noticed that all terms in \eqref{objective_fun_split} are symmetric. Discretizing the two first terms will involve the introduction of mass matrices on $\Gn^\pm$. On needs to keep the symmetry of the discrete form. Since the covariance matrix $\CC$ is positive definite, it is also the same for its inverse and we can write
$$
\CC = Q D Q^{-1}, \qquad \CC^{-1} = QD^{-1}Q^{-1},  
$$
where $Q$ is the matrix which columns are the eigenvectors of $\CC$ and $D$ the diagonal matrix of the eigenvalues of $\CC$. This diagonalized form allows us to write $\CC^{-1} = \CC^{-\frac{1}{2}}\CC^{-\frac{1}{2}}$, where
$$
\CC^{-\frac{1}{2}} = QD^{-\frac{1}{2}}Q^{-1}.
$$
In view of this, we rewrite the continuous cost function $J$ as
\begin{equation}\label{objective_fun_split_sym}
\begin{split}
 J(\t)&=\frac{1}{2}\int_{\Gn^+} (\u^+-\u_{d}^+)^\top{\CC^{-\frac{1}{2}}\CC^{-\frac{1}{2}}} (\u^+-\u_{d}^+) \dd \Gn^+ +\frac{1}{2}\int_{\Gn^-} (\u^--\u_{d}^-)^\top{\CC^{-\frac{1}{2}}\CC^{-\frac{1}{2}}} (\u^--\u_{d}^-) \dd \Gn^-\\
 &+\frac{\alpha_0}{2}\ds\int_{\Gc}\lvert \t\rvert^2\dd \Gc+\frac{\alpha_1}{2}\ds\int_{\Gc} \lvert\nabla \t\rvert^2\dd \Gc.
\end{split}
\end{equation}

As in Section \ref{discrete-model}, we will identify the previous functions with their values at the mesh nodes. Define a set of finite element basis functions $\boldsymbol\chi_i^\pm$ (with 3 degrees of freedom at each mesh node) defined on $\Gn^\pm$. Then, define the ground mass matrix
\begin{equation}\label{ground-mass}
M_G=\left( \begin{array} {ll}
M_G^+ & 0\\
0& M_G^-
\end{array}\right),
\end{equation}
with $\ds \left[M_G^\pm\right]_{ij}=\int_{\Gn}\boldsymbol\chi_i^\pm\cdot \boldsymbol\chi_j^\pm\dd\Gn$.
Let us also define the matrices $M_{F_0}$ and $M_{F_1}$ as
$$
\left[M_{F_0}\right]_{ij}=\ds\int_{\Gc}\boldsymbol{\theta_i}\cdot\boldsymbol{\theta_j}\dd\Gc, \qquad \left[M_{F_1}\right]_{ij}=\ds\int_{\Gc}\nabla\boldsymbol{\theta_i}\cdot \nabla\boldsymbol{\theta_j}\dd\Gc.
$$
Recall that $\boldsymbol\theta_j$ are the finite element basis functions defined on $\Gc$.

Finally, in view of the domain decomposition method, the mesh nodes located on the ground will be on either side of the crack. Therefore, the measured displacement and the restriction to the ground nodes of the computed displacement (denoted $\u_G$) can be written as
$$
\u_d = \left( \begin{array} {l}
\upd \\
\uned 
\end{array}\right), \qquad
\u_G = \left( \begin{array} {l}
\upG \\
\uneG 
\end{array}\right). 
$$
Consider then the reduction matrix $\mathcal{O}_R$ such that 
$$
\mathcal{O}_R \X = \u_G,
$$
where $\X$ is the solution of \eqref{Discrete-Problem}.

The discrete version of the cost function $J$ defined by \eqref{objective_fun_split_sym} is then
\begin{equation}\label{discrete_min_prob}
J_\mathtt{d}(\t) = \frac{1}{2}(\mathcal{O}_R\X-\u_{d})^\tra\CC^{-\frac{1}{2}} M_G\CC^{-\frac{1}{2}}(\mathcal{O}_R\X-\u_{d})+\frac{\alpha_0}{2}
 (\t^\top M_{F_0} \t)+\frac{\alpha_1}{2}
 (\t^\top M_{F_1} \t).
\end{equation}

The optimality conditions for the minimizer of the discrete cost function $J_\mathtt{d}$ are obtained via the characterization of the saddle point of the following (discrete) Lagrangian function:
\begin{equation*}
\la_\mathtt{d} (\X,\t, \boldsymbol\phi)=J_\mathtt{d}(\t)-\langle \mathrm{K}\X-(L_{\small\Omega}\f+L_C\t),\boldsymbol\phi \rangle.
\end{equation*}
Computing the partial derivative of $\la_\mathtt{d}$ and then cancelling them, lead to the discrete counterpart
of the adjoint problem \eqref{problem_adjoint} as:
\begin{equation*}
 \mathrm{K} \boldsymbol\phi=\mathcal{O}_R^\top\CC^{-\frac{1}{2}}M_G\CC^{-\frac{1}{2}}(\mathcal{O}_R\X-\u_{d}),
\end{equation*}
which rewrites, denoting $\u_G = \mathcal{O}_R\X$,
\begin{equation}\label{discrete_adjoint}
 \mathrm{K} \boldsymbol\phi=\mathcal{O}_R^\top\CC^{-\frac{1}{2}}M_G\CC^{-\frac{1}{2}}(\u_G - \u_{d}),
\end{equation}
where $K$ is the (symmetric) matrix of System \eqref{Discrete-Problem}.
The gradient of $J_\mathtt{d}$ at any point $\t$ is then:
 \begin{equation}\label{discrete_gradient}
 \nabla J_\mathtt{d}(\t) =\alpha_0 M_{F_0}\t+\alpha_1 M_{F_1}\t+L_C^\top \boldsymbol\phi.
 \end{equation}
\subsection{Practical minimization algorithm}

In a previous work (see \cite{BCDK2021}), we presented two methods to minimize a simpler version of our cost function $J_\mathtt{d}$ : the conjugate gradient algorithm and a quasi-Newton method (low storage BFGS). These two techniques are natural choices due to the quadratic structure of the cost function. They were studied in depth in terms of computational performance and accuracy. The BFGS method appeared to be converging faster when the mesh of the domain $\Omega$ is rather coarse. When the mesh becomes finer, the number of unknowns in the problem increases and the two methods tend to give similar results. In the framework of applications which we are interested in, we aim at fast processes, that is we will most of the time consider meshes that are rather coarse except in the neighborhood of the crack $\Gc$. Therefore we will use a BFGS algorithm (see \cite{BROYDEN70, Fletcher70, Goldfarb70, Shanno70}) in our numerical tests. It involves the construction of a sequence of approximation of the inverse of the Hessian matrix of the cost function $J_\mathtt{d}$. We use a limited storage version of the algorithm called L-BFGS (see \cite{Nocedal80}). The method is presented in Algorithm~\ref{IP_DDM1}, with the necessary adaptations to our problem.

As usual with this type of methods sequences are built and will be denoted as follows:
\begin{itemize}
\item $\t^k$: traction force vector,
\item $\X^k$: solution of the state equation \eqref{Discrete-Problem},
\item $\u^k$ : displacement field {\bf on the ground},
\item $\phi^k$: solution of the adjoint state equation \eqref{discrete_adjoint},
\item $\g^k$: gradient of the cost function $J_\mathtt{d}$ at point $\t^k$,
\item $\d^k, \bw^k$: displacement directions for the optimization,
\item $\rho^k$: optimal displacement step.
\end{itemize}
The number $k\geq 0$ represents the iteration number.

Notice that the underlying quadratic form in the cost function allow to compute explicitely the optimal step size at each iteration, via formula \eqref{step_size}. The detailed computation is presented in the Appendix.

In the algorithm, the domain decomposition technique is used to solve the discrete state and adjoint state equations \eqref{Discrete-Problem} and \eqref{discrete_adjoint}. Notice that the matrix $K$ is symmetric and does not change during the iterative process, which means that the computation cost due to matrix factorization is reduced.

The algorithm iterates until the gradient of $J_\mathtt{d}$ becomes
\textit{sufficiently small}, as precised in the algorithm below.

\begin{algorithm}[htb]
\caption{Algorithm to minimize $J_\mathtt{d}$}\label{IP_DDM1}
\begin{algorithmic}
 \REQUIRE $\t^0$, $\u_d$ and $\epsilon>0$
 \bigskip    
 \STATE $k \leftarrow 0$  
 \STATE Solve \eqref{Discrete-Problem} with $\t=\t^0$ $\rightarrow$ $\X^0$
 \STATE $u^0 \leftarrow \mathcal{O}_R \X^0$
 \STATE Solve \eqref{discrete_adjoint} with $\u = \u^0$ $\rightarrow$ $\boldsymbol\phi^0$
 \STATE Initial gradient: $\g^0 \leftarrow \alpha_0 M_{F_0}\t^0+\alpha_1 M_{F_1}\t^0+L_C^\top \boldsymbol\phi^0$
 \STATE Initial direction: $\d^0 \leftarrow  -\g^0$
 \STATE $H^0 \leftarrow \mathbf{I}_C$
   
  \bigskip
  
\WHILE{$\lVert \g^{k+1}\rVert_{\Gc} <\epsilon\Vert \g^0\Vert_{\Gc}$}
\STATE Solve \eqref{Discrete-Problem} with $\t=\t^k$ $\rightarrow$ $\W^k$
\STATE $\bw^k \leftarrow \mathcal{O}_R \W^k$
\STATE Compute step size $\rho^k$ by \eqref{step_size} with $\u=\u^k$, $\t=\t^k$, $\d=\d^k$, $\bw=\bw^k$

\bigskip

\STATE  $\t^{k+1} = \t^k+\rho^k\d^k$ 
\STATE $\u^{k+1} = \u^k+\rho^k\bw^k$ 

\bigskip

\STATE Solve \eqref{discrete_adjoint} with $\u=\u^{k+1}$ $\rightarrow$ $\boldsymbol\phi^{k+1}$
\STATE   $\g^{k+1}$ in \eqref{gradient} with $\t^{k+1}$ and $\boldsymbol\phi^{k+1}$
\STATE Compute $H^{k+1}$ via formula \eqref{BFGS_Update}
\STATE $\d^{k+1}=-H^{k+1}\g^{k+1}$ 

\bigskip

\STATE $k \leftarrow k+1$
\ENDWHILE
\end{algorithmic}
\end{algorithm}

\subsection{Numerical tests and validation with synthetic data}

\begin{center}
\begin{figure}[htp!]\hspace*{-0.5cm}   
\subfigure[Topo f\/lat mesh]{\includegraphics[scale=0.16]
{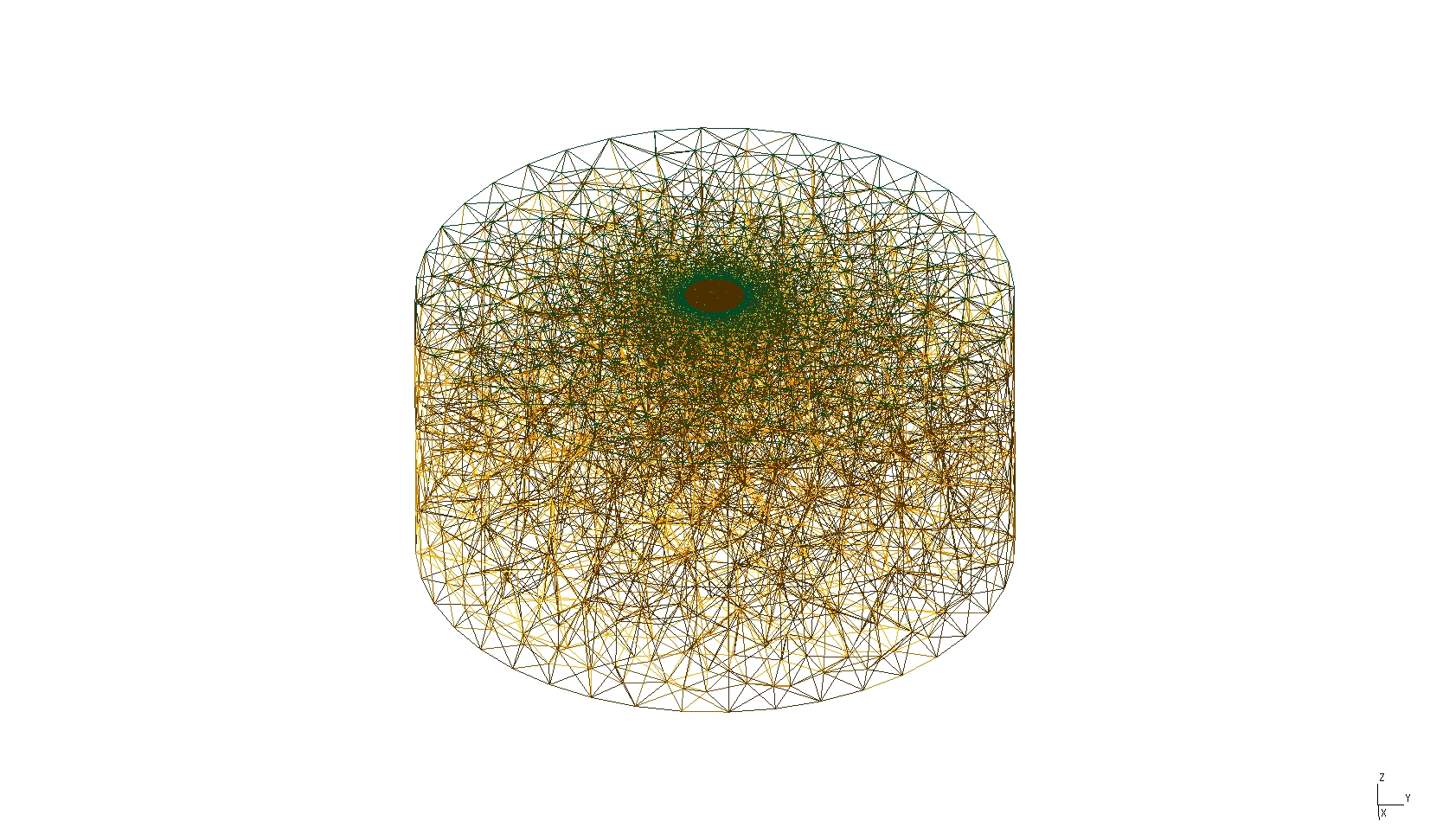}}\hspace*{-0.5cm}  
\subfigure[Circular fracture]{\includegraphics[scale=0.25]
{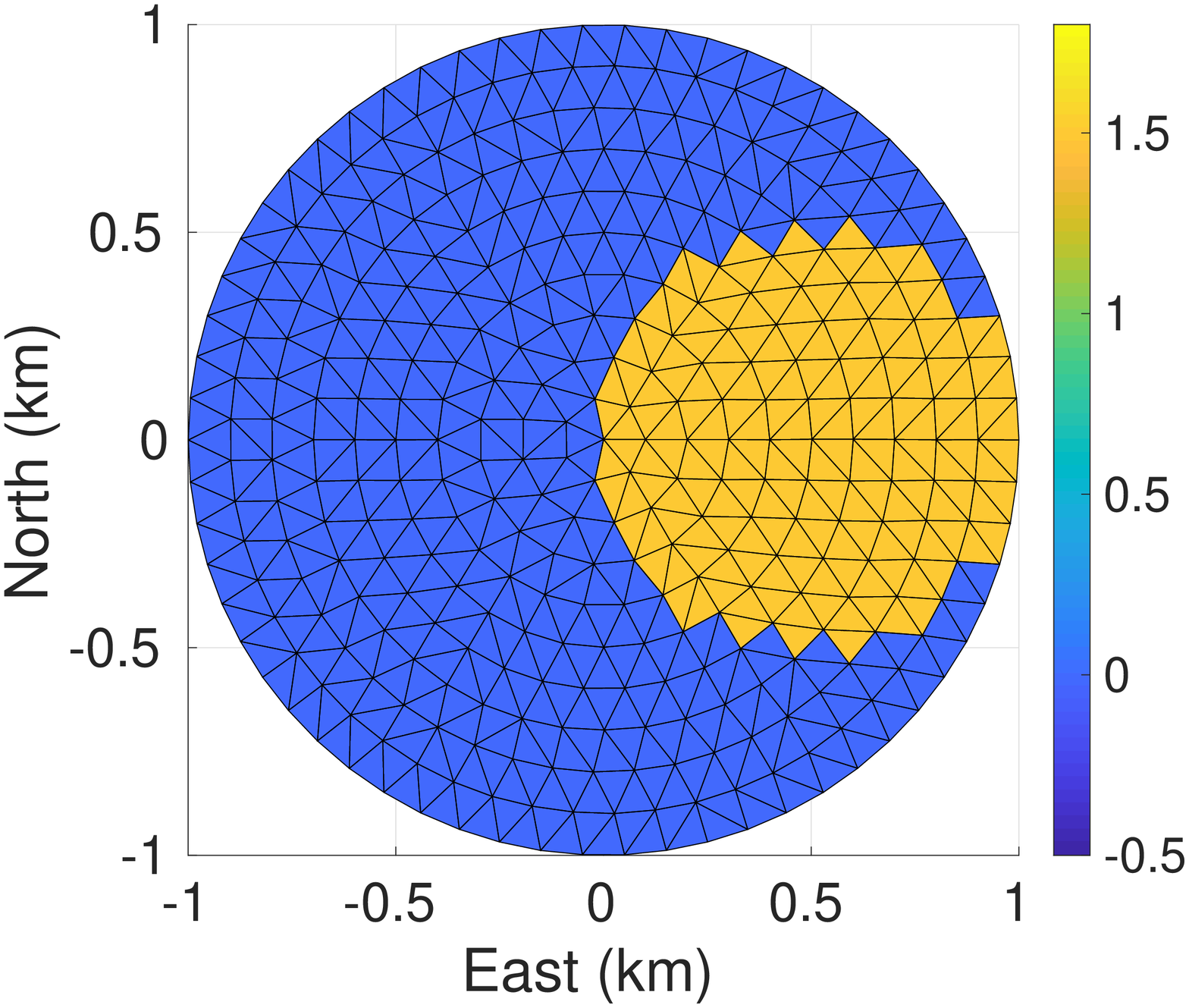}
}\\\hspace*{-0.8cm} 
\subfigure[Surface synthetic displacements]
{\includegraphics[scale=0.19]
{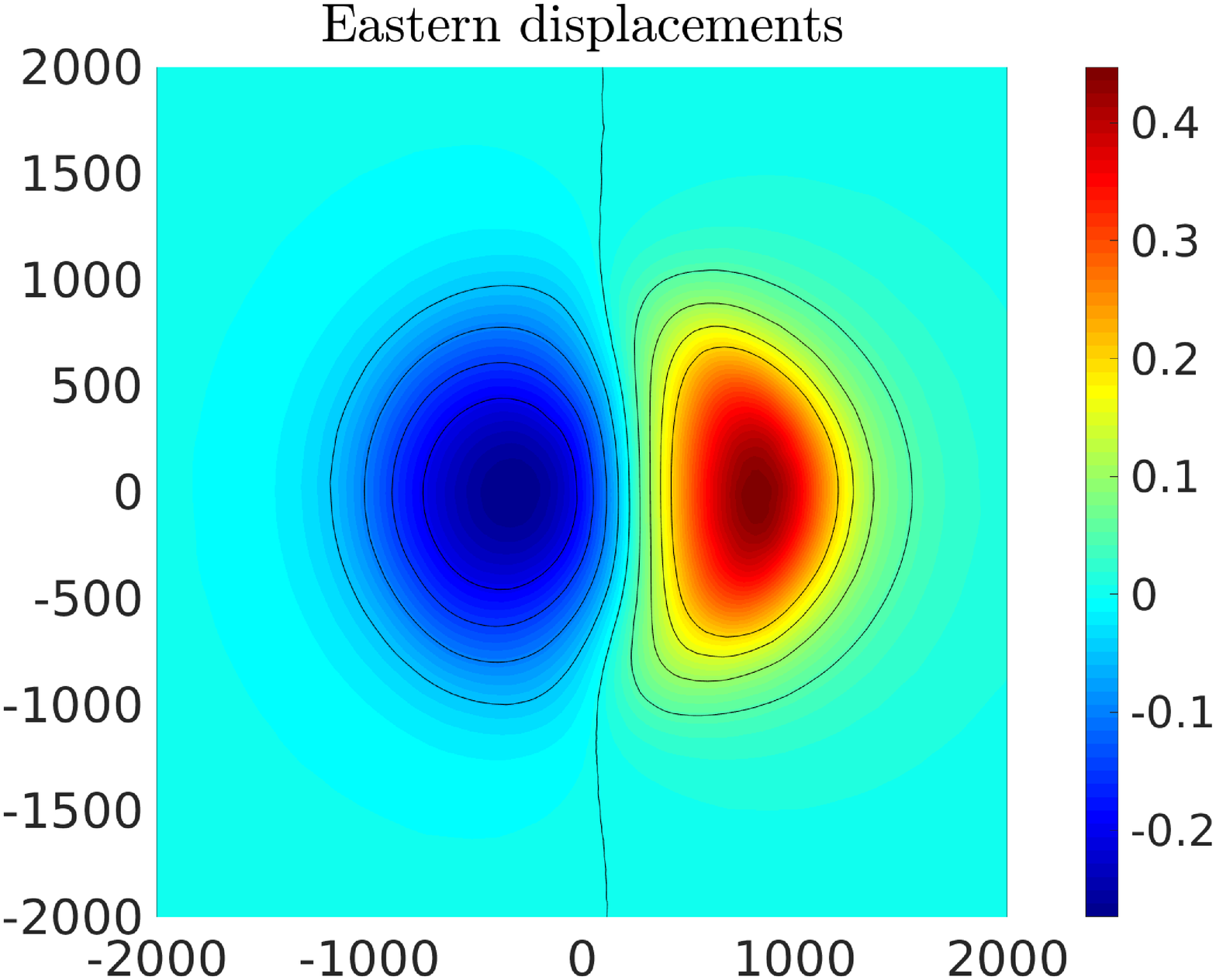}}
\hspace*{-0.5cm} 
\subfigure
{\includegraphics[scale=0.19]
{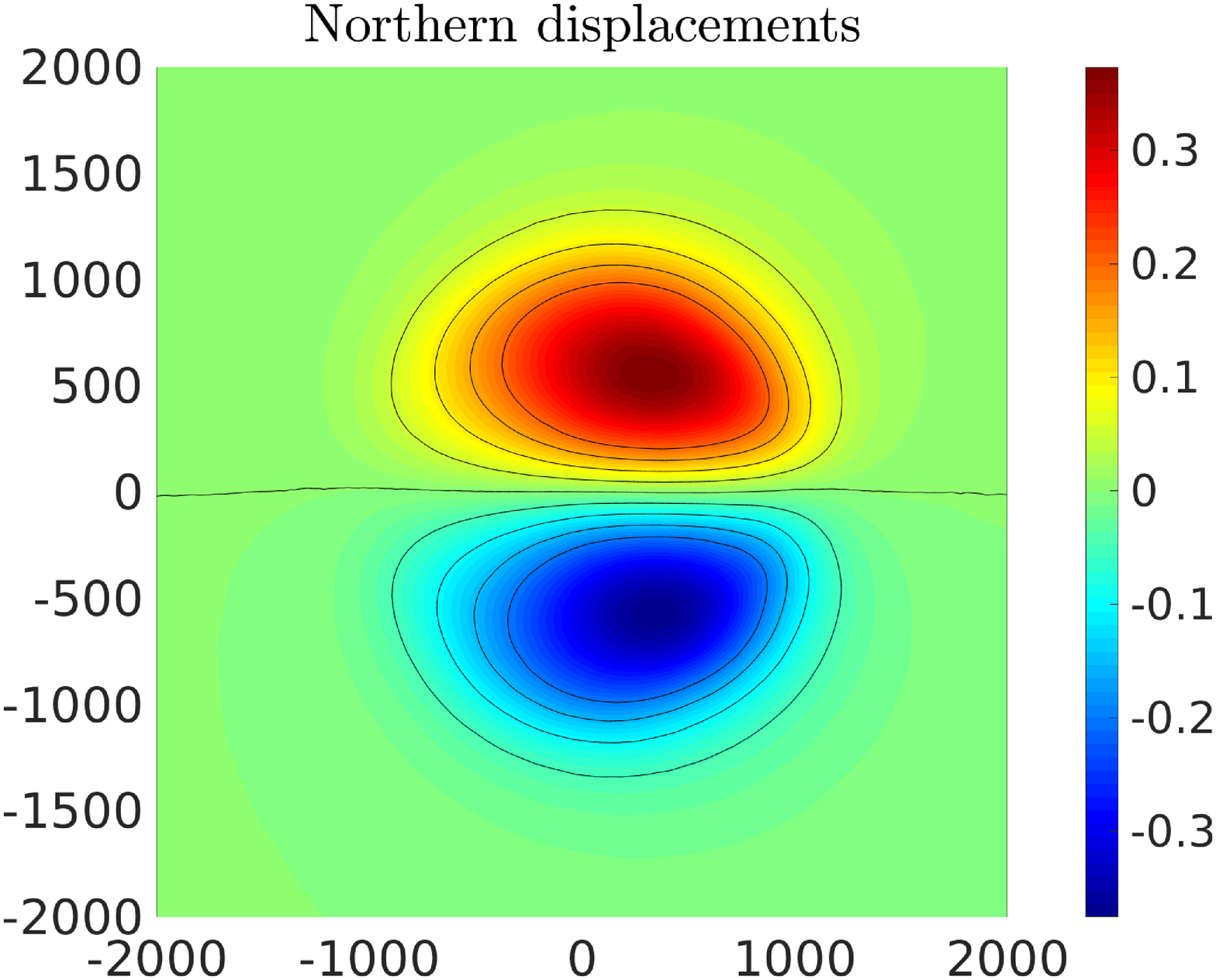}}
\hspace*{-0.5cm} 
\subfigure
{\includegraphics[scale=0.19]
{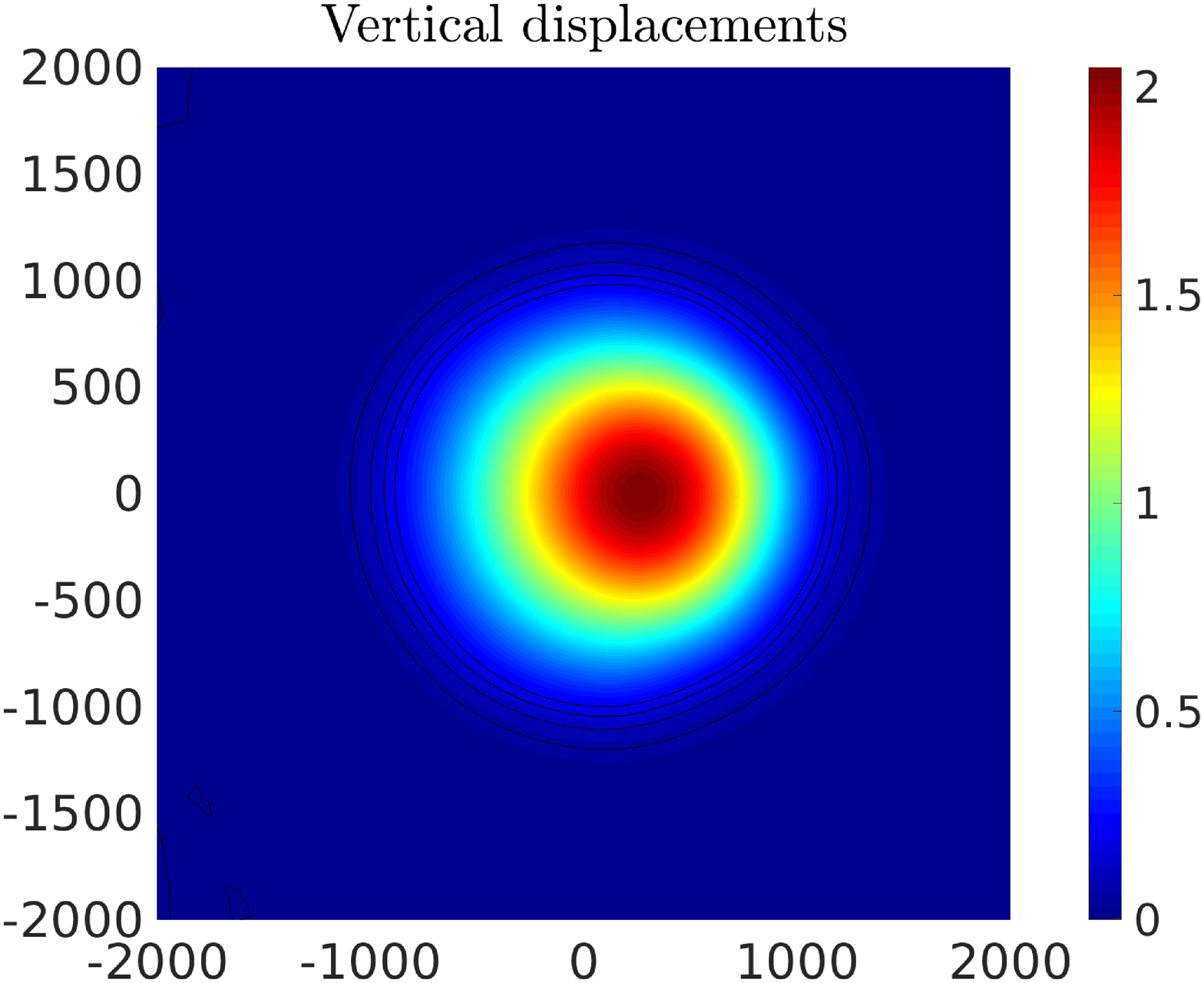}}
\caption{Configuration of the synthetic test corresponding to a horizontal circular fracture beneath a f\/lat topography. (a) Progressive mesh used. The mesh has a radius of 100 km. The fine mesh region has a 1.5 km radius; the intermediate mesh size goes from 1.5 to 15 km radius; further away the mesh gets coarser. (b) The source is a pressurized disk with a 1 km radius  located at different depths. The yellow patch is submitted to a normal traction $\t_{exact} = (0,0,1.5)$ MPa and the blue patch has a null traction $\t_{exact} = (0,0,0)$. (c) Surface synthetic displacements obtained by this pressure.}\label{topo_flat}
\end{figure}
\end{center}

The numerical application is performed by a synthetic test via simulation of a f\/lat volcano as follows: 
A semi-infinite elastic domain at center with radius $100$ km and extending down to $[0, −20]$ km (see Figure \ref{topo_flat}(a)), with the elasticity parameters: the Young modulus $E = 5000.0$ MPa and Poisson ratio $\nu = 0.25$.
A horizontal circular fracture is defined with radius $r = 1$ km and located at $(x, y, z) = (0, 0, −0.3)$ km in the topo f\/lat.
The mesh generation and level set implementation is explained in \cite{BCDK2020, BCDK2021}.
The ground surface is free to move and the other sides is f\/ixed i.e. satisfy the boundary conditions on $\Gn$ and $\Gd$ respectively. We generated a synthetic solution $\u_d$ associated to a discontinuous traction $\t^\pm = (0,0,\pm1.5) $ MPa, applied on a part of the fracture $\Gc$, as shown in Figure \ref{topo_flat}(b) (the yellow patches) and $\t^\pm = (0,0,0)$ on the rest of the fracture (the blue patches). In some numerical experiments, this traction is interpreted as a pressure $\p^\pm = 1.5$ MPa applied on the yellow zone.  
Finding a robust and ef\/f\/icient method to compute appropriate
regularization parameters $\alpha_0$ and $\alpha_1$ for given mesh and noisy data, is not straightforward. The admissible solution $\t$ is dependent on the chosen $\alpha_0$ and $\alpha_1$ and it is crucial to obtain a good approximation of the solution to the optimization problem.
For choosing the optimal parameters, a series of tests is organized by using the graphical tools. Indeed, three important criteria are compared by varying $\alpha_0$ and $\alpha_1$ as follows:
we set $\alpha_0 \in \{ 1.0\ee-12, 1.0\ee-11, \ldots, 1.0\ee0\}$
and $\alpha_1\in \{1.0\ee-5, 1.0\ee-4, \ldots, 1.0\ee5\}$.
Then we compare: 
A relative ground error defined by:
\begin{equation*}
\text{E}_\u =\frac{\ds \int_{\Gn} \lvert \u -\u_d\rvert^2\dd \Gn}{\ds\int_{\Gn} \lvert \u_d\rvert^2\dd \Gn}\times 100.
\end{equation*}   
A relative error of traction $\t$ on the fracture source, defined by:
\begin{equation*}
\text{E}_\t =\frac{\ds \sum \lvert \t -\t_{exact}\rvert^2}{\ds\sum \lvert \t_{exact}\rvert^2\dd \Gc}\times 100.
\end{equation*}   
Note that an inevitable numerical error is already produced to compute the traction $\t$ on the fracture source. Using the non-conformal mesh to implement the fracture by level set method, is the reason of this error which can be reduced by using a finer mesh around the fracture.\\ 
finally, the iteration number for a given $\epsilon = 1.0\ee-14$ in Algorithm~\ref{IP_DDM1} to satisfying the convergence criteria
\begin{equation*}
\frac{\ds \int_{\Gc} \lvert \g^{k+1}\rvert^2\dd \Gc}{\ds\int_{\Gc} \lvert \g^{0}\rvert^2\dd \Gc}< \epsilon .
\end{equation*}

\begin{center}
\begin{figure}[htp!]\hspace*{-0.2cm}   
\subfigure{\includegraphics[scale=0.18]
{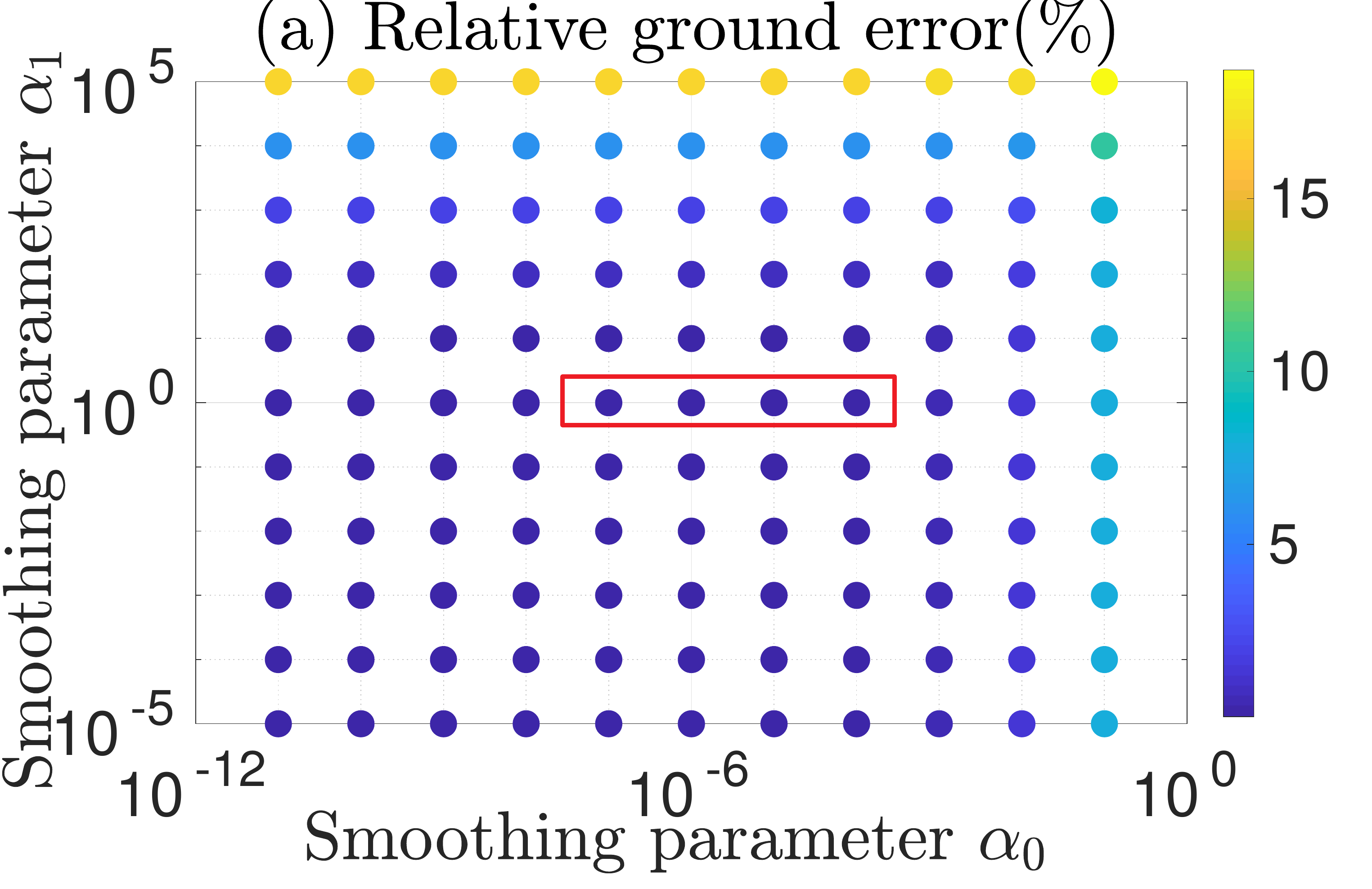}}\hspace*{-0.3cm}  
\subfigure{\includegraphics[scale=0.17]
{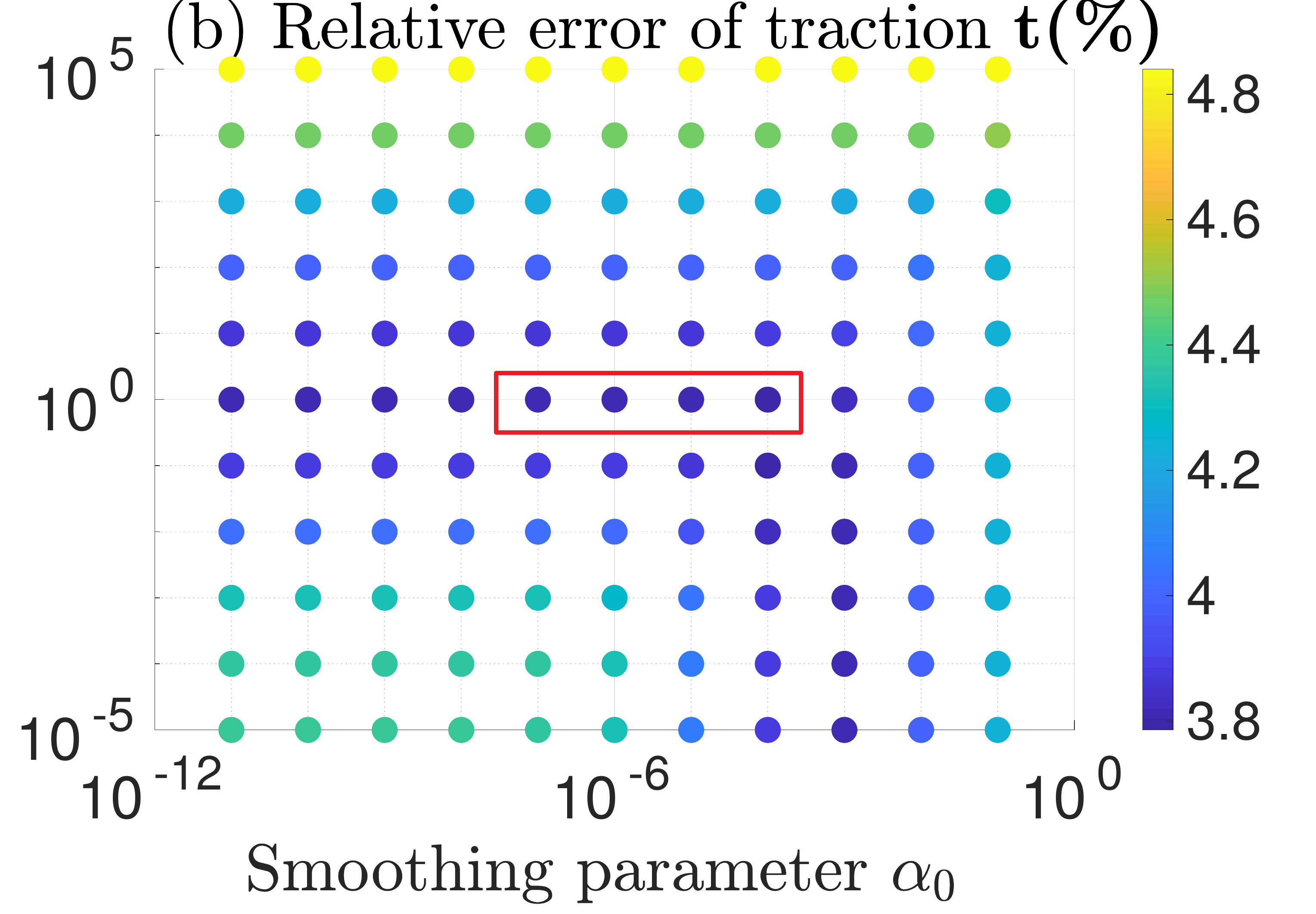}}
\hspace*{-0.45cm} 
\subfigure{\includegraphics[scale=0.18]
{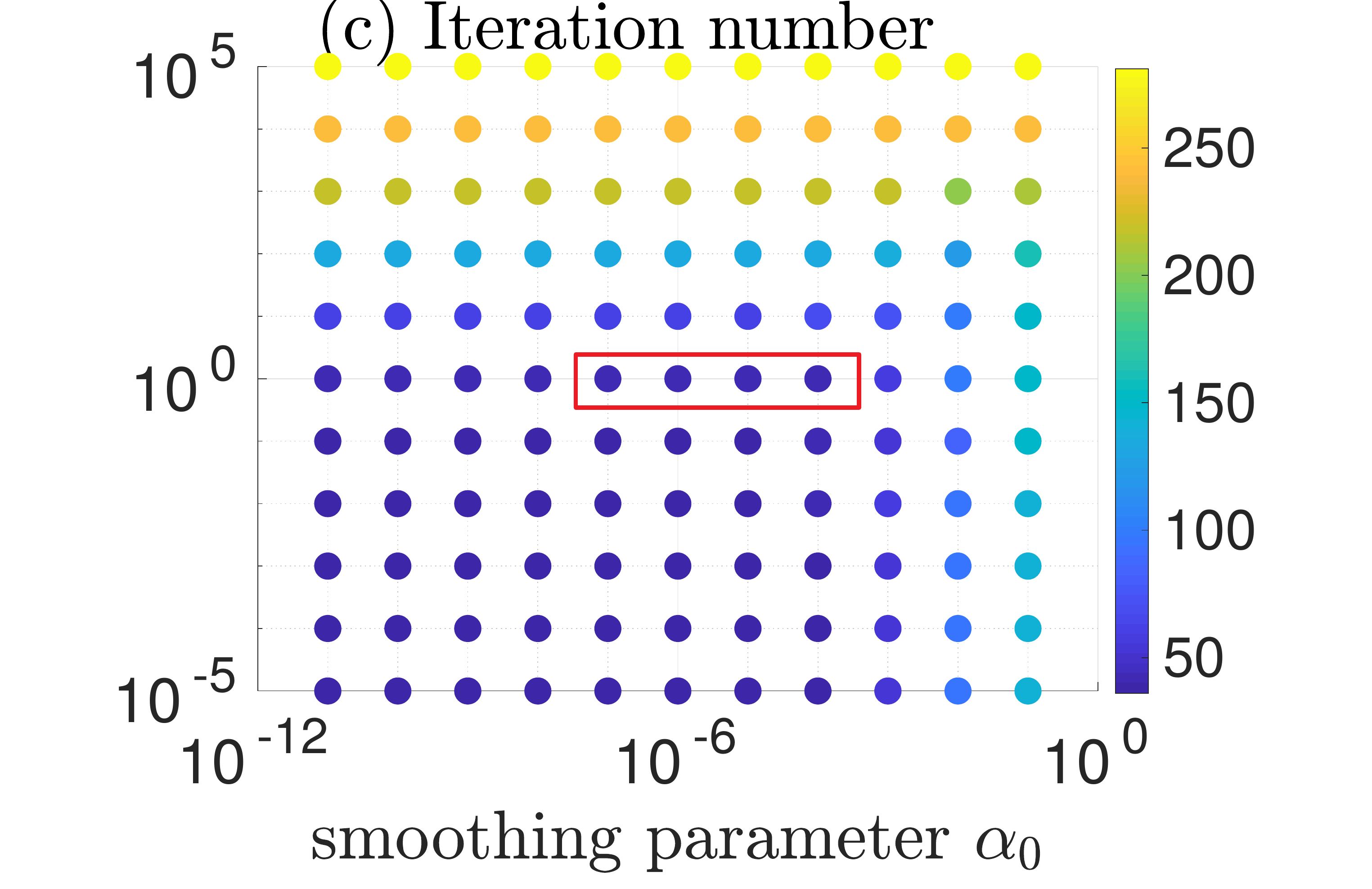}}
\caption{Systematic exploration of the smoothing parameters $\alpha_0$ and  $\alpha_1$ for minimizing the cost function in equation \eqref{min_prob}. The source is located at  $-0.3$ km depth beneath the flat topography. The acceptable combination of smoothing parameters is found by comparing (a) relative ground error, (b) relative error of traction on the disk and (c) iteration number. The source has $291$ unknowns. The best compromises are indicated by
the red boxes.}\label{optim_alpha_param1}
\end{figure}
\end{center}
\begin{figure}[htp!]
\centering
 \includegraphics[scale=0.5]{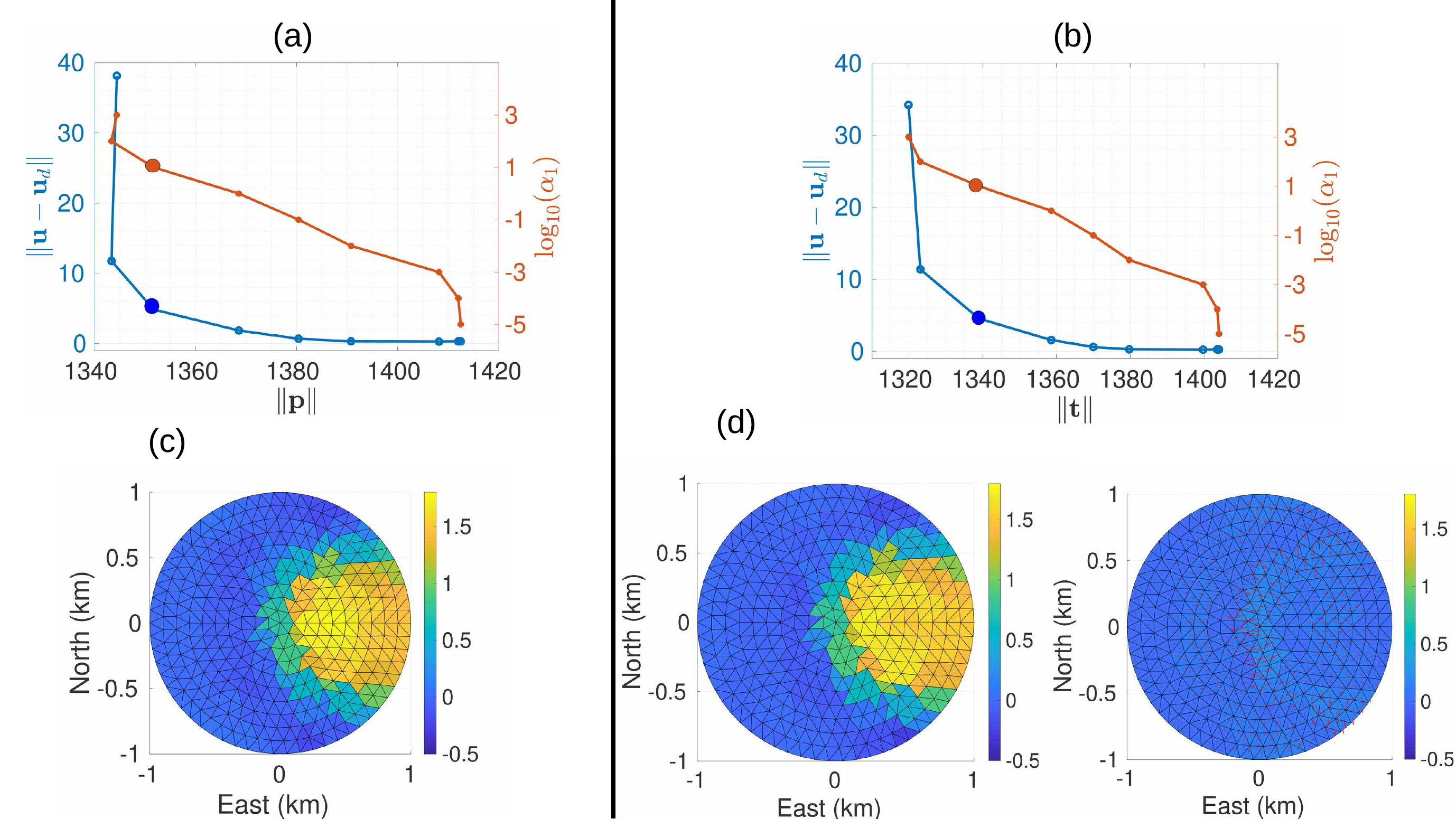}
 \caption{L-curves used to find $\alpha_1$ representing the best compromise between the data fit (equation \eqref{misfit}) and the smoothing (equation \eqref{smoothing}) in the cost function (equation \eqref{min_prob}). The fracture is located at $-0.3$ km.  (a) L-curve when solving for pressure $\p$. The larger points indicate the best compromise.  (b) L-curve when solving for traction $\t$. The larger points indicate the best compromise.  (c) Fracture pressure corresponding to the best $\alpha_1=1.0\ee1$.   (d) Normal and (e) tangential tractions projected on the fracture corresponding to the best $\alpha_1=1.0\ee1$. Here, we set $\alpha_0 = 1.0\ee-7 $.}\label{L_curve_300}
\end{figure}

The numerical tests presented in Figure \ref{optim_alpha_param1},   
show that, $\alpha_1 = 1$ can be chosen as an acceptable parameter. Moreover, it seems, the above criteria are less sensitive to $\alpha_0$ between $1.0\ee-7$ and $1.0\ee-4$.
Accordingly, we employ L-curve, another graphical tool,
to trade of\/f between two criteria:
the misfit  
\begin{equation}\label{misfit}
\lVert \u-\u_{d} \rVert=\big(\int_{\Gn} \lvert\u-\u_{d}\lvert^2 \dd \Gn\big)^{\frac{1}{2}},
\end{equation} 
and the norm of the traction or smoothing
\begin{equation}\label{smoothing}
\lVert \t \rVert=\big(\ds\int_{\Gc}\lvert \t\rvert^2\dd \Gc)^{\frac{1}{2}}.
\end{equation}
The L-curves are depicted by setting $\alpha_0 = 1.0\ee-7$ and varying $\alpha_1$. Closest point to the origin shows the best compromise between misf\/it and smoothing, and it will be the acceptable $\alpha_1$. In particular case when we aim to f\/ind the pressure, the traction $\t$  may replaced by the pressure $\p$. In Figure \ref{L_curve_300}, for the above fracture and mesh, two L-curves corresponding to both  pressure $\p$ and traction $\t$ are illustrated.  

\begin{center}
\begin{figure}[htp!]\hspace*{-0.5cm}   
\subfigure[$\alpha_1=1.0\ee{-1}$]{\includegraphics[scale=0.205]
{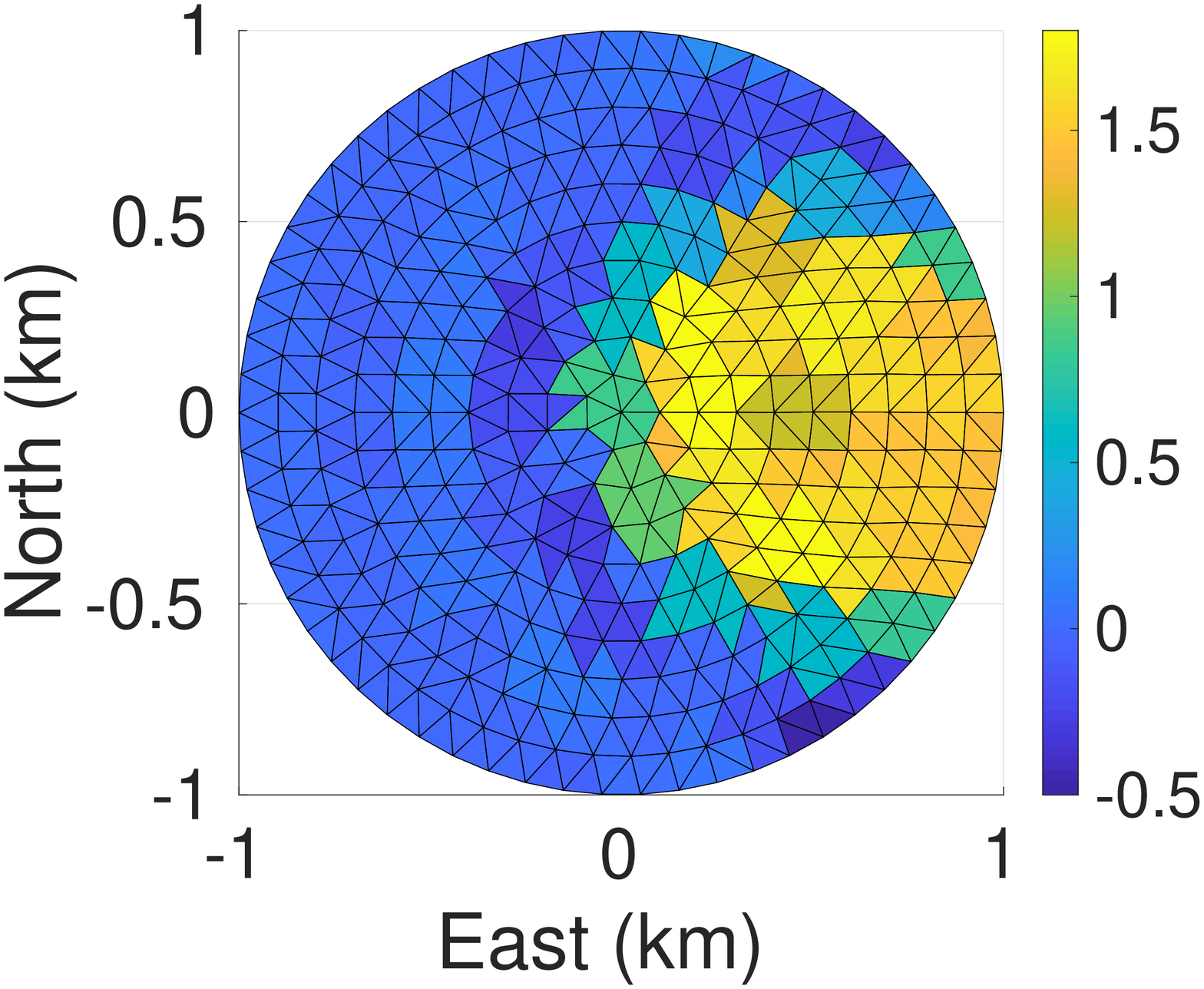}}\hspace*{-0.2cm}  
\subfigure[$\alpha_1=1.0\ee0$]{\includegraphics[scale=0.205]
{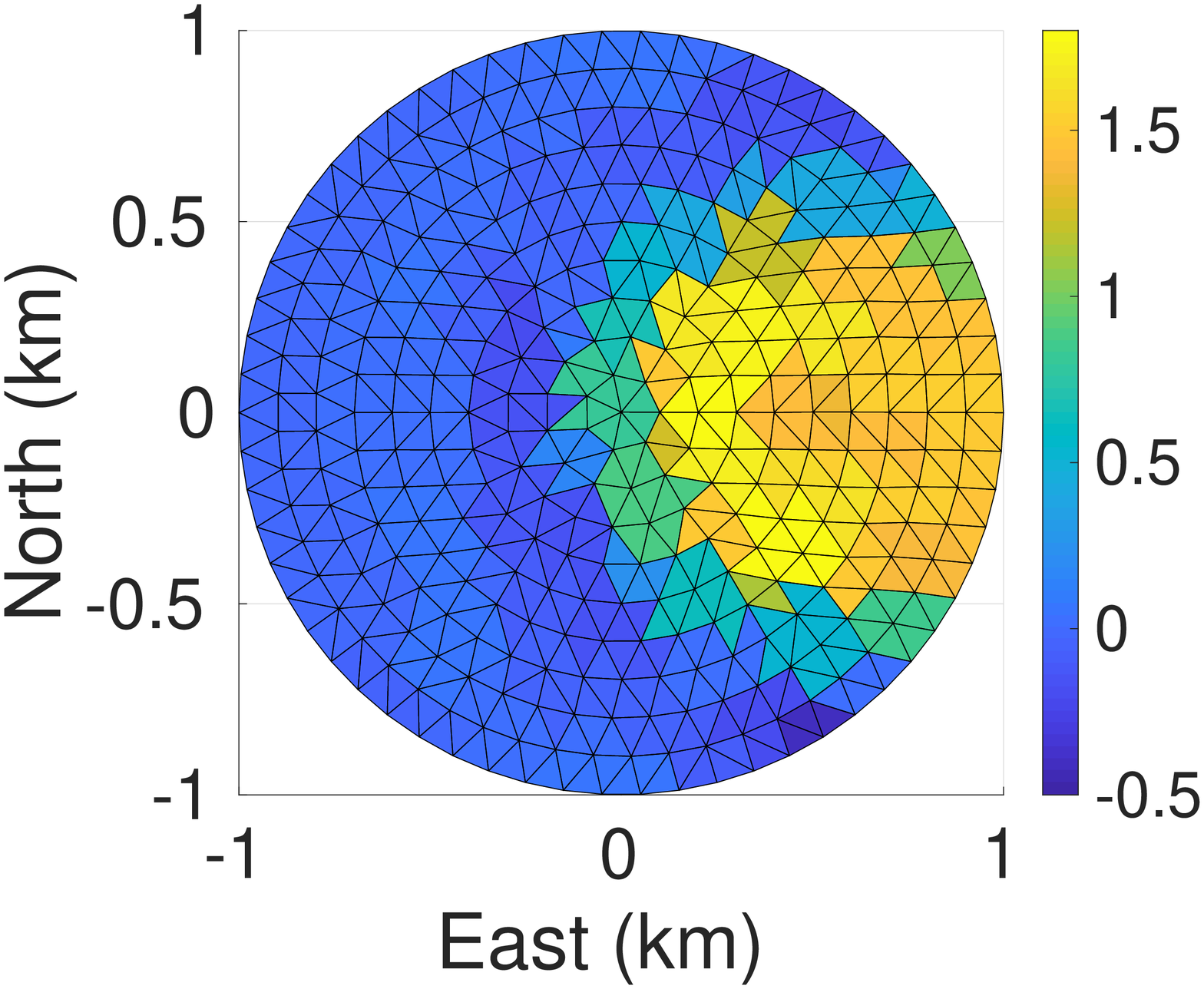}}
\hspace*{-0.2cm}  
\subfigure[$\alpha_1=1.0\ee2$]{\includegraphics[scale=0.205]
{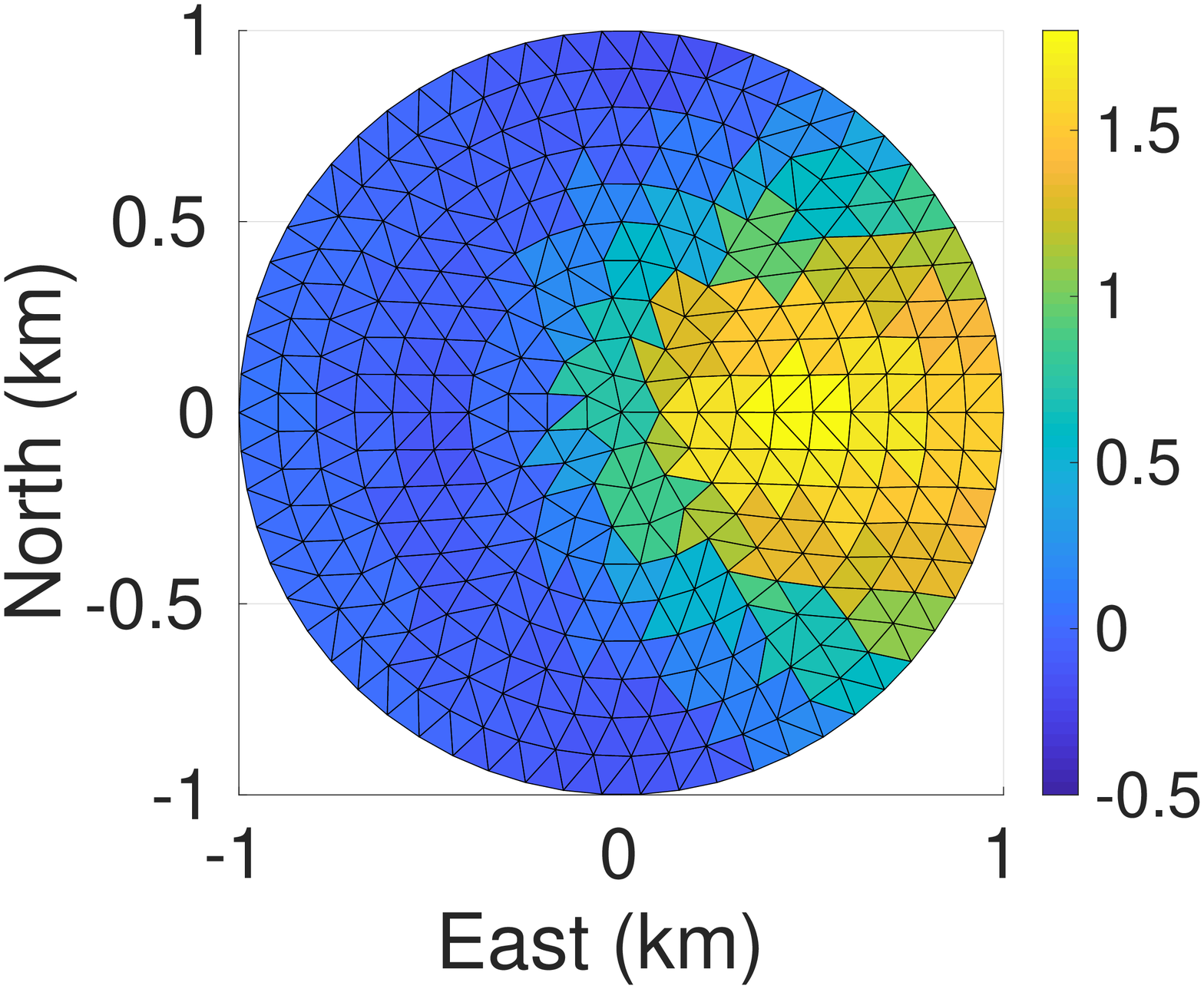}}
\caption{Fracture pressure located at $-0.3$ km, corresponding to (a) $\alpha_1=1.0\ee{-1}$, (b) $\alpha_1=1.0\ee0$ and (c) $\alpha_1=1.0\ee2$}\label{solution_source_300}
\end{figure}
\end{center}

\begin{center}
\begin{figure}\hspace*{-0.45cm}   
\subfigure[$\alpha_1=1.0\ee0$]{\includegraphics[scale=0.2]
{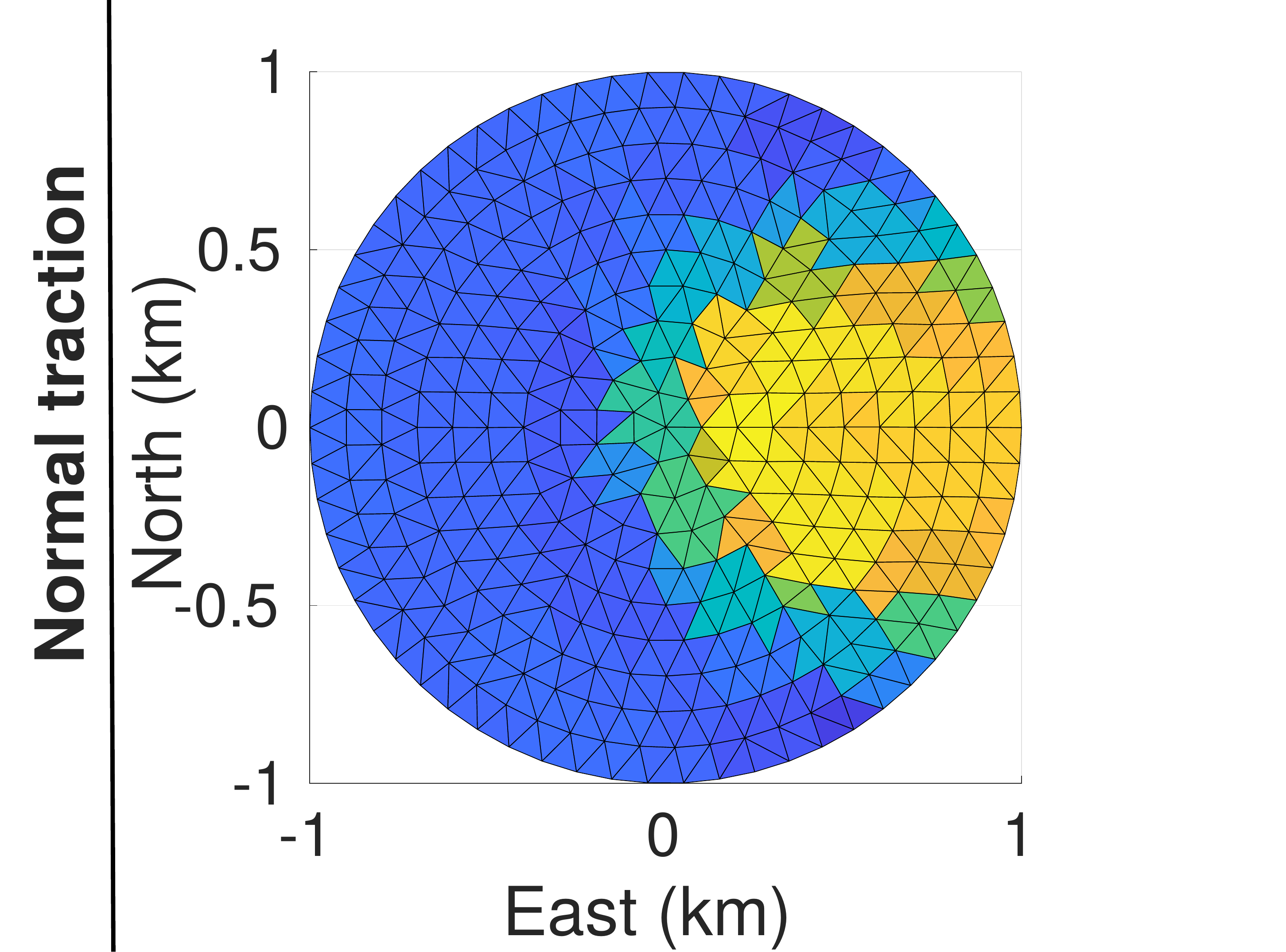}}\hspace*{-1.0cm}  
\subfigure[$\alpha_1=1.0\ee1$]{\includegraphics[scale=0.2]
{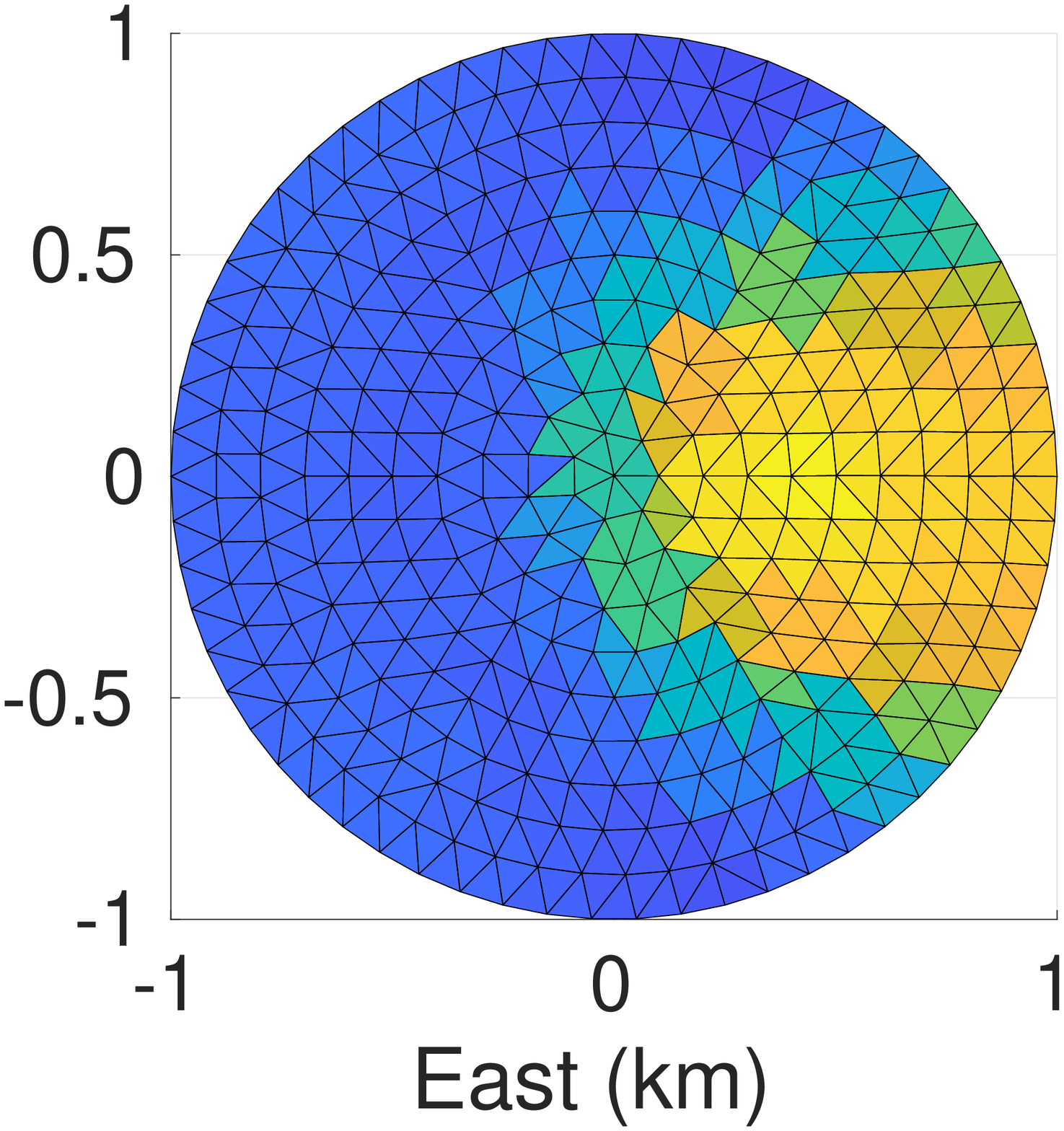}}
\hspace*{-1.0cm}  
\subfigure[$\alpha_1=1.0\ee2$]{\includegraphics[scale=0.2]
{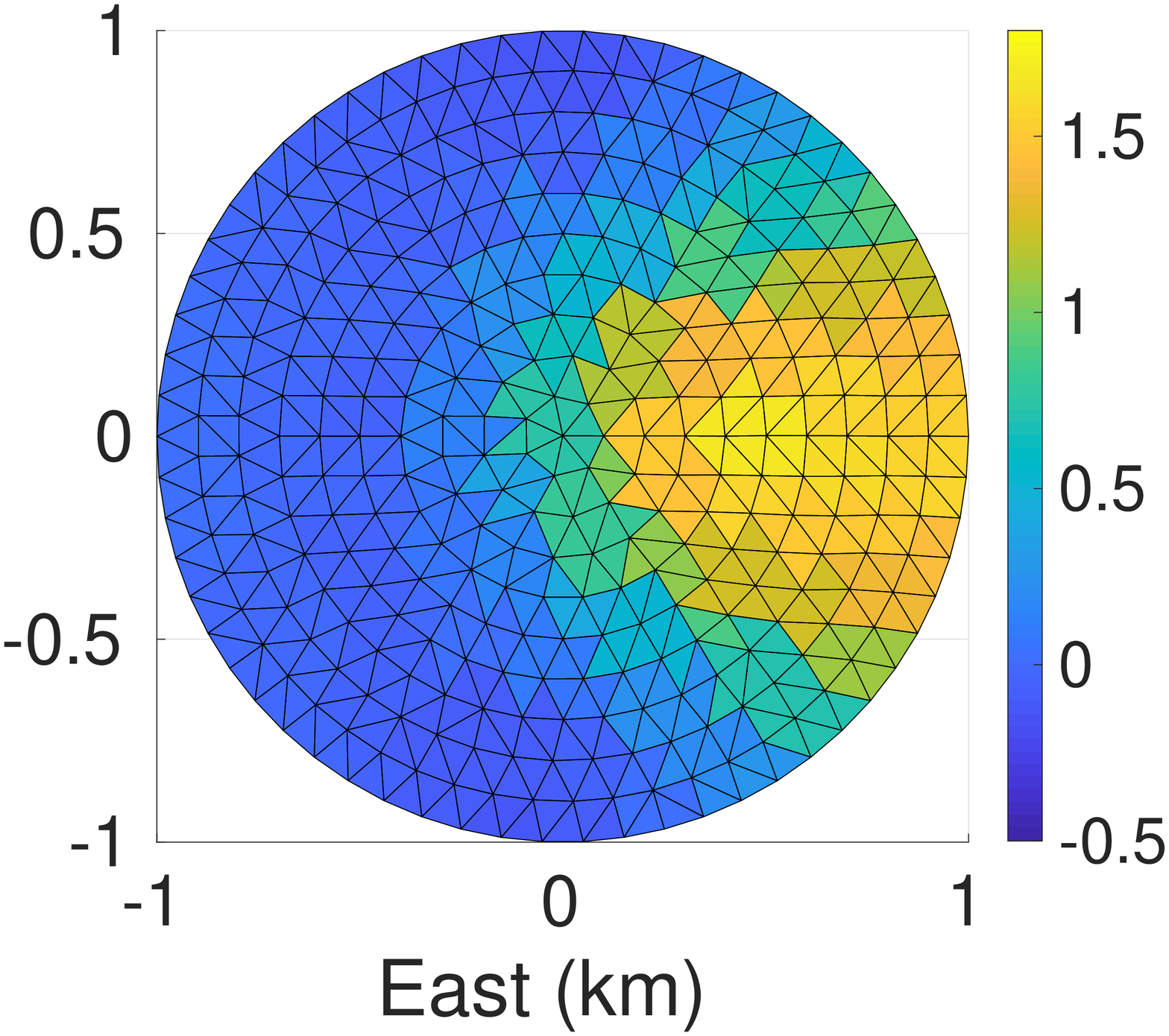}}\\\hspace*{-0.5cm} 
\subfigure{\includegraphics[scale=0.20]
{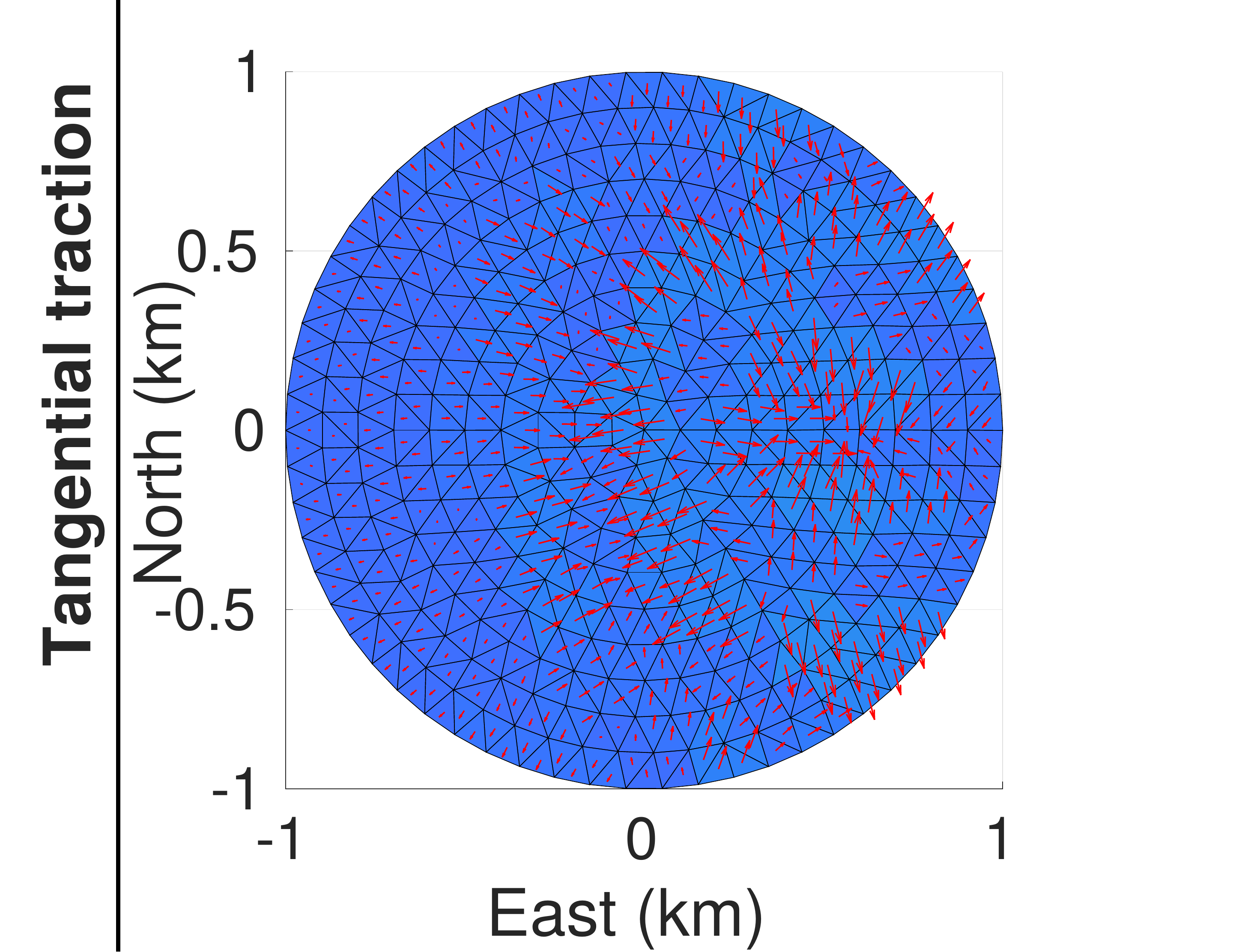}}\hspace*{-1.0cm}  
\subfigure{\includegraphics[scale=0.20]
{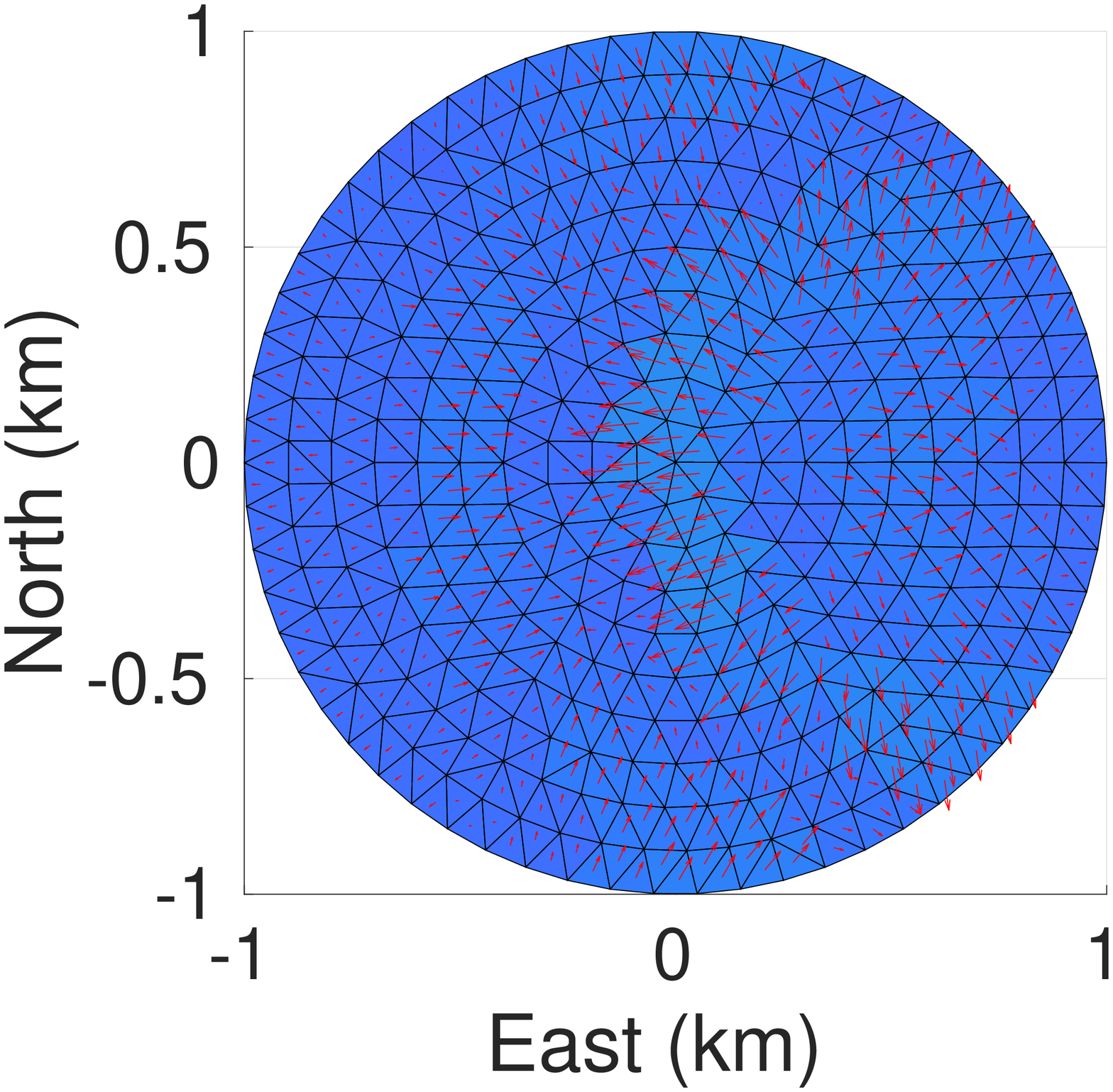}}
\hspace*{-1.0cm}  
\subfigure{\includegraphics[scale=0.20]
{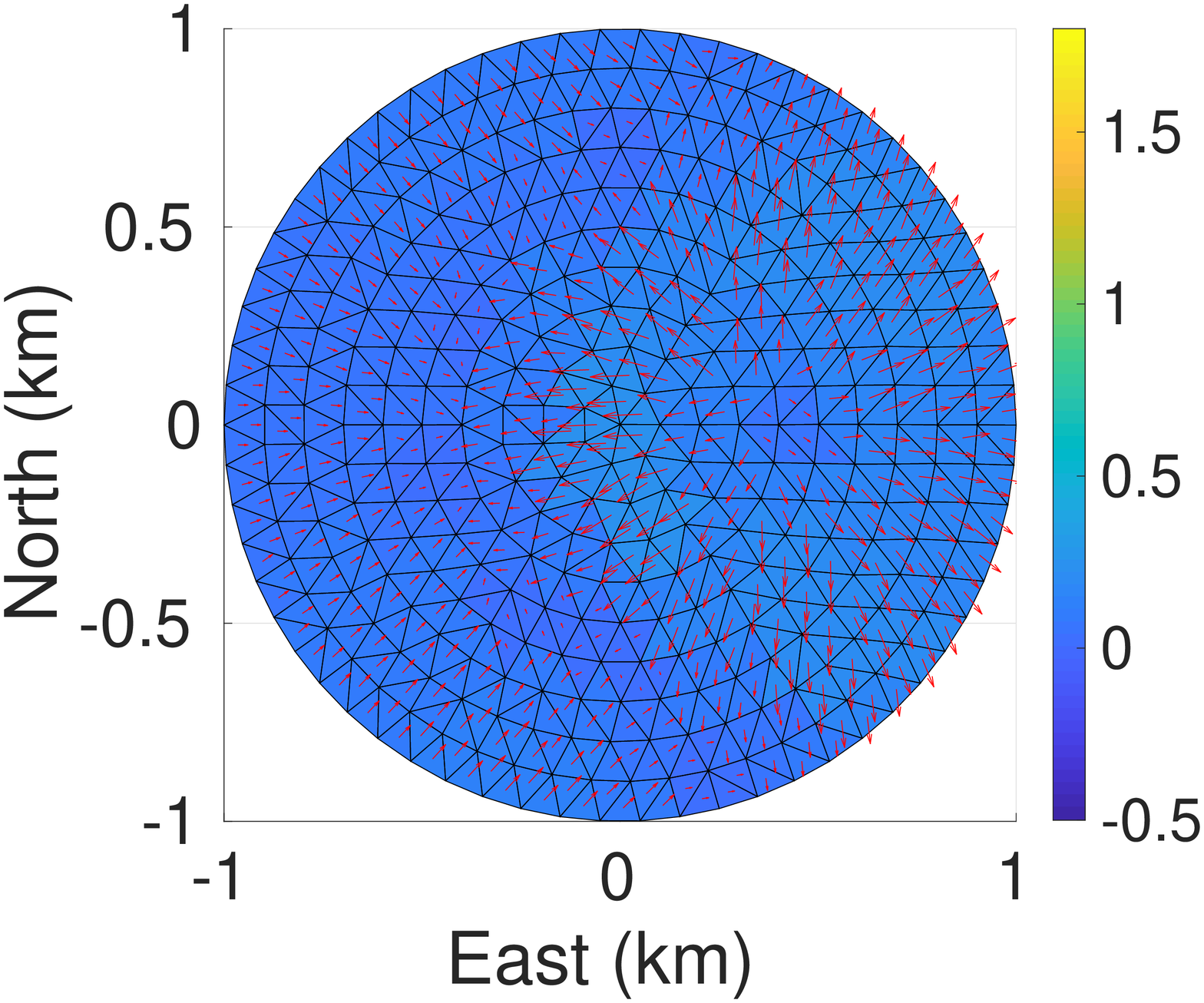}}
\caption{Normal and tangential tractions projected on the fracture located at $-0.3$ km,  corresponding to (a) $\alpha_1=1.0\ee0$, (b) $\alpha_1=1.0\ee1$ and (c) $\alpha_1=1.0\ee2$. The directions of the tangetial stress are shown by red vectors.}\label{fig:solution_source_traction_300}
\end{figure}
\end{center}

Fracture pressure corresponding to $\alpha_1=1.0\ee1$ as an appropriate choice is presented in Figure \ref{L_curve_300}(c).
As shown in Figures 
\ref{solution_source_300} and \ref{fig:solution_source_traction_300}, 
in 3D realistic volcanoes where we interested in large-scale problem, the values around the $1.0\ee1$ are yet acceptable.
Normal and tangential tractions are depicted in Figure \ref{fig:solution_source_traction_300}. Note that we suppose $(0,0,-1)$ as the unit normal vector $\n$.   

%

\begin{center}
\begin{figure}[htp!]\hspace*{-0.5cm}  
\subfigure{\includegraphics[scale=0.17]
{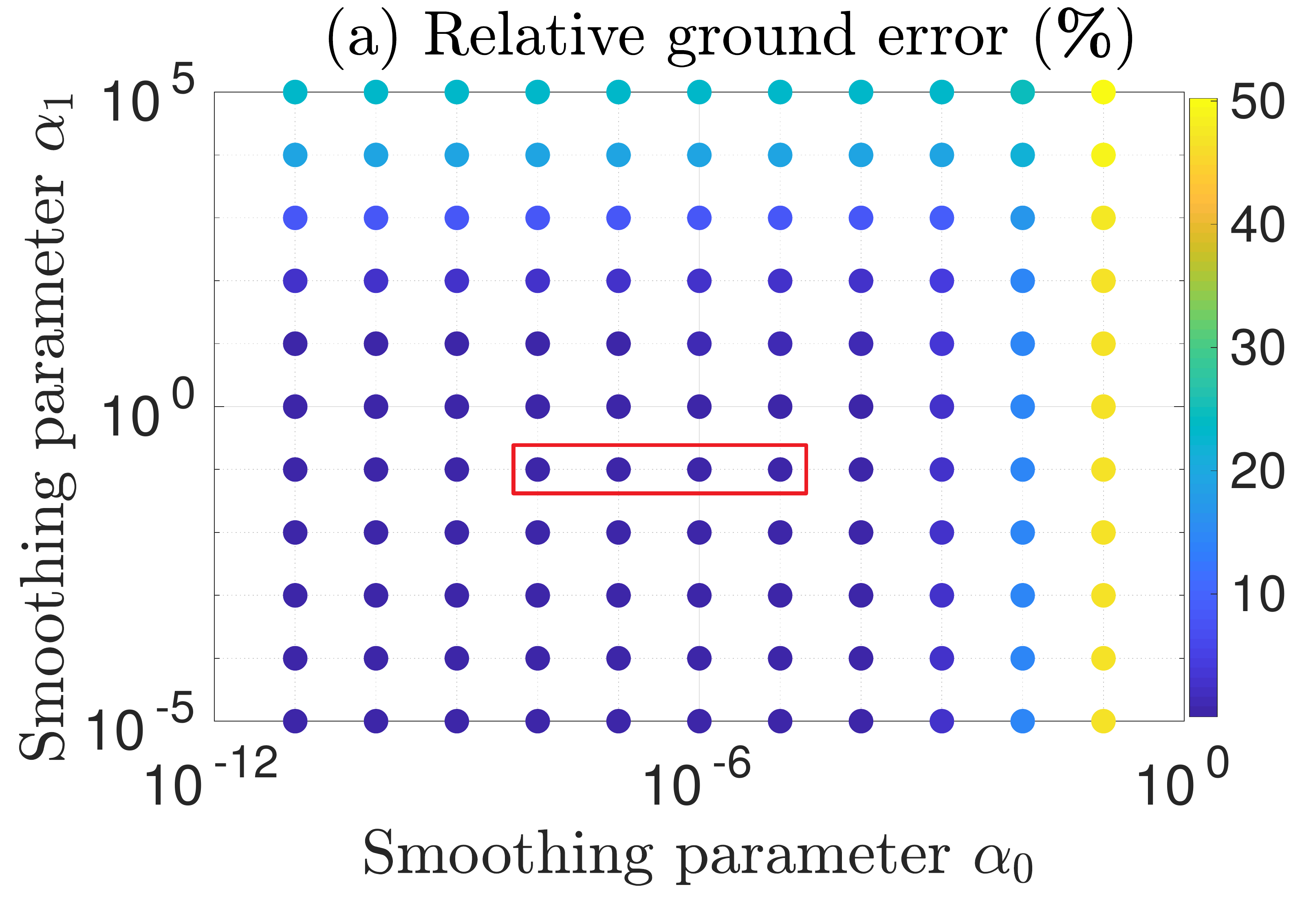}}
\hspace*{-0.26cm} 
\subfigure{\includegraphics[scale=0.17]
{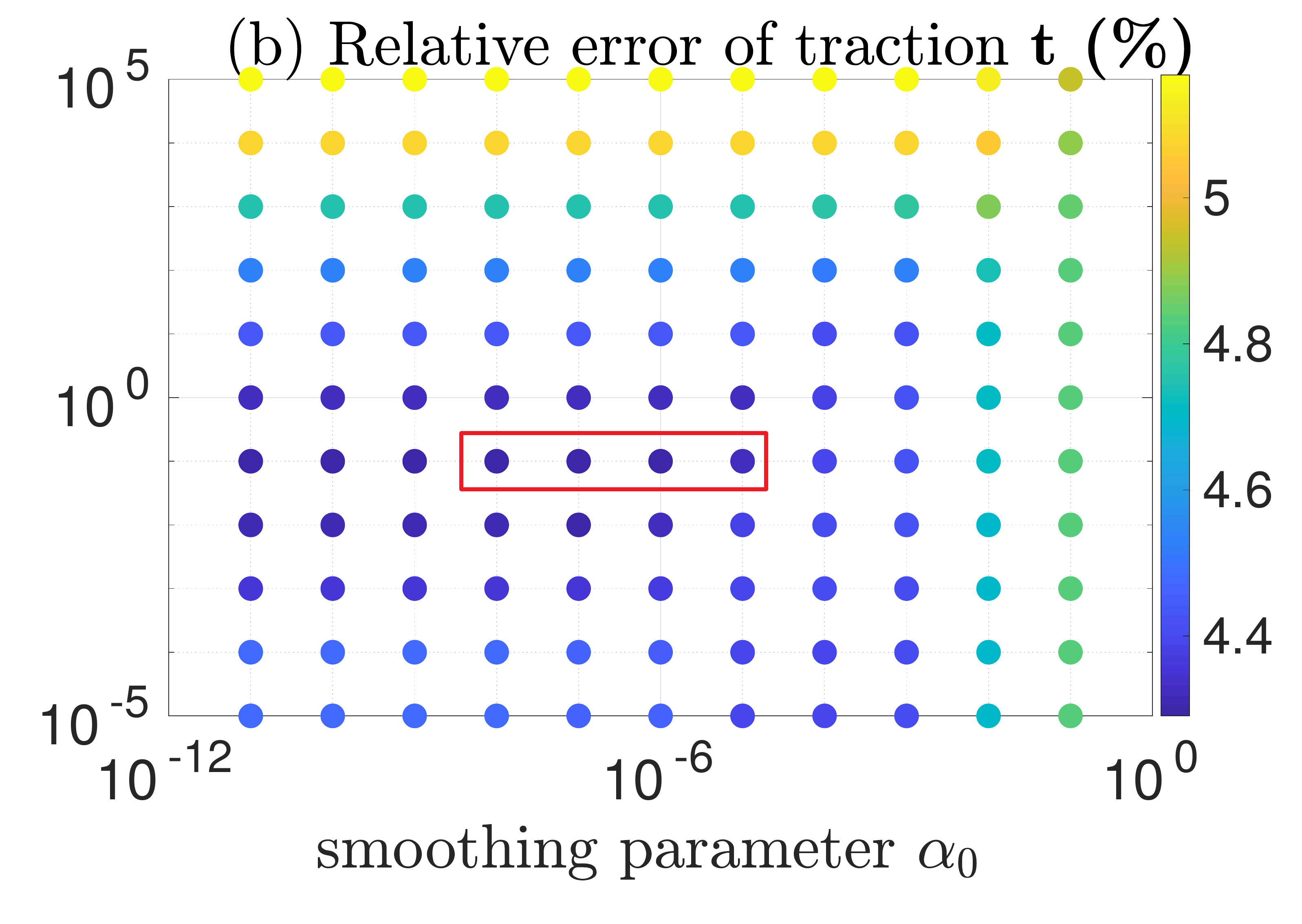}}
\hspace*{-0.3cm} 
\subfigure{\includegraphics[scale=0.17]
{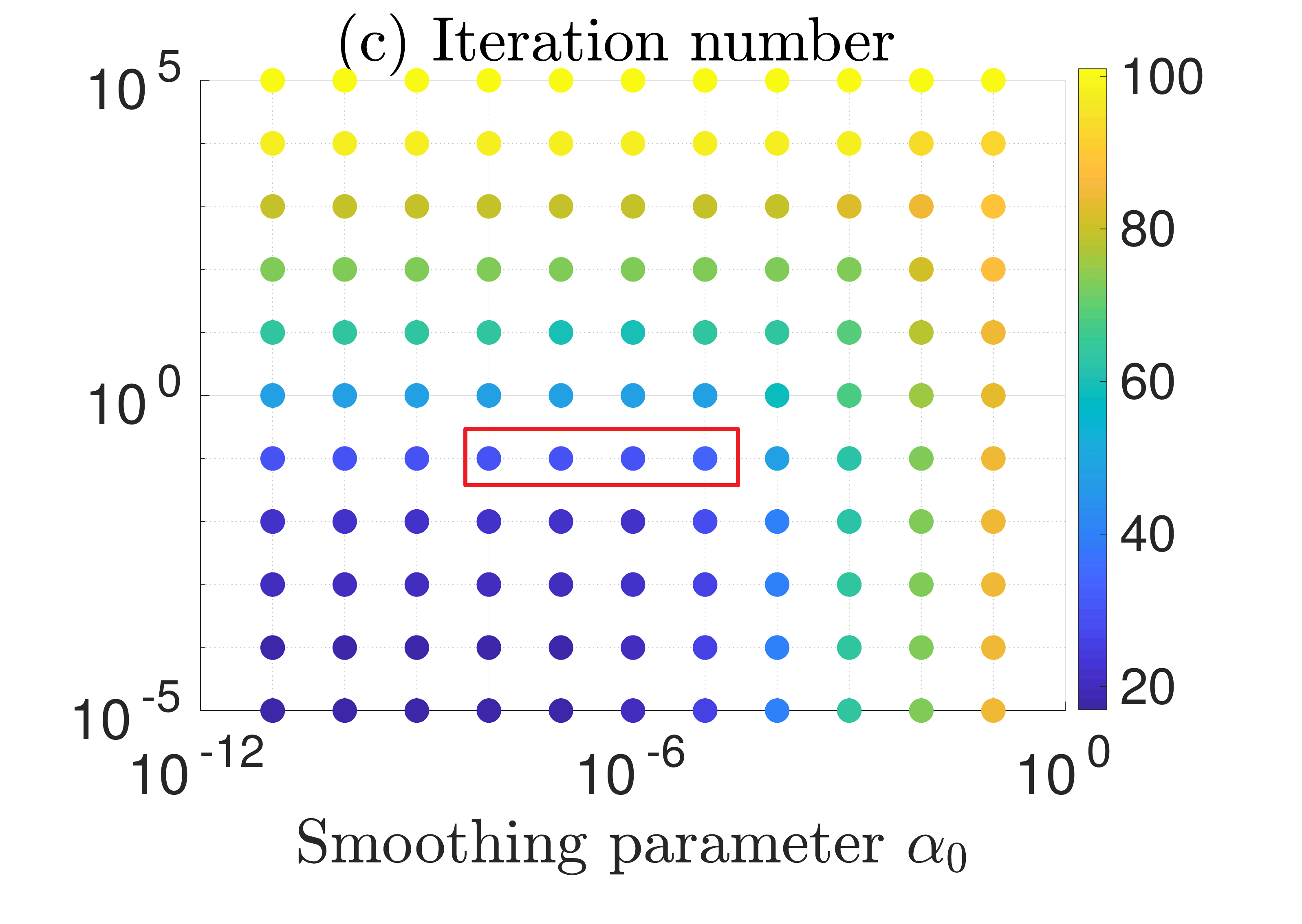}}
\caption{Systematic exploration of the smoothing parameters $\alpha_0$ and  $\alpha_1$ for minimizing the cost function in equation \eqref{min_prob}. The source is located at  $-0.9$ km depth beneath the flat topography. The acceptable combination of smoothing parameters is found by comparing (a) relative ground error, (b) relative error of traction on the disk and (c) iteration number. The source has $107$ unknowns. The best compromises are indicated by
the red boxes.}\label{optim_alpha_param2}
\end{figure}
\end{center}

\begin{figure}[htp!]
\centering
 \includegraphics[scale=0.5]{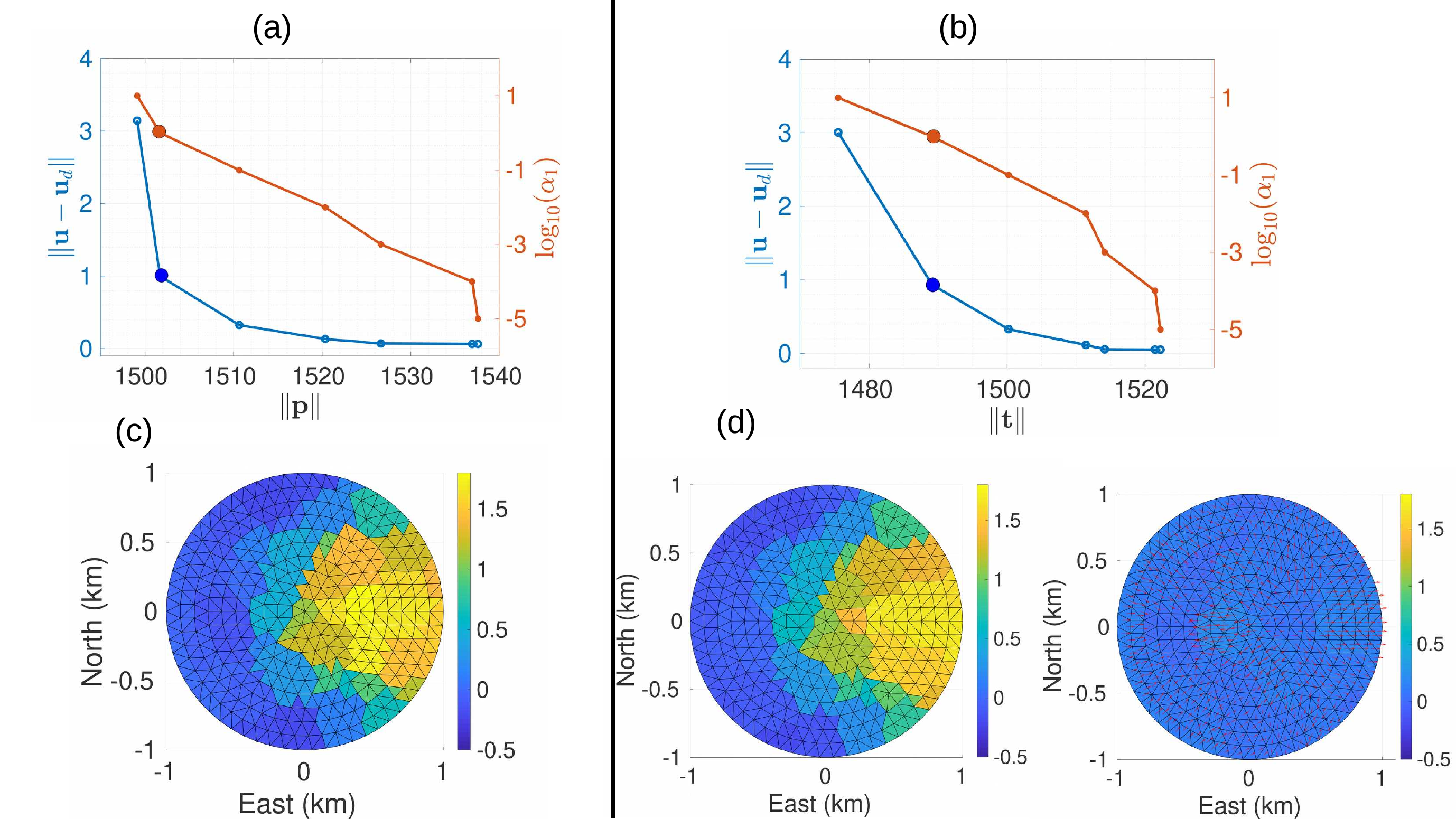}
 \caption{L-curves used to find $\alpha_1$ representing the best compromise between the data fit (equation \eqref{misfit}) and the smoothing (equation \eqref{smoothing}) in the cost function (equation \eqref{min_prob}). The fracture is located at $-0.9$ km.  (a) L-curve when solving for pressure $\p$. The larger points indicate the best compromise.  (b) L-curve when solving for traction $\t$. The larger points indicate the best compromise.  (c) Fracture pressure corresponding to the best $\alpha_1=1.0\ee1$.   (d) Normal and (e) tangential tractions projected on the fracture corresponding to the best $\alpha_1=1.0\ee0$. Here, we set $\alpha_0 = 1.0\ee-7 $.}\label{L_curve_900}
\end{figure}

\begin{center}
\begin{figure}[htp!]\hspace*{-0.5cm}   
\subfigure[$\alpha_1=1.0\ee{-1}$]{\includegraphics[scale=0.205]
{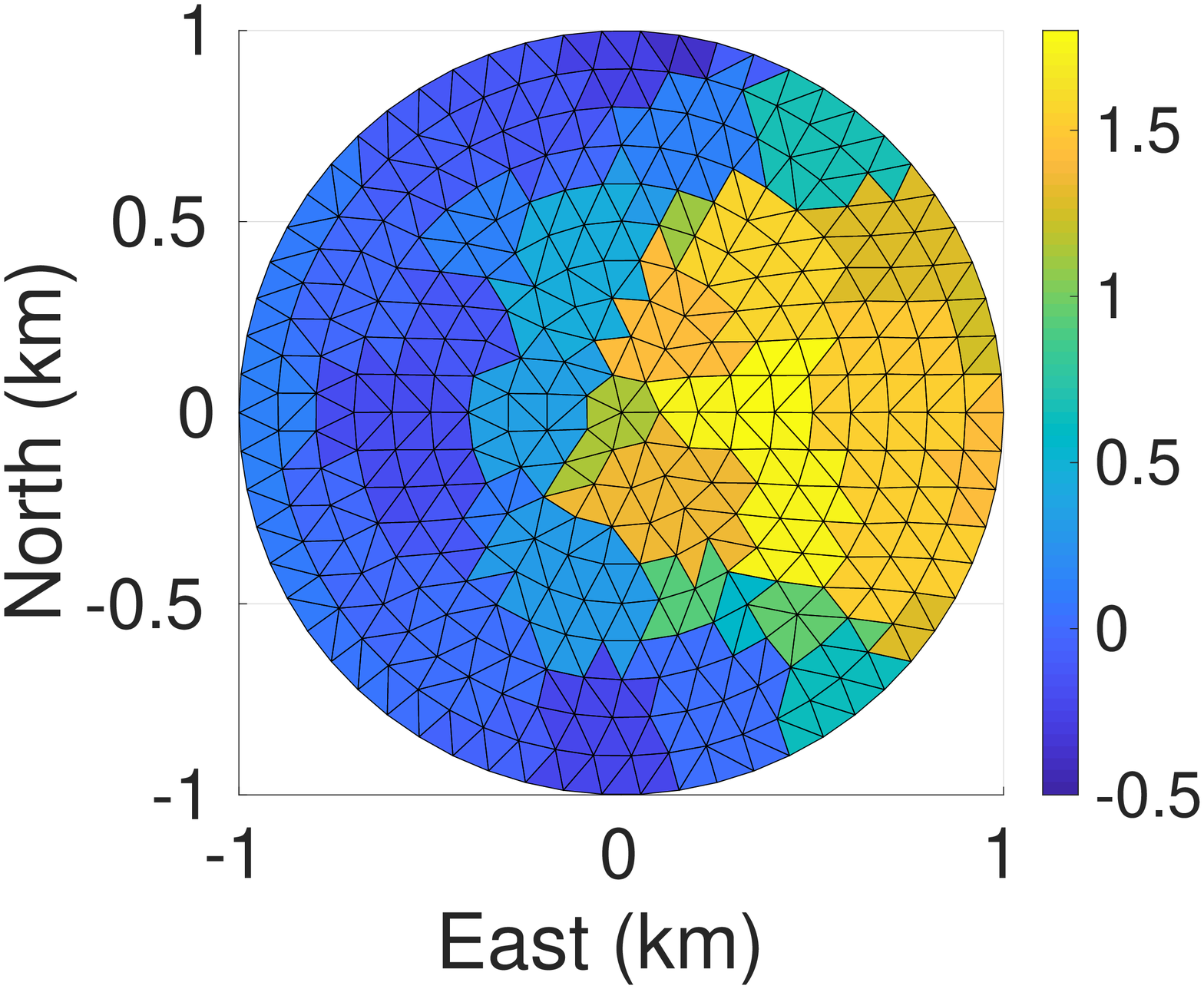}}\hspace*{-0.2cm}  
\subfigure[$\alpha_1=1.0\ee1$]{\includegraphics[scale=0.205]
{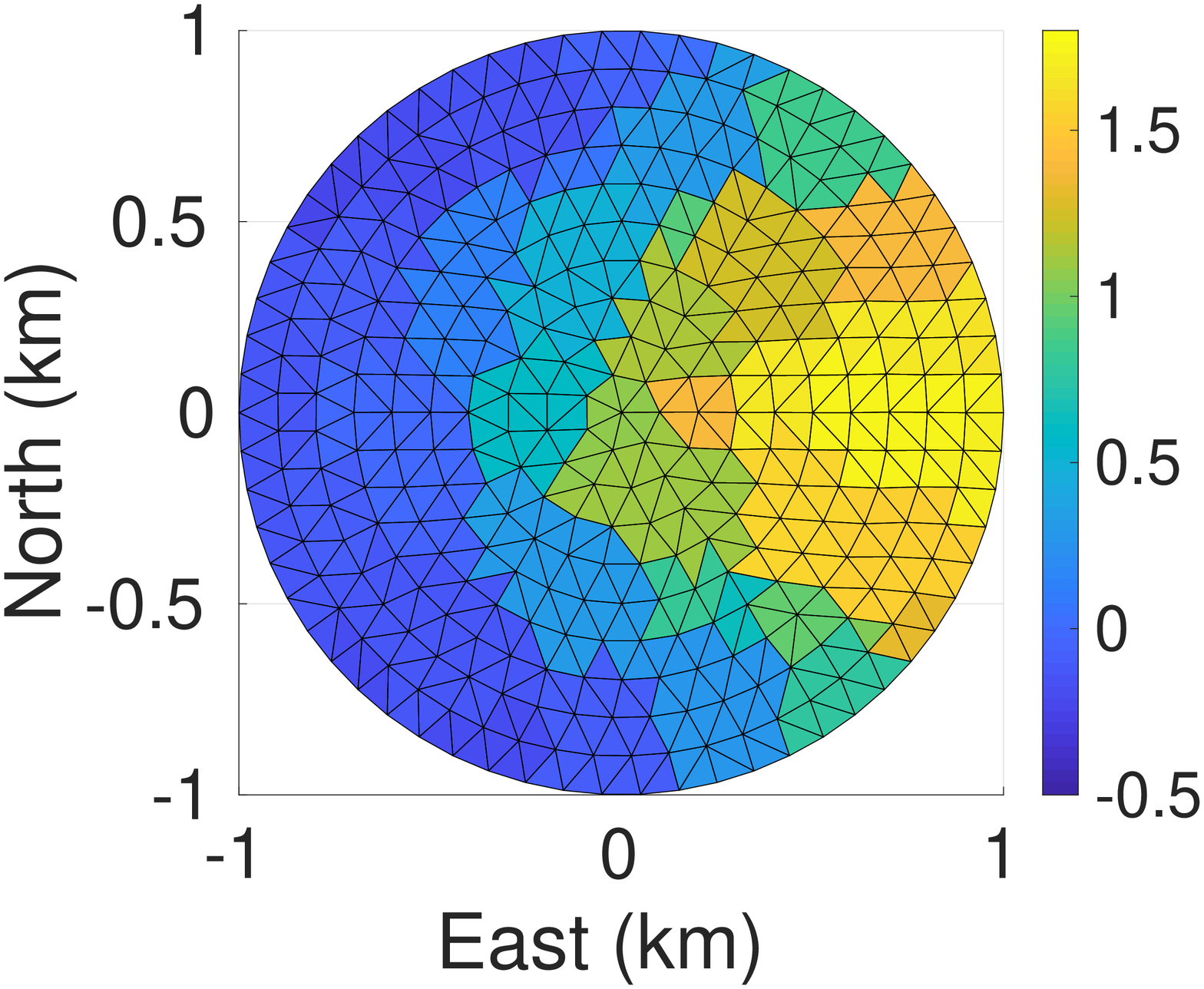}}
\hspace*{-0.2cm}  
\subfigure[$\alpha_1=1.0\ee2$]{\includegraphics[scale=0.205]
{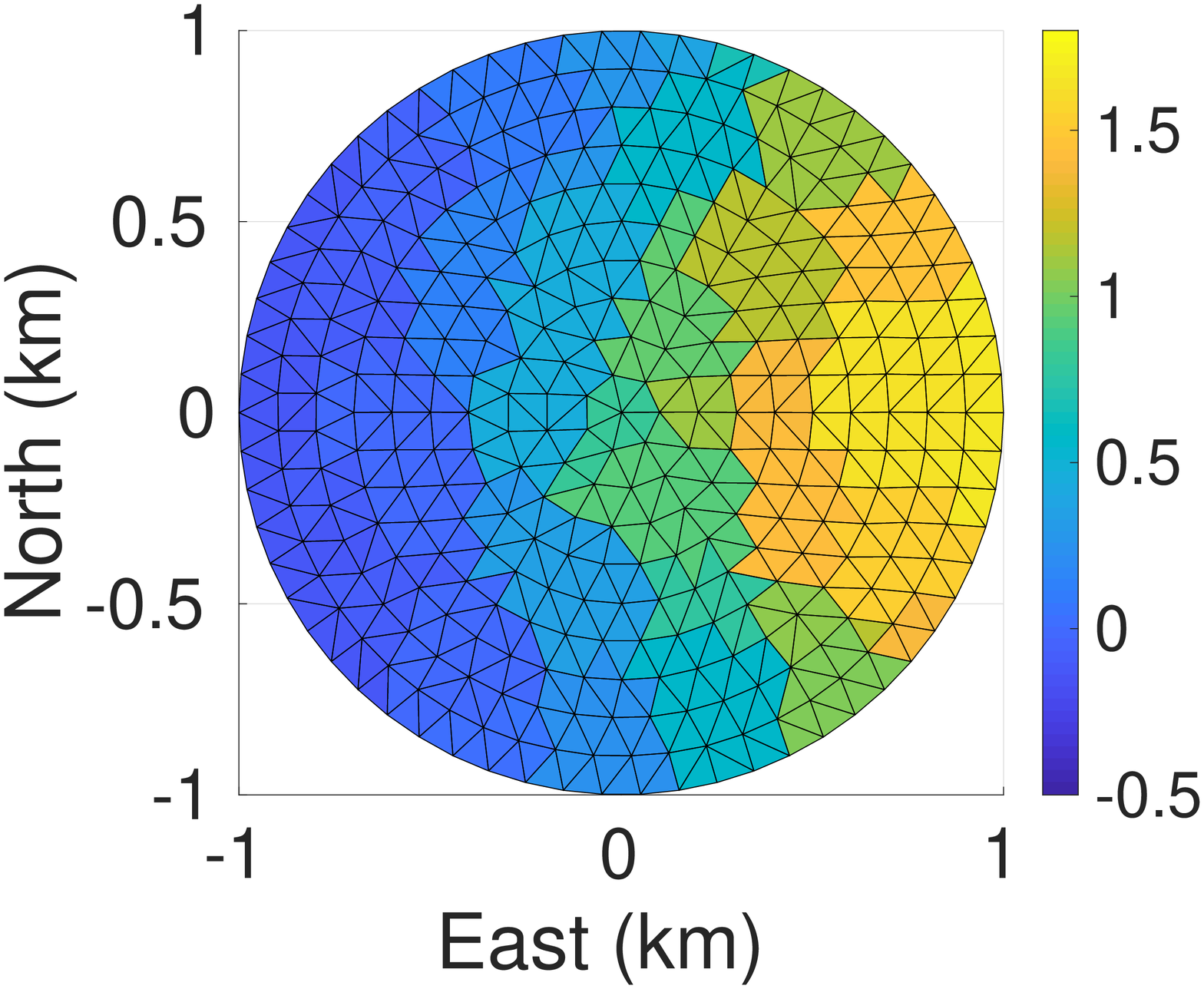}}
\caption{Fracture pressure located at $-0.9$ km, corresponding to (a) $\alpha_1=1.0\ee{-1}$, (b) $\alpha_1=1.0\ee1$ and (c) $\alpha_1=1.0\ee2$.}\label{solution_source_900}
\end{figure}
\end{center}

\begin{center}
\begin{figure}[htp!]\hspace*{-0.5cm}   
\subfigure[$\alpha_1=1.0\ee{-1}$]{\includegraphics[scale=0.18]
{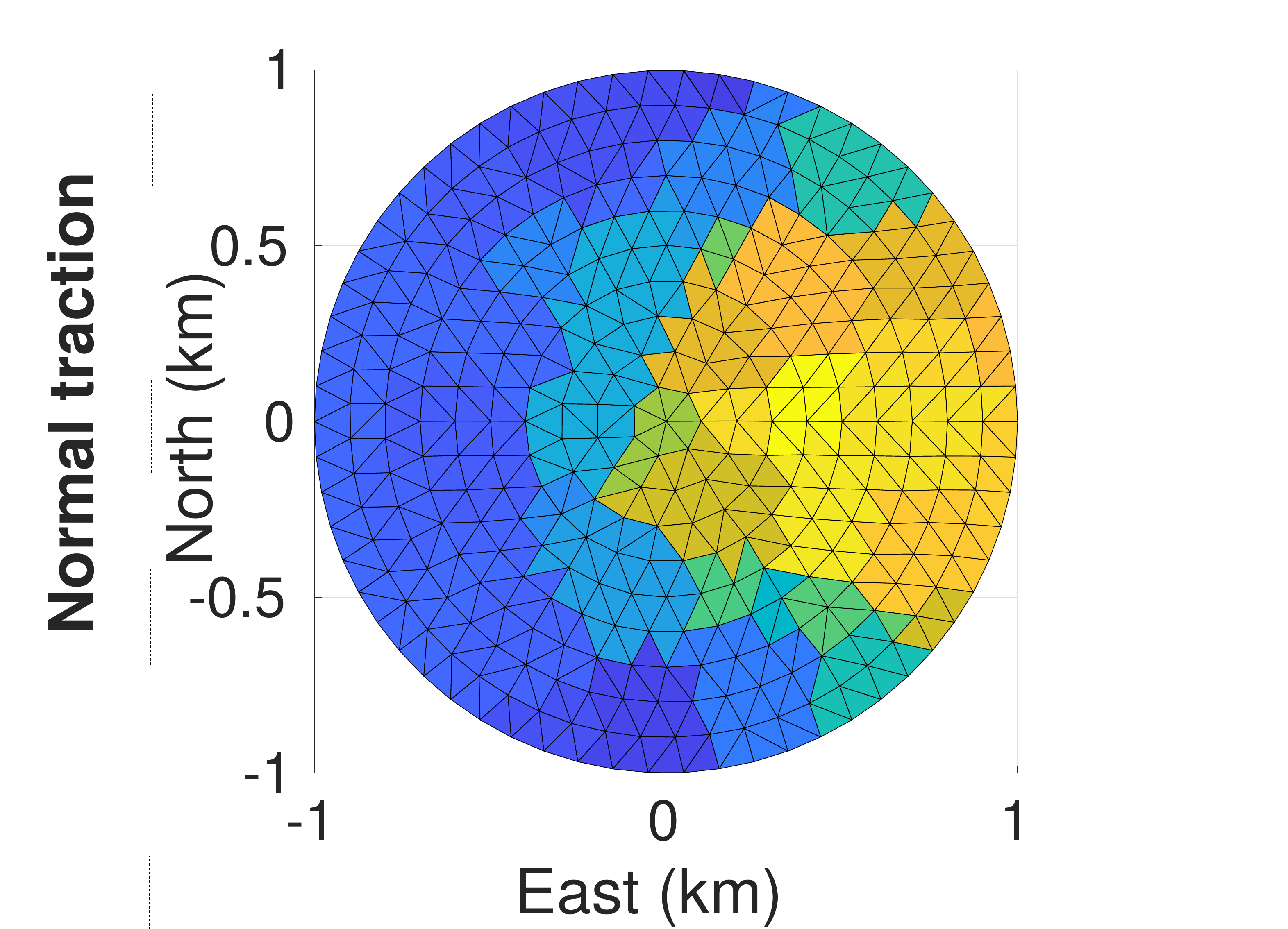}}\hspace*{-0.9cm}  
\subfigure[$\alpha_1=1.0\ee0$]{\includegraphics[scale=0.18]
{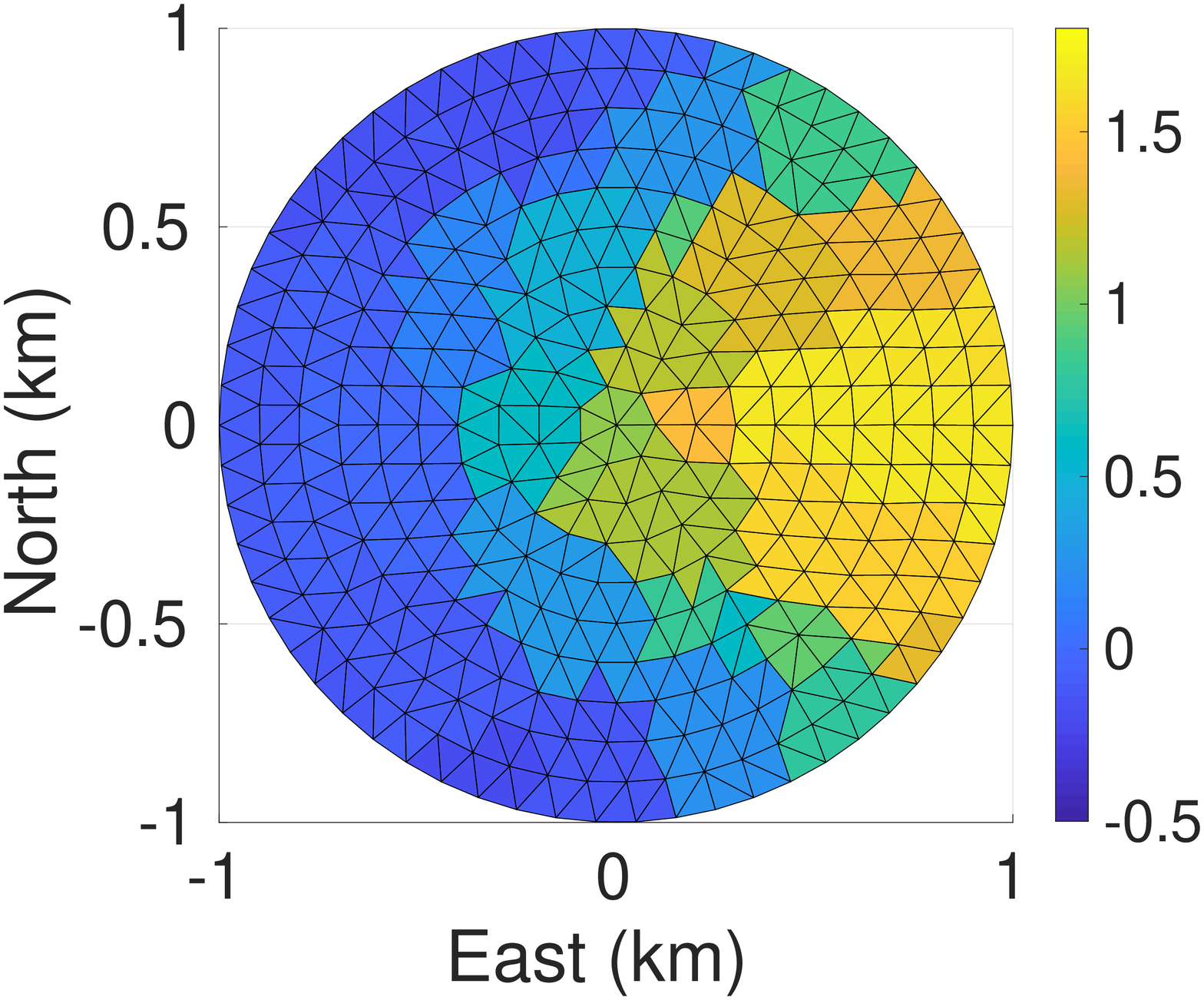}}
\hspace*{-0.9cm}  
\subfigure[$\alpha_1=1.0\ee1$]{\includegraphics[scale=0.18]
{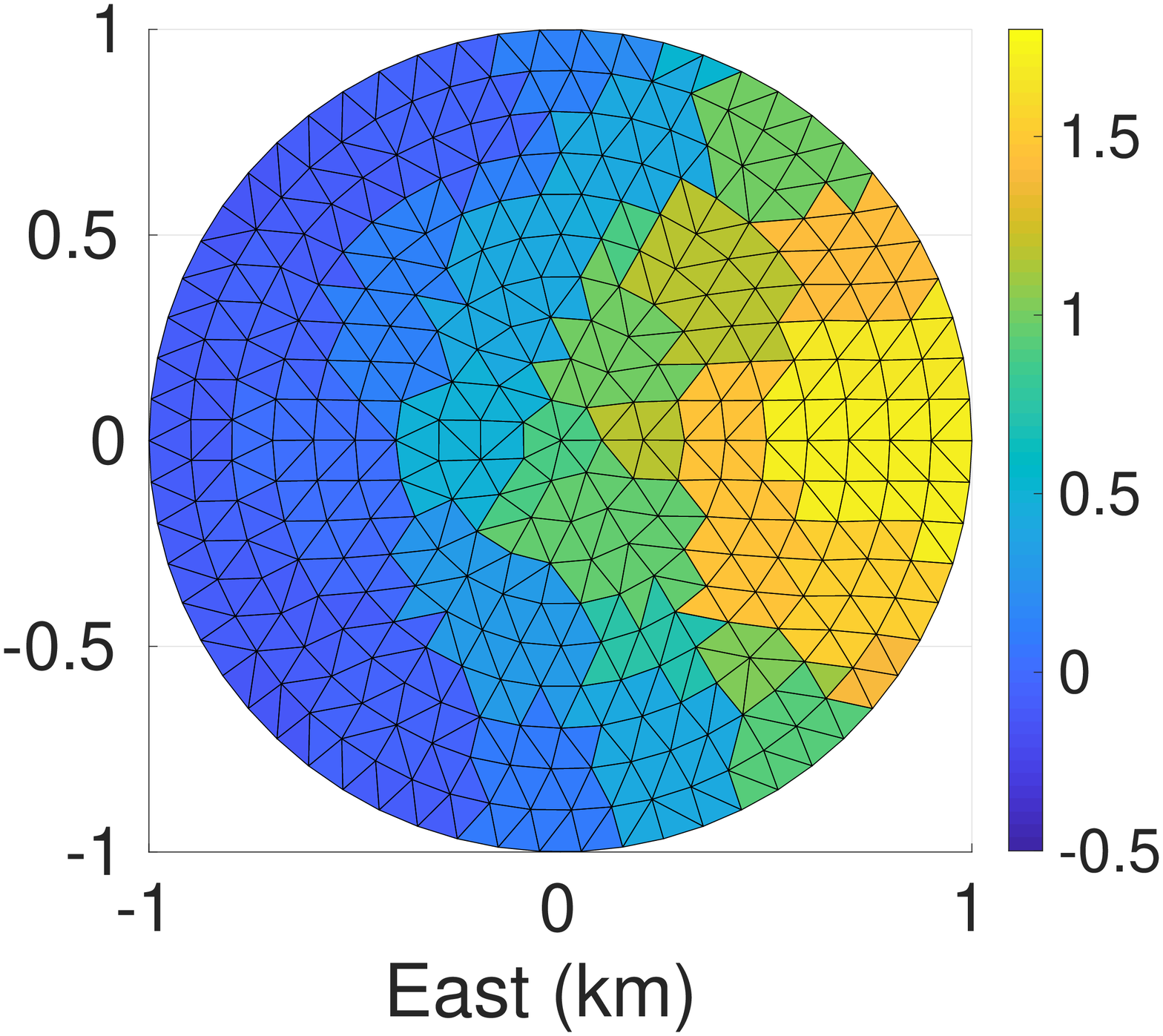}}\\\hspace*{-0.5cm} 
\subfigure{\includegraphics[scale=0.18]
{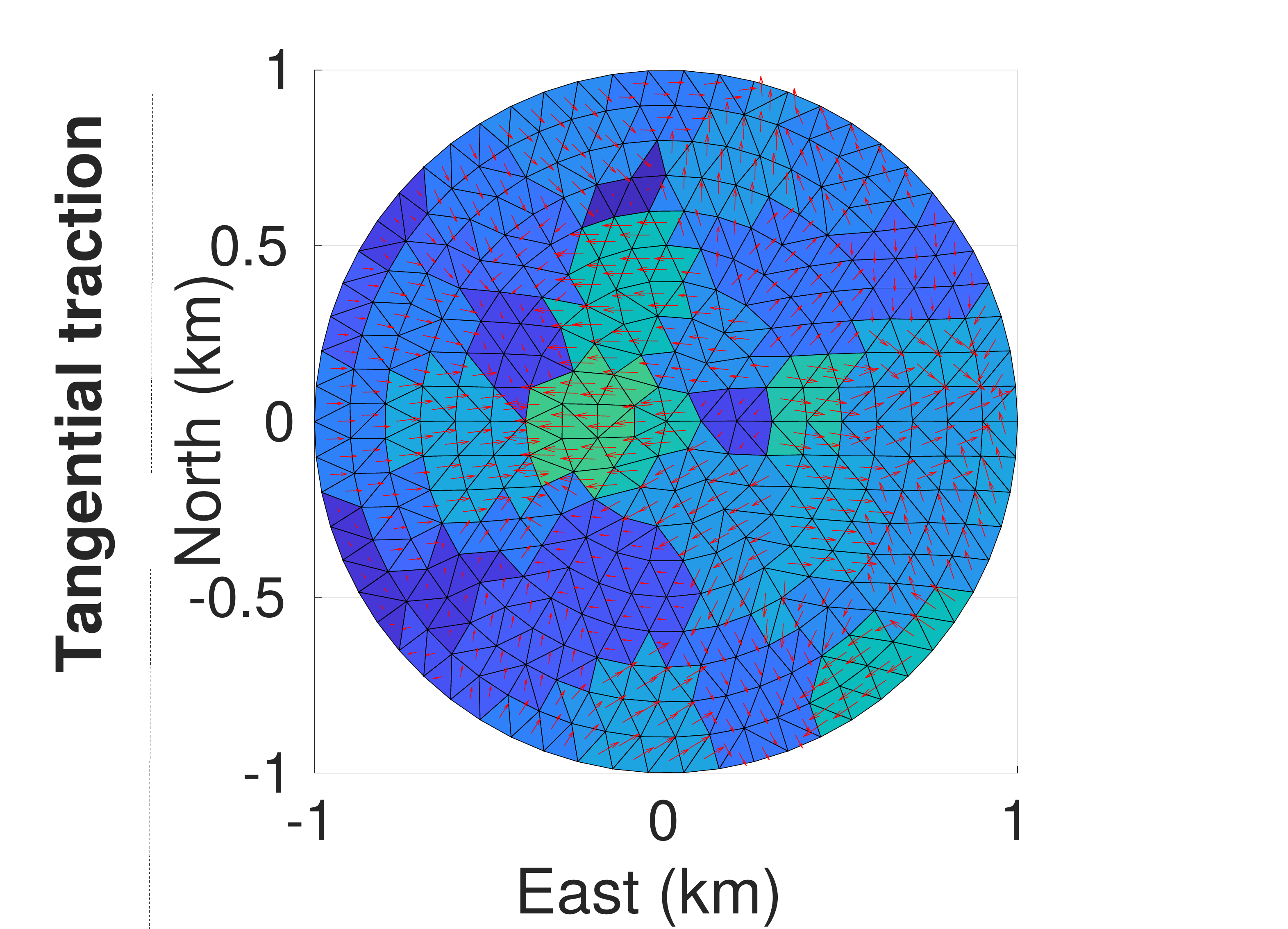}}\hspace*{-0.9cm}  
\subfigure{\includegraphics[scale=0.18]
{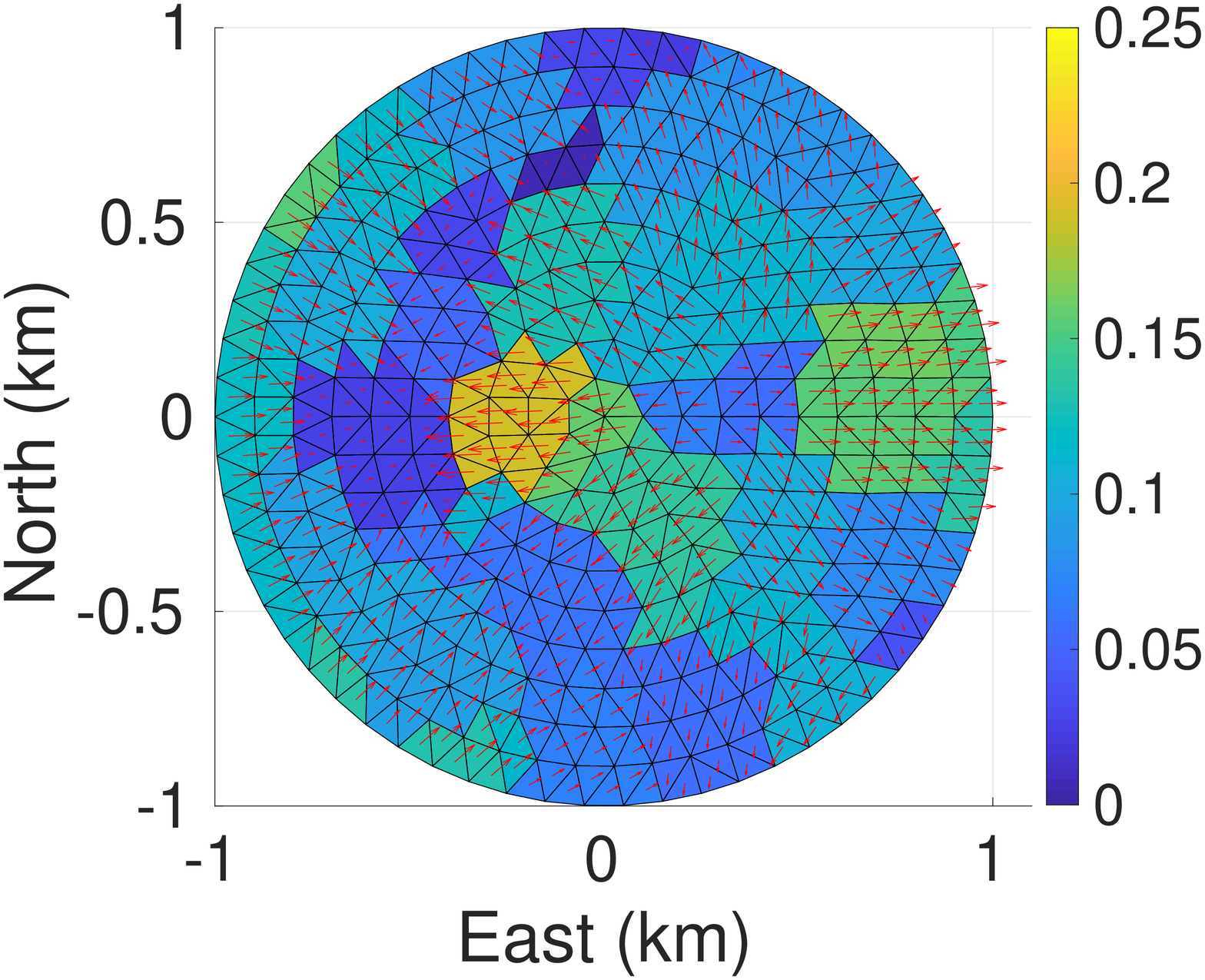}}
\hspace*{-0.9cm}  
\subfigure{\includegraphics[scale=0.18]
{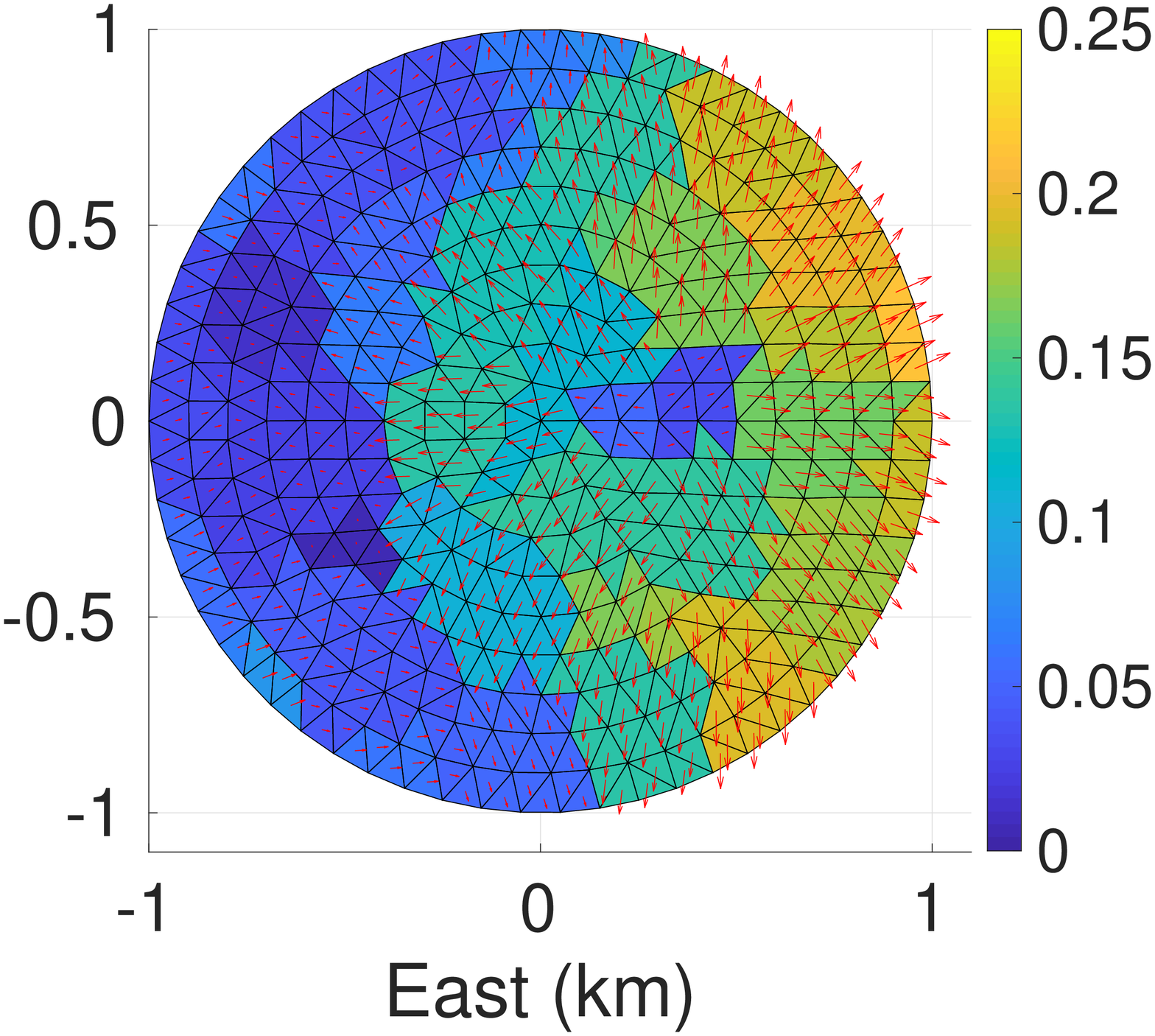}}

\caption{Normal and tangential tractions projected on the fracture located at $-0.9$ km,  corresponding to (a) $\alpha_1=1.0\ee{-1}$, (b) $\alpha_1=1.0\ee0$ and (c) $\alpha_1=1.0\ee1$. The directions of the tangetial stress are shown by red vectors.
}\label{solution_source_traction_900}
\end{figure}
\end{center}

We then aim to consider the impact of the fracture depth 
to the approximated admissible solution. To do that, the numerical experiments are repeated for the fracture at $-0.9$ km of depth.
The best combination for $\alpha_1=1.0\ee-1$ and $\alpha_0$ between $1.0\ee-7$ and $1.0\ee-4$, in red box are shown in Figure \ref{optim_alpha_param2}. The L-curves depicted for $\p$ and $\t$ in Figure \ref{L_curve_900}(a, b) show that the closest point to the origin for the $0.9$ km is $\alpha_1 = 1.0\ee0$. The normal stress projected on the fracture presented in Figure \ref{L_curve_900}(c) compare to exact solution confirms an acceptable approximated solution. 
As previous case the values around the $1.0\ee0$ for $\alpha_1$ are acceptable (see Figures \ref{solution_source_900} and \ref{solution_source_traction_900}). A comparison of the iteration number presented in Table \ref{iterations_comparison}, indicates  that the number of iterations is augmented for the greater values of $\alpha_1$.          

%

\begin{table}[htbp]
\footnotesize
\centering
\begin{tabular}{|c|c|c|c|c|}
\hline
Depths& $\alpha_1$& Unknown& Number of iteration & Number of unknowns\\
\hline
\multirow{2}{*}{$-0.3$ km}&\multirow{2}{*}{$1.0\ee1$ }& Pressure $\p$ &$71$&291\\
&  &Traction $\t$ &$84$&873\\
 \hline
\multirow{2}{*}{$-0.9$ km}&\multirow{2}{*}{$1.0\ee0$ }& Pressure $\p$ &$48$&107\\
&  &Traction $\t$ &$67$&321\\
 \hline 
\end{tabular}
\caption{Comparison between the number of iterations for two depths -0.3 and -0.9 km and when the unknowns are pressure $\p$ and traction $\t$ for $\alpha_0= 1.0\ee{-7}$.}\label{iterations_comparison}
\end{table}

As seen in the numerical simulations, we may trust to the L-curves results to find an appropriate $\alpha_1$ for more complicated geometry of fracture and mesh, which is clearly help us to computational cost.

\section{Taking into account the earth-satellite directions}

\subsection{Mathematical framework and new optimization problem}

Before going further, let us first remark that the previous section can easily be adapted to the case where the measurements of the displacement field are performed only on a part of the ground $\widetilde{\Gn}\subset\Gn$.
The cost function to be minimized becomes
$$
J(\t)=\frac{1}{2}\int_{\widetilde{\Gn}} (\u-\u_{d})^\top{\CC^{-1}} (\u-\u_{d}) \dd \Gn 
+\frac{\alpha_0}{2}\ds\int_{\Gc}\lvert \t\rvert^2\dd \Gc+\frac{\alpha_1}{2}\ds\int_{\Gc} \lvert\nabla \t\rvert^2\dd \Gc.
$$
Rewriting the optimality conditions and adapting the algorithms is straightforward. This is what is actually implemented in our software.

However, assuming the displacement field to be known in cartesian coordinates is not realistic for the applications. In this section we aim at adapting our work to the case where measurements are provides by satellite radar interferometry
(as e.g. in\cite{FuCaDu05}). We will first state precisely what measurements are made and then adapt our methods to this new framework.

Let $\bu_d = (u_{x},u_{y},u_{z})^\top$ be a displacement field on $\Gn$, written in cartesian coordinates. A satellite will aim at the ground to measure a displacement, and the resulting measurement will be in the form
$\p^\top \bu_d$ where the aiming direction (also called earth-satellite direction) $\p = (p_{x},p_{y},p_{z})$ is a unit vector oriented from the ground to the satellite. Let then $N\geq 1$ be the number of satellites, each associated to a direction $\p_i = (p_{ix},p_{iy},p_{iz})$ for $i = 1,\ldots, N$. We can build a matrix $\bP:\mathbb{R}^3\rightarrow\mathbb{R}^N $ as $\bP=(\p_1,\p_2,\cdots,\p_N)^\top $, so that the actual measurement is
\begin{equation*}
\bR_d = \bP \bu_d.
\end{equation*}
The vector field $\bR_d \in \mathbb{R}^N$ is defined on $\Gn$, and the matrix $\bP$ also operates on the solutions of System \eqref{problem1-1}--\eqref{problem1-4}. Hence, given the field $\bR_d$, our new optimization problem consists now in finding the traction vector $\t$ minimizing the following cost function:
\begin{equation}\label{min_prob_esd}
 J(\t)=\frac{1}{2}\int_{\Gn} (\bP\u-\bR_{d})^\top{{\CC}^{-\frac{1}{2}}{\CC}^{-\frac{1}{2}}} (\bP\u-\bR_{d}) \dd \Gn 
+ \frac{\alpha_0}{2}\ds\int_{\Gc}\lvert \t\rvert^2\dd \Gc+\frac{\alpha_1}{2}\ds\int_{\Gc} \lvert\nabla \t\rvert^2\dd \Gc,
\end{equation} 
where $\alpha_0>0$ and $\alpha_1>0$ are the regularization parameters.
The optimality conditions for this problem are derived in the same way than in Section \ref{math-form}. We will not go into details about this and focus on the discrete version of the problem.

Using the definitions of Sections \ref{discrete-model} and \ref{discrete-problem}, and in view of \eqref{discrete_min_prob} and \eqref{min_prob_esd} the discrete cost function becomes
\begin{equation}\label{objectiv_discrete}
J_\mathtt{d}(\t)=\frac{1}{2}(\bP\mathcal{O}_R\X-\mathbf{R}_d)^\tra\CC^{-\frac{1}{2}} M_G\CC^{-\frac{1}{2}}(\bP\mathcal{O}_R\X-\mathbf{R}_d)+\frac{\alpha_0}{2}
 (\t^\top M_{F_0} \t)+\frac{\alpha_1}{2}
 (\t^\top M_{F_1} \t).
\end{equation}
where $X$ is the solution of the discrete system \eqref{Discrete-Problem}. This finite dimensional problem then boils down to finding the saddle point of the following Lagrangian
\begin{equation*}
\la_\mathtt{d} (\X,\t, \boldsymbol\phi)=J_\mathtt{d}(\t)-\langle \mathrm{K}\X-(L_{\small\Omega}\f+L_C\t),\boldsymbol\phi \rangle.
\end{equation*}
Computing the partial derivative of $\la_\mathtt{d}$ and then cancelling them, leads to the discrete adjoint problem \eqref{problem_adjoint}:
\begin{equation}\label{dis_adjoint_esd}
 \mathrm{K} \boldsymbol\phi=\mathcal{O}_R^\top\bP^\top\CC^{-\frac{1}{2}}M_G\CC^{-\frac{1}{2}}(\bP\mathcal{O}_R\X-\mathbf{R}_d),
\end{equation}
and the gradient of $J_\mathtt{d}$ is:
 \begin{equation}\label{dis_grad_esd}
 \nabla J_\mathtt{d} =\alpha_0 M_{F_0}\t+\alpha_1 M_{F_1}\t+L_C^\top \boldsymbol\phi.
 \end{equation}
The minimization Algorithm \ref{IP_DDM1} can then be adapted to the new adjoint state and gradient to obtain the numerical approximation of the traction $\t$ on the crack by using the surface measurements provided by radar interferometry $\br_d$.

\subsection{Applications to synthetic test}
The numerical applications here are proceeded in three dif\/ferent steps:
First, we aim to the adaptation of the theoretical results in Earth-Satellite directions. For the sake of simplicity, we assume the covariance is an identity matrix. Next, in order to reduce the observed data, the identity covariance matrix is reduced to the nodal mesh points. We then consider a problem with dense covariance matrix adapted to a limited number of observed data. 

\begin{center}
\begin{figure}[htp!]\hspace*{-1.1cm}   
\subfigure[S4]{\includegraphics[scale=0.2]
{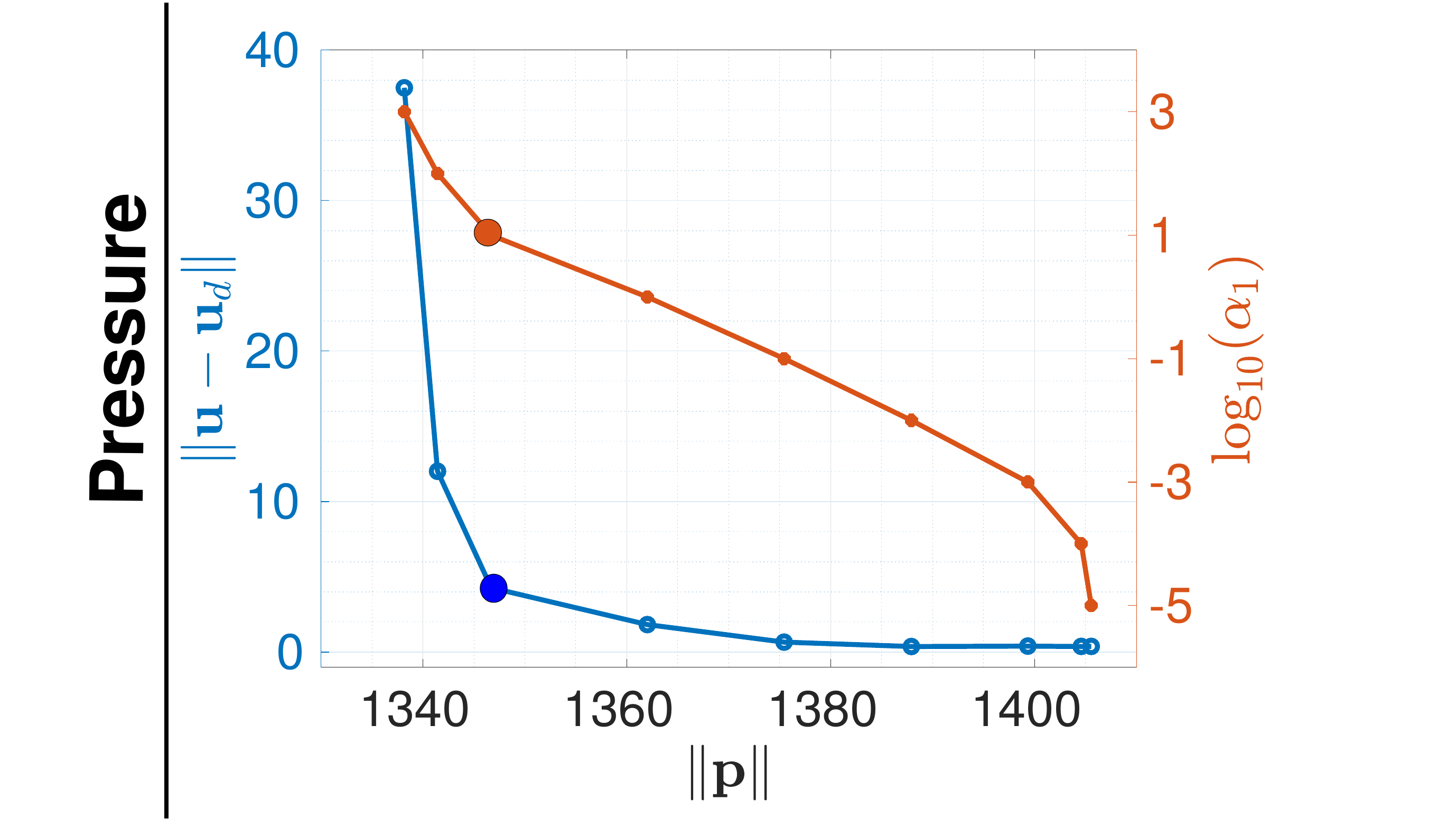}}\hspace*{-0.8cm}  
\subfigure[S4 S6]{\includegraphics[scale=0.13]
{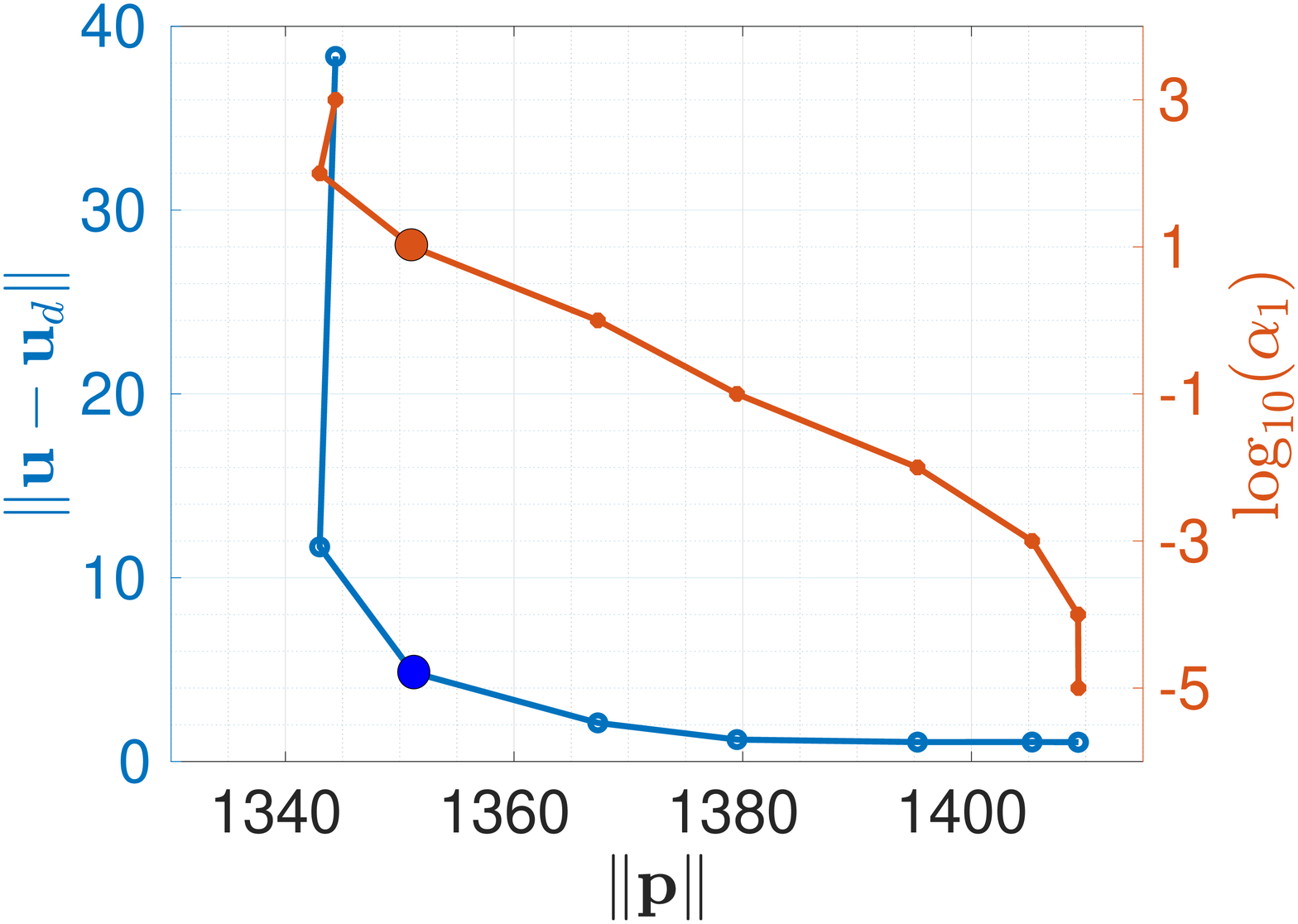}}
\hspace*{-0.2cm}  
\subfigure[S4 S6 TSXA]{\includegraphics[scale=0.13]
{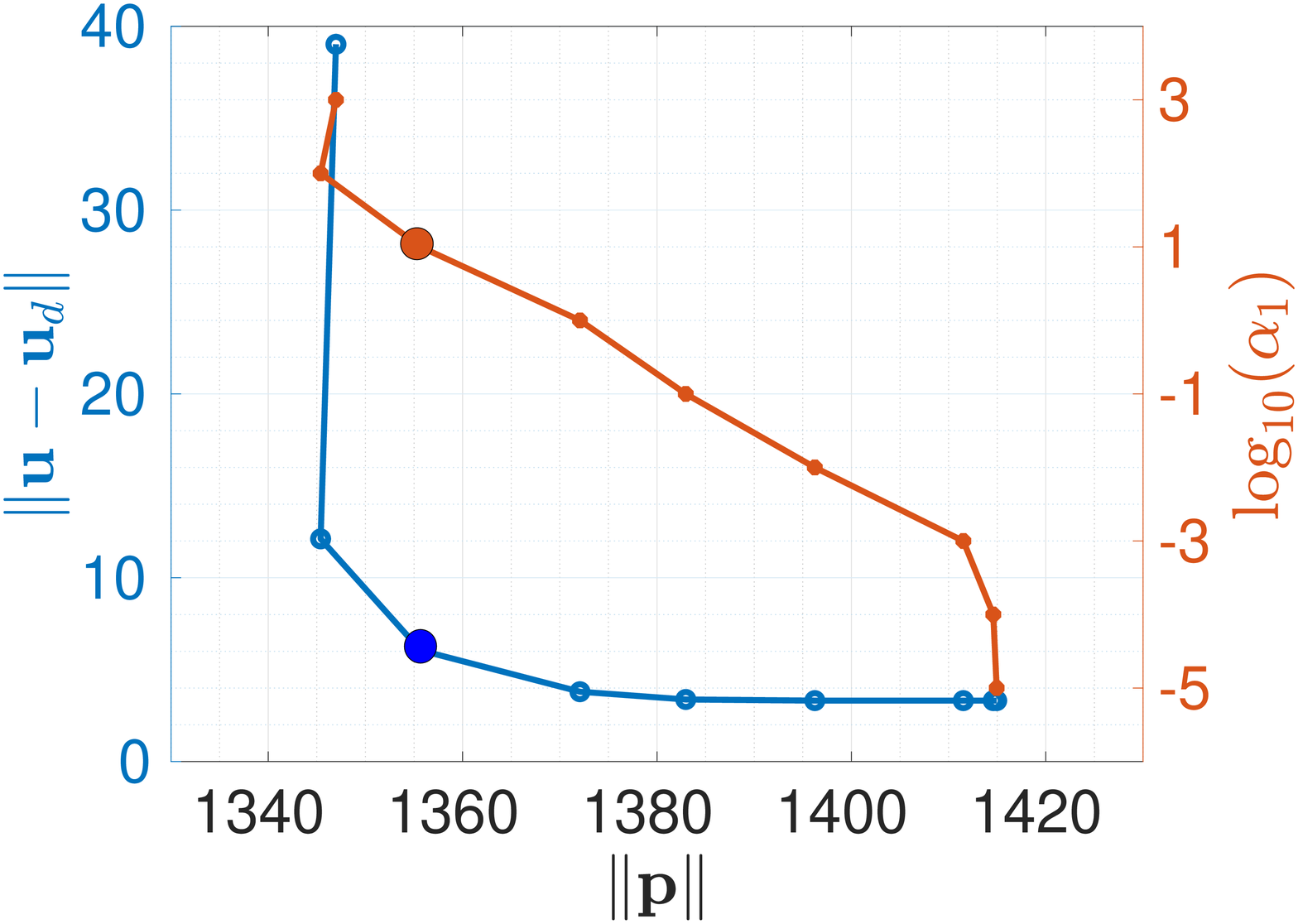}}
\hspace*{-0.2cm}  
\subfigure[S4 S6 TSXA TSXD]{\includegraphics[scale=0.13]
{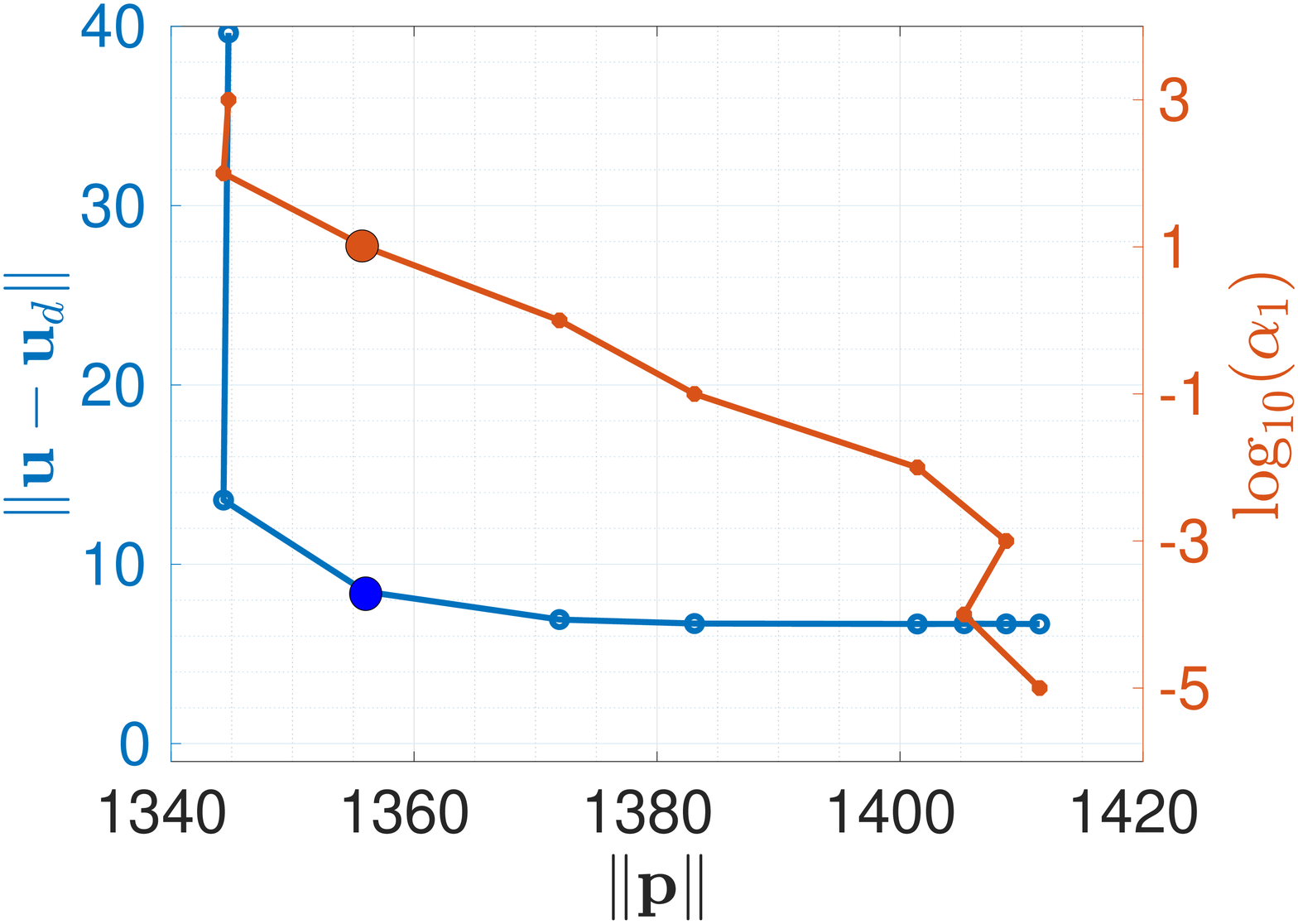}}\\\hspace*{-1.1cm} 
\subfigure{\includegraphics[scale=0.2]
{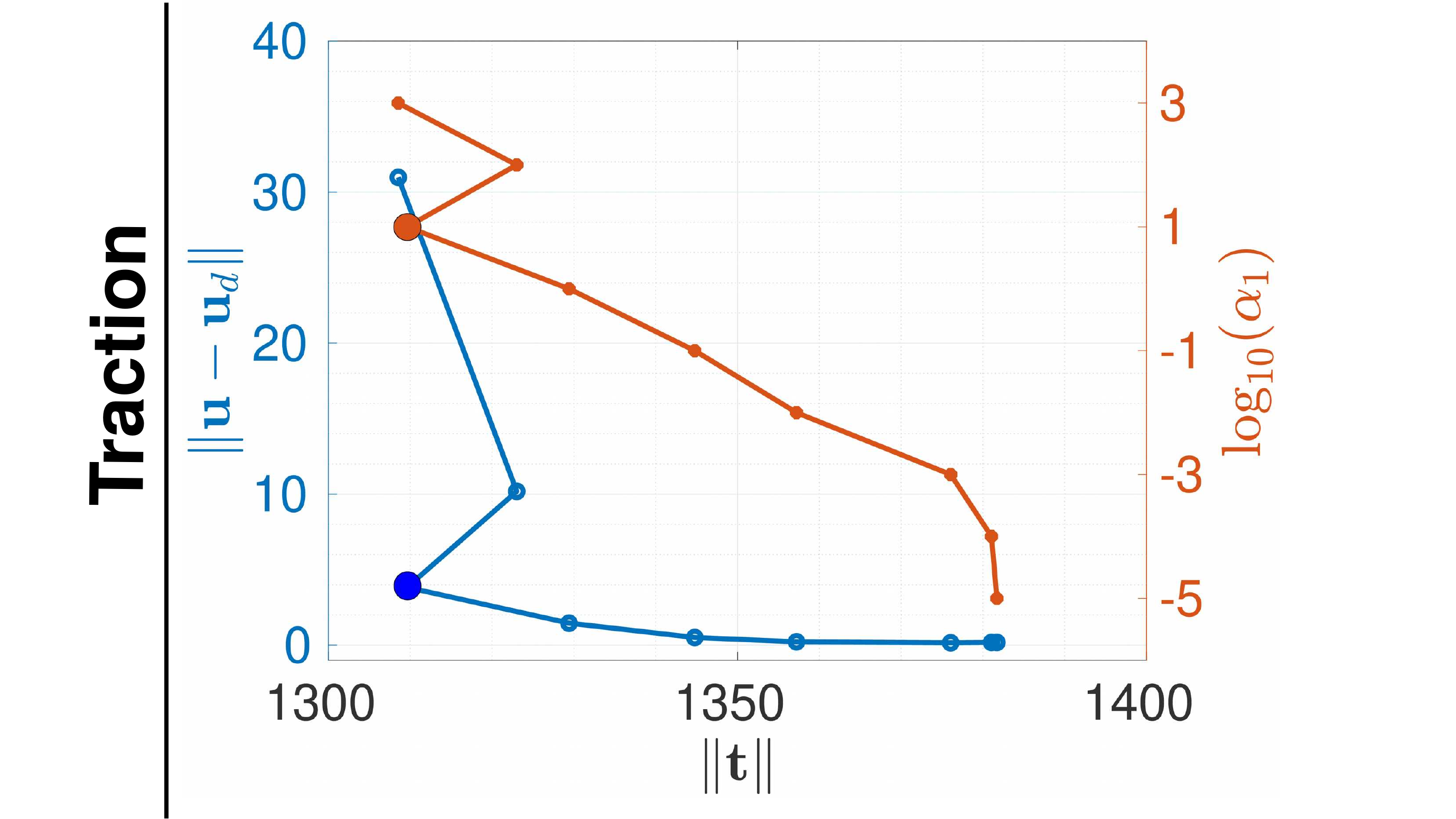}}\hspace*{-0.8cm}  
\subfigure{\includegraphics[scale=0.13]
{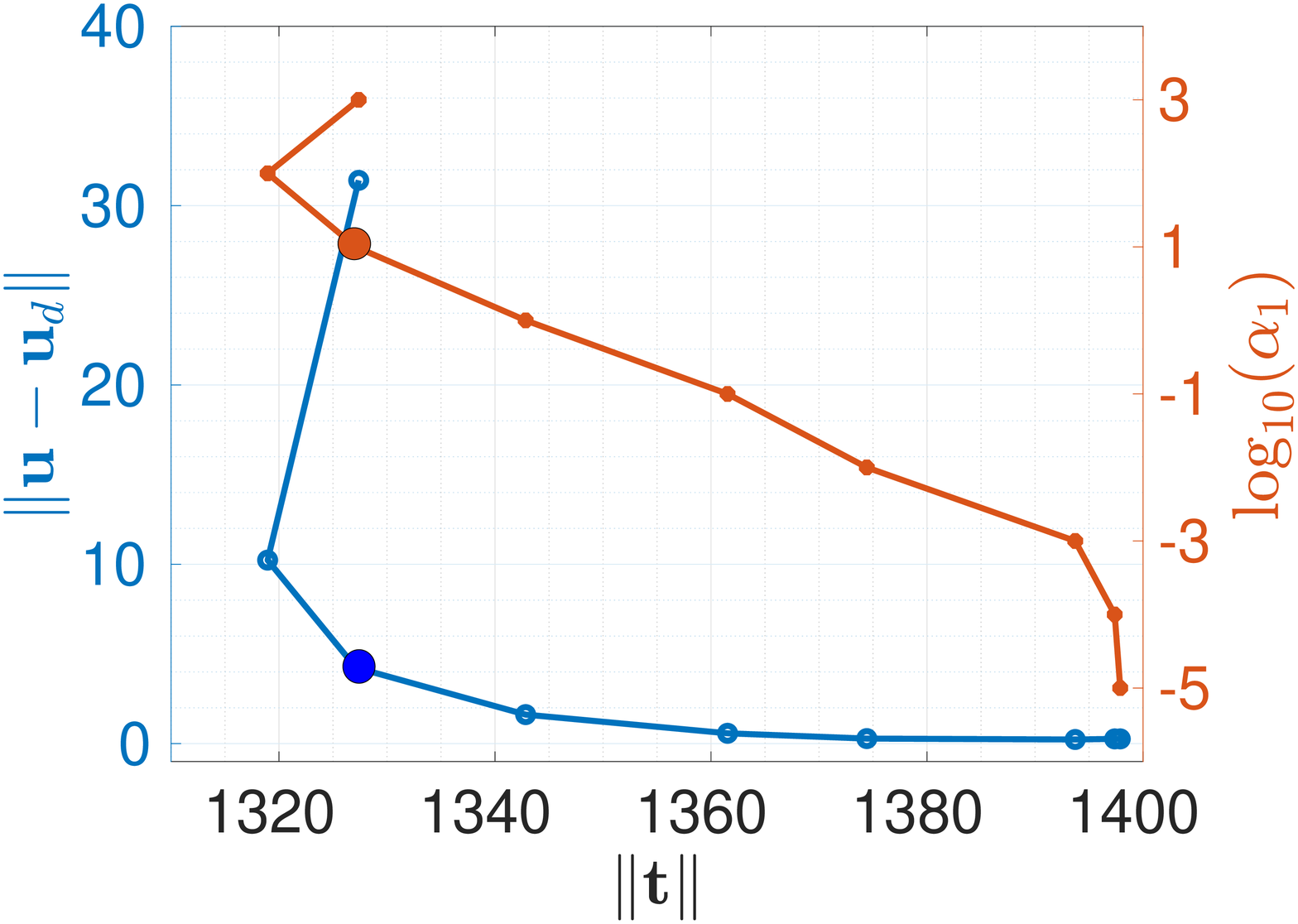}}
\hspace*{-0.2cm}  
\subfigure{\includegraphics[scale=0.13]
{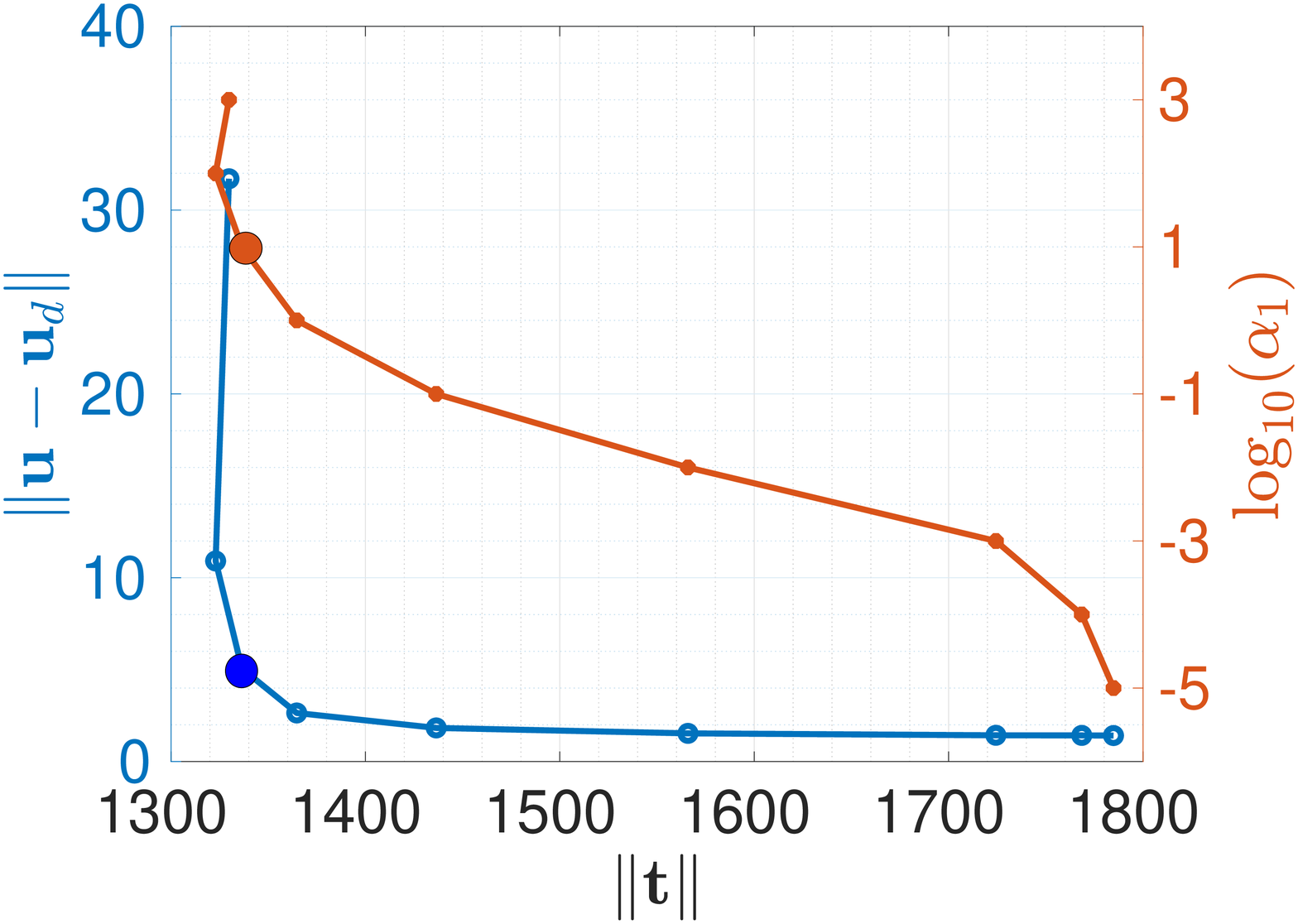}}
\hspace*{-0.2cm}  
\subfigure{\includegraphics[scale=0.13]
{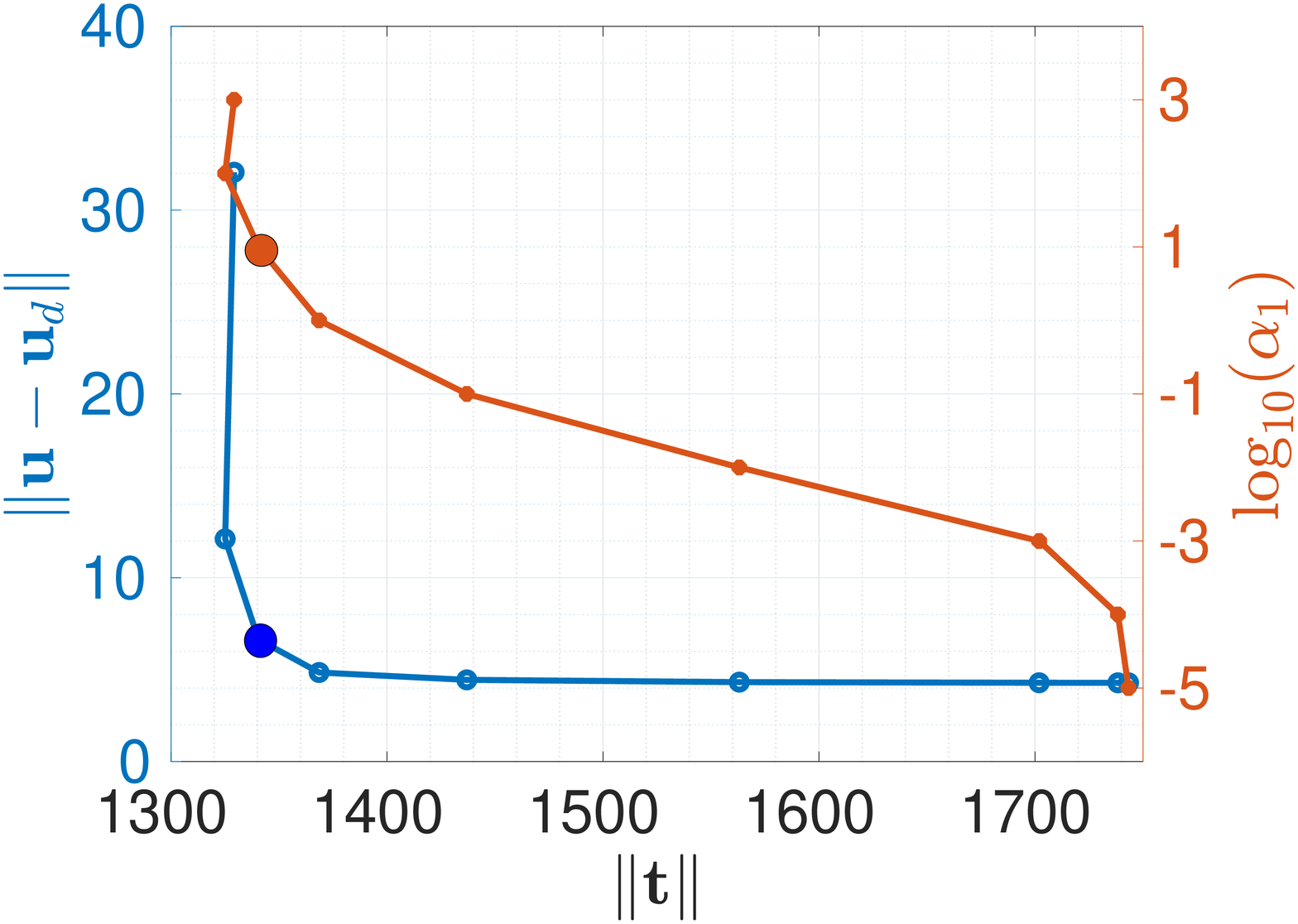}}

\caption{L-curves used to find $\alpha_1$ representing the best compromise between the data fit (equation \eqref{misfit}) and the smoothing (equation \eqref{smoothing}) in the cost function (equation \eqref{min_prob_esd}) taking into account the earth-satellite directions with identity covariance matrix. The fracture is located at $-0.3$ km, when solving for pressure $\p$ and traction $\t$. The larger points indicate the best compromise. (a) in S4 (b) in S4 and S6 (c) in S4, S6 and TSXA (d) in S4, S6, TSXA and  TSXD directions. Here, we set $\alpha_0 = 1.0\ee-7 $.}\label{L_curve_sat_identityCov_300}
\end{figure}
\end{center}

\begin{center}
\begin{figure}[htp!]\hspace*{-0.5cm}   
\subfigure[$\alpha_1=1.0\ee{1}$]{\includegraphics[scale=0.205]
{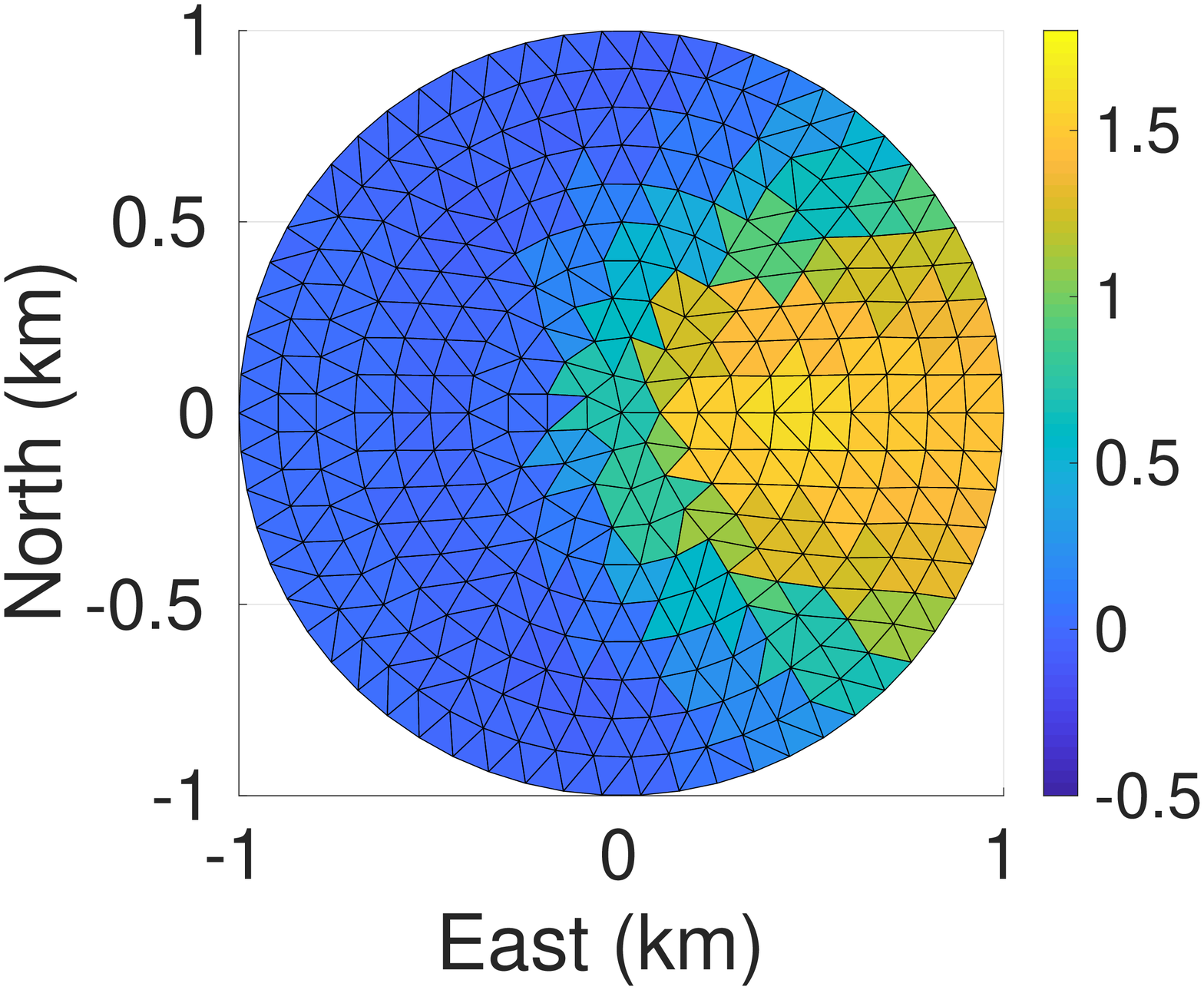}}\hspace*{-0.2cm}  
\subfigure[$\alpha_1=1.0\ee2$]{\includegraphics[scale=0.205]
{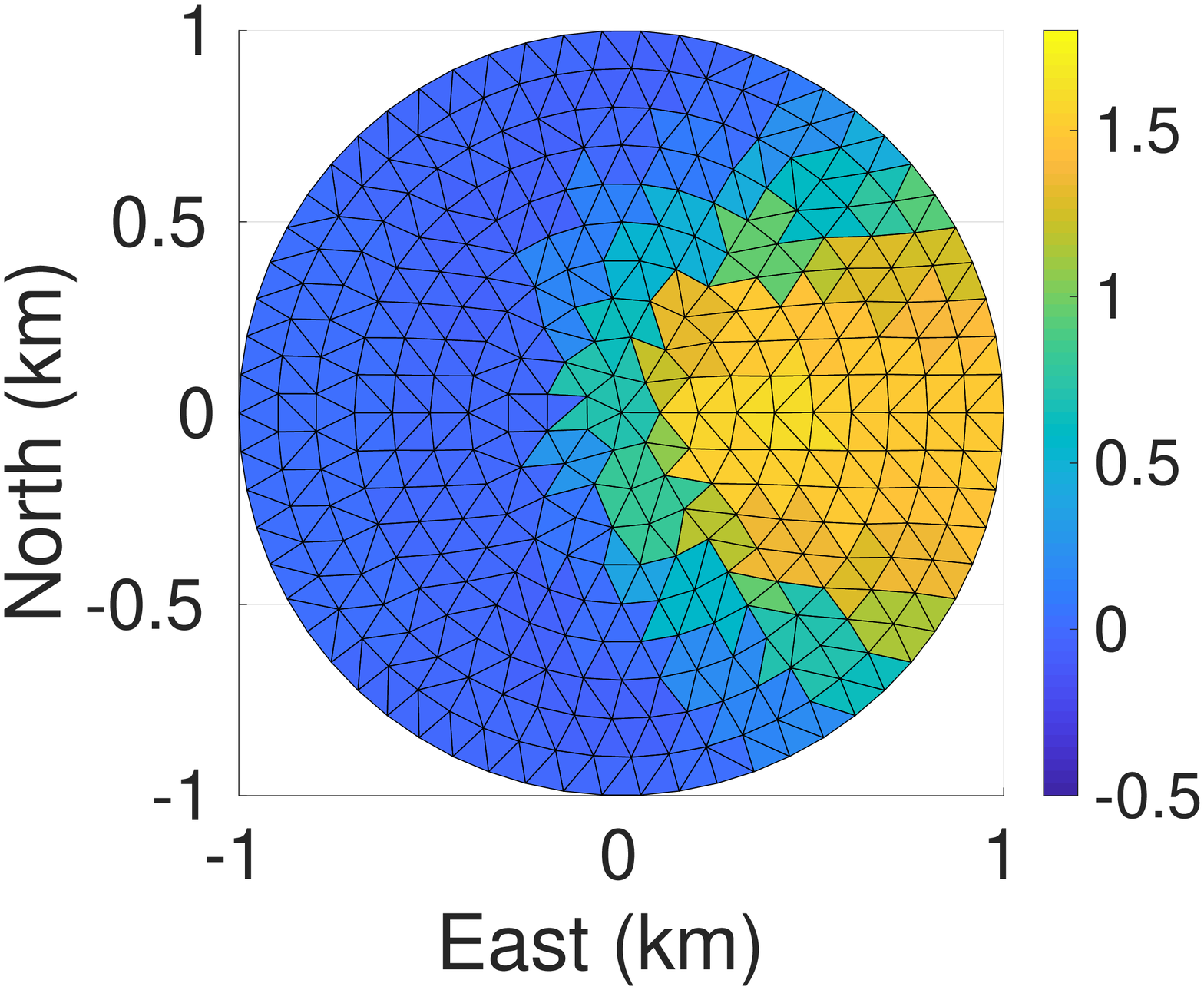}}
\hspace*{-0.2cm}  
\subfigure[$\alpha_1=1.0\ee2$]{\includegraphics[scale=0.205]
{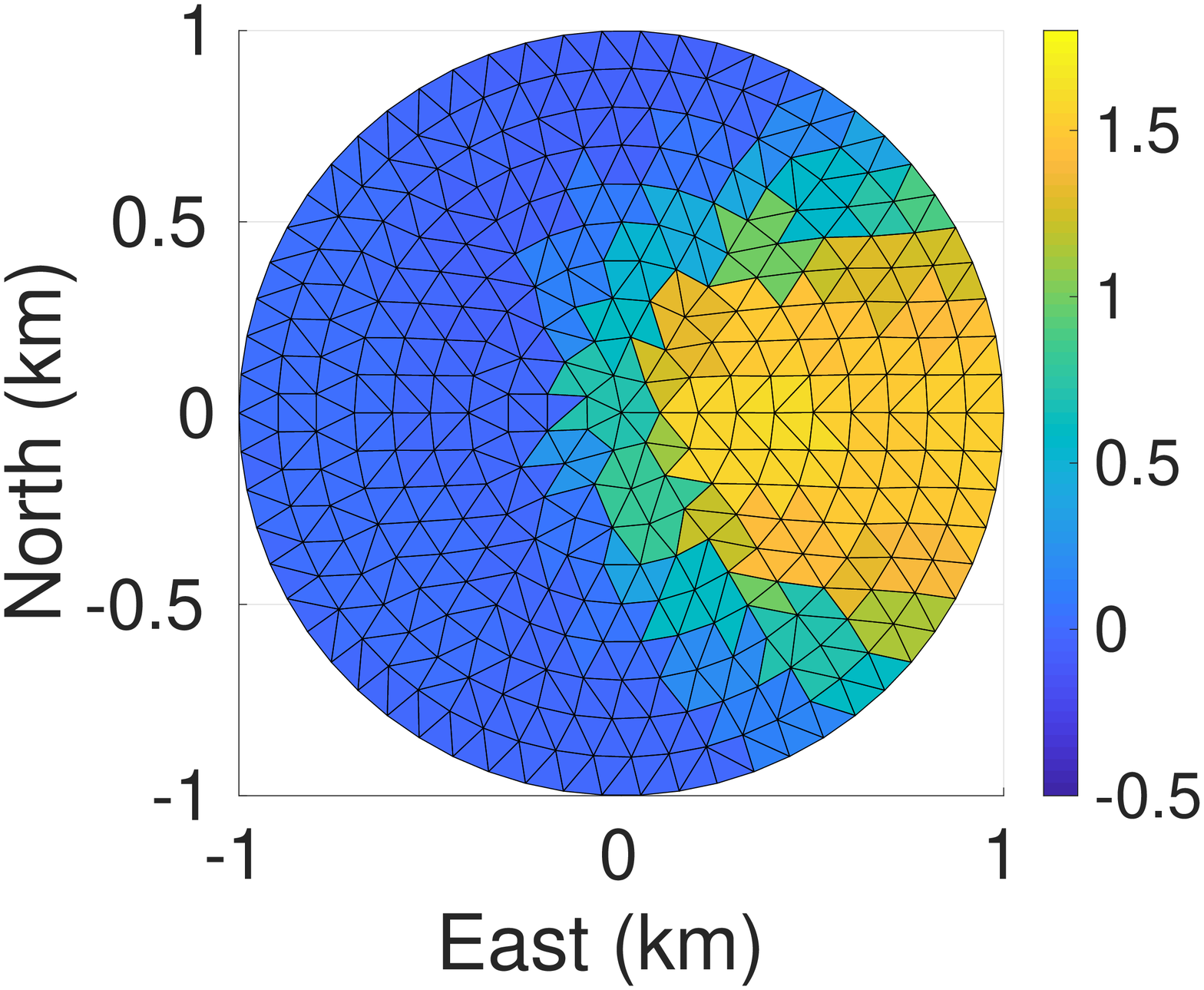}}\\\hspace*{-0.5cm} 
\subfigure{\includegraphics[scale=0.205]
{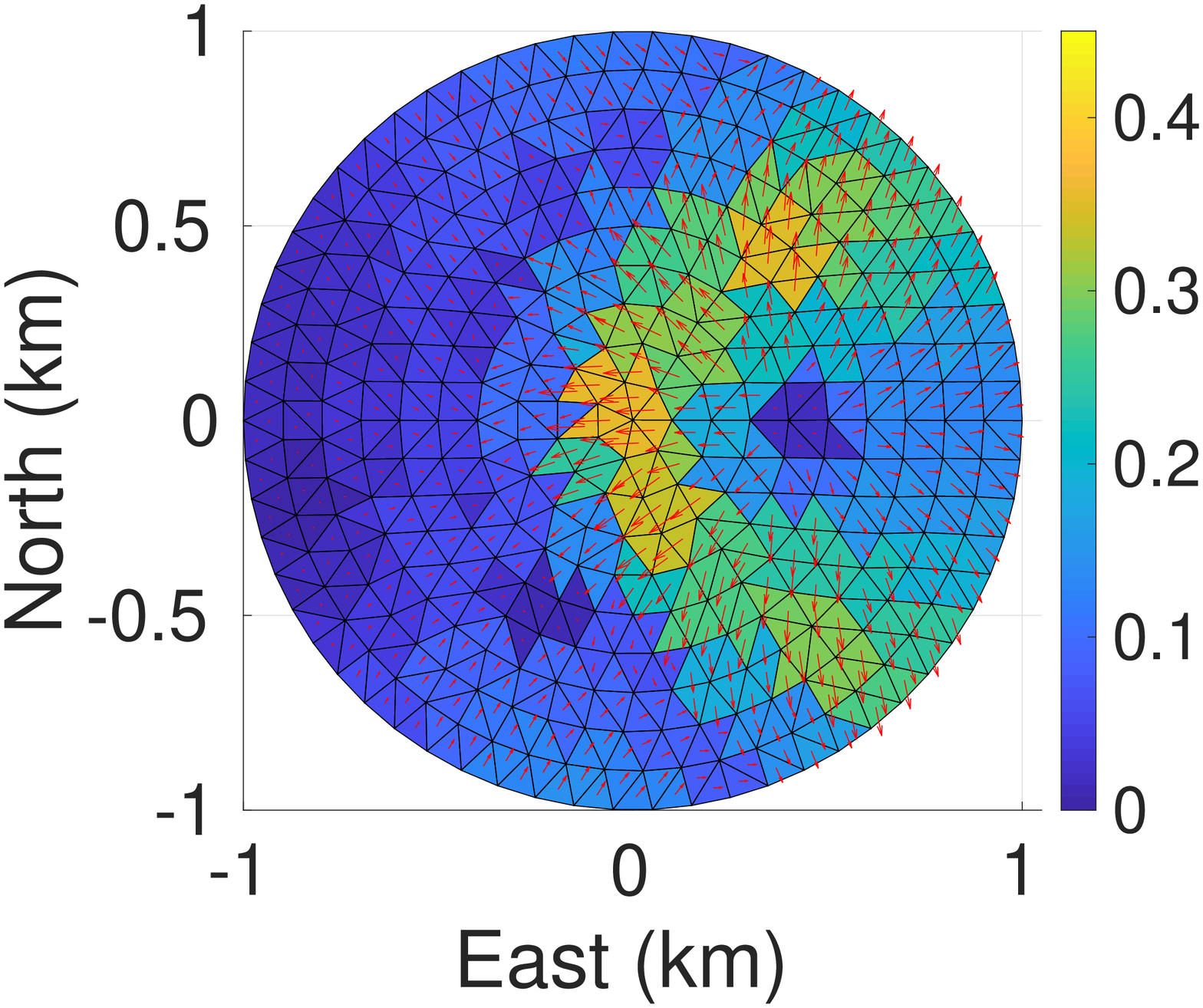}}\hspace*{-0.2cm}  
\subfigure{\includegraphics[scale=0.205]
{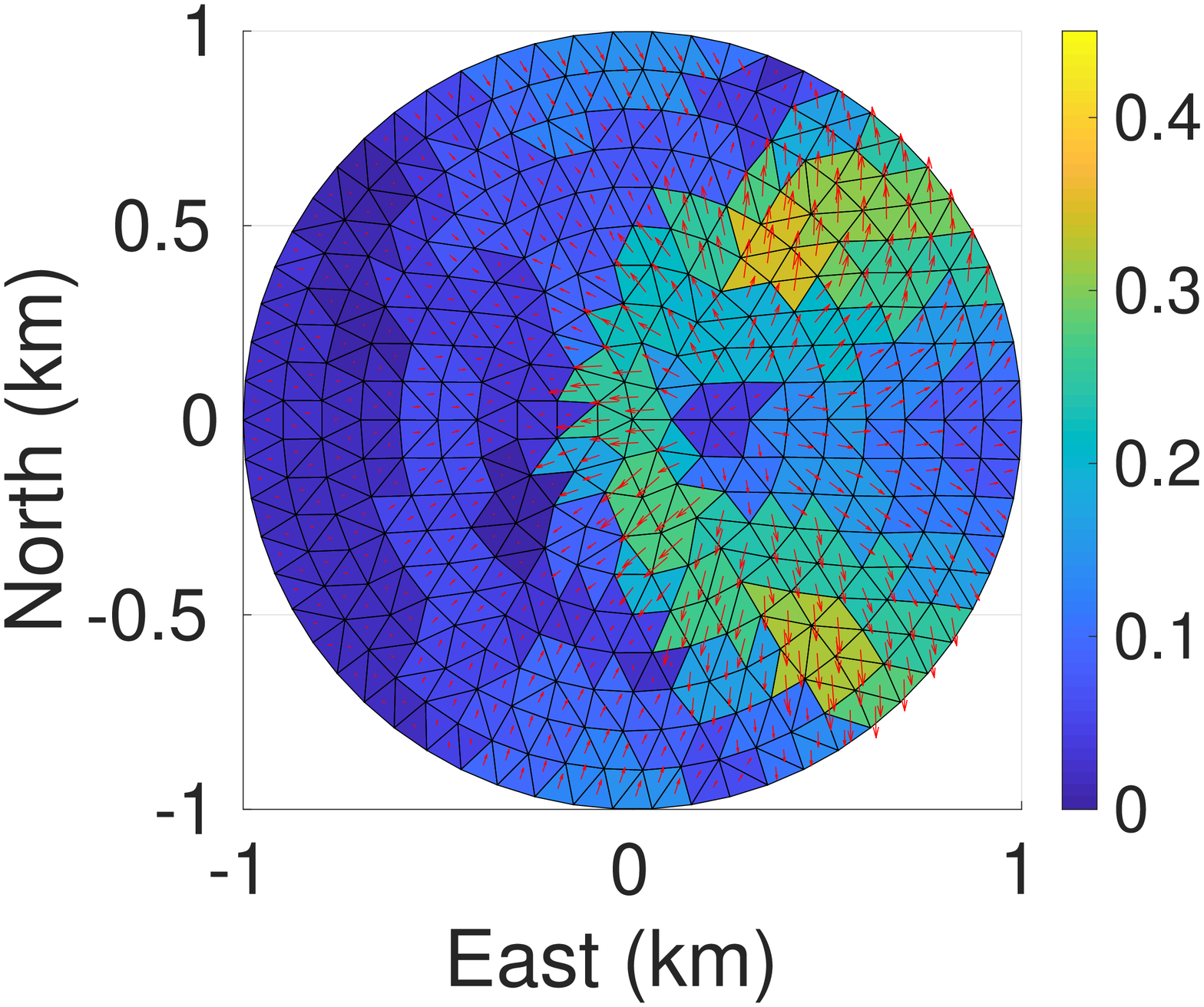}}
\hspace*{-0.2cm}  
\subfigure{\includegraphics[scale=0.205]
{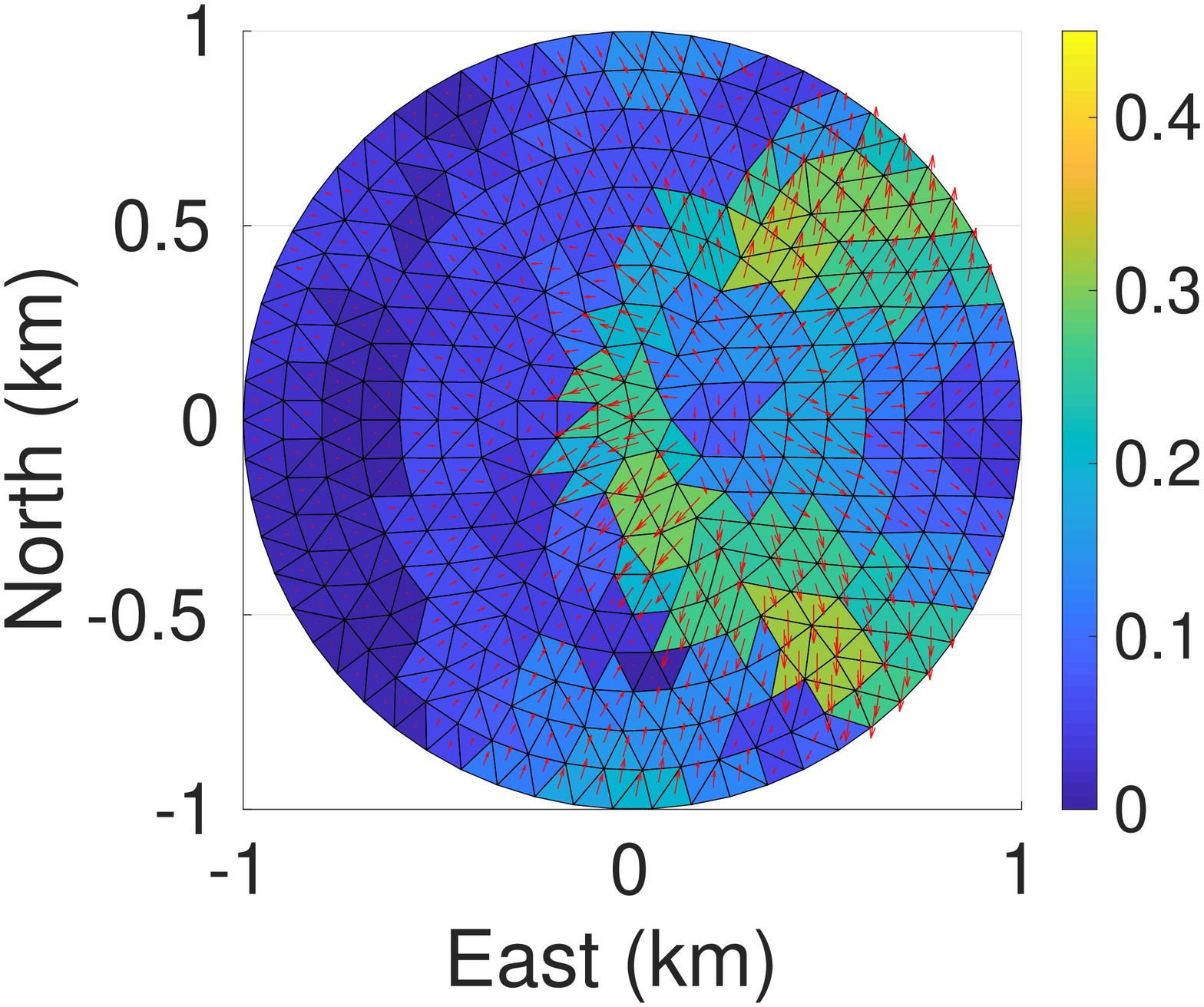}}

\caption{Fracture stress located at $-0.3$ km, f\/irst row normal stress and the second row tangential stress, corresponding to (a) $\alpha_1=1.0\ee1$ for S4 direction, (b) $\alpha_1=1.0\ee2$ S4, S6 directions and (c) $\alpha_1=1.0\ee2$ S4, S6, TSXA directions with identity covariance matrix. The directions of the tangential stress are shown by red vectors. }\label{solution_source_t_sat_identityCov_300}
\end{figure}
\end{center}

The vector directions are chosen by using InSAR satellite directions, listed in Table \ref{vectors_sat}.
The numerical tests are performed with one to four radar looks to confirm the theoretical results. 
We set the  same configuration as previous section for the problems in Cartesian coordinate. The horizontal circular fracture beneath the flat topography located at $0.3$ km (see Figure \ref{topo_flat}).
We still take $\alpha_0 = 10^{-7}$ and by depicting the L-curves, we find the best $\alpha_1$ with the best compromise between misf\/it and the traction norm $\t$ or pressure $\p$ (see Figure \ref{L_curve_sat_identityCov_300})
Normal and tangential tractions projected on the fracture corresponding to the best $\alpha_1=1$ are presented in Section supplementary Material, Figure \ref{solution_source_t_sat_identityCov_300}.
A comparison between the iteration numbers listed in Table \ref{iterations_comparison_sat} for the observed synthetic data provided by different number of radar looks, show that 
the use of more radar looks leads to a decrease in the number of iteration.

\begin{table}[htbp]
\footnotesize
\centering
\begin{tabular}{|c c l|}
\hline
\textbf{S4:}& D R 151 0.056 & [0.56 -0.14 0.80]
\\
\hline
\textbf{S6:}& A R 144 0.056 &[-0.66 -0.17 0.73]\\
 \hline
\textbf{TSX:}& A R 036 0.031  &[-0.54 0.11 0.83]\\
\hline
\textbf{TSX:}& D R 036 0.031 & [0.59 -0.13 0.80]\\
\hline
\end{tabular}
\caption{InSAR Unit vector directions}\label{vectors_sat} 
\end{table}

\begin{table}[htbp]
\footnotesize
\centering
\begin{tabular}{|c|c|c|c|c|}
\hline
radar look & $\alpha_1$& Unknown& Number of iteration & Number of unknowns\\
\hline
\multirow{2}{*}{S4}&\multirow{2}{*}{$1.0\ee1$ }& Pressure $\p$ &$93$&291\\
&  &Traction $\t$ &$108$&873\\
 \hline
\multirow{2}{*}{S4 S6}&\multirow{2}{*}{$1.0\ee1$ }& Pressure $\p$ &$59$&291\\
&  &Traction $\t$ &$80$&873\\
 \hline 
 \multirow{2}{*}{S4 S6 TSXA}&\multirow{2}{*}{$1.0\ee1$ }& Pressure $\p$ &$51$&291\\
&  &Traction $\t$ &$58$&873\\
 \hline
  \multirow{2}{*}{S4 S6 TSXA TSXD}&\multirow{2}{*}{$1.0\ee1$ }& Pressure $\p$ &$50$&291\\
&  &Traction $\t$ &$58$&873\\
 \hline
\end{tabular}
\caption{Comparison between the number of iterations 
taking into account the earth-satellite directions with identity covariance matrix in InSAR unit vector directions. The fracture is located at $-0.3$ km, when solving for pressure $\p$ and traction $\t$. Here, we set $\alpha_0= 1.0\ee{-7}$.}\label{iterations_comparison_sat}
\end{table}

\begin{center}
\begin{figure}[htp!]\hspace*{-0.5cm}   
\subfigure[]{\includegraphics[scale=0.18]
{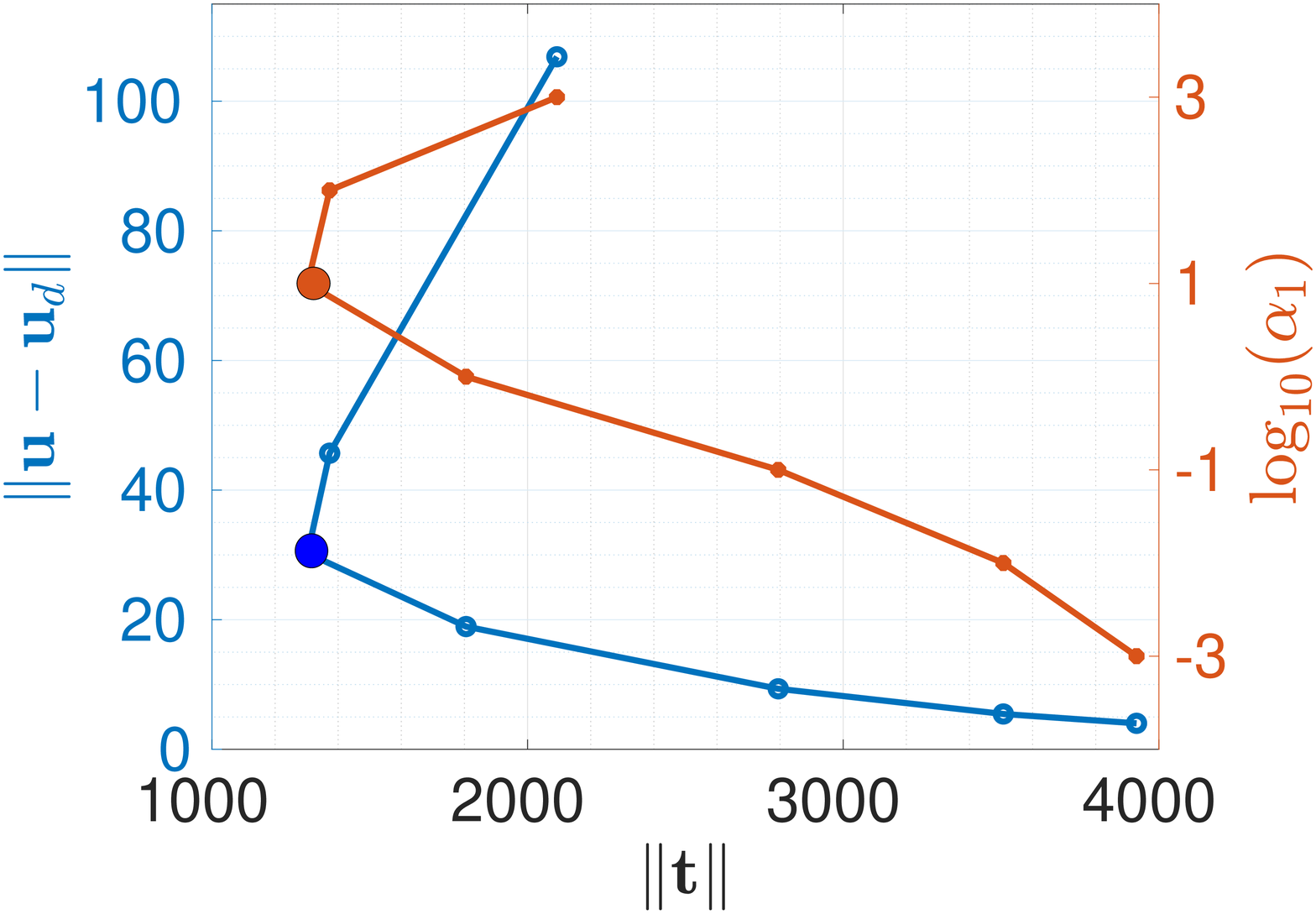}}\hspace*{+0.2cm}  
\subfigure[Normal stress]{\includegraphics[scale=0.2]
{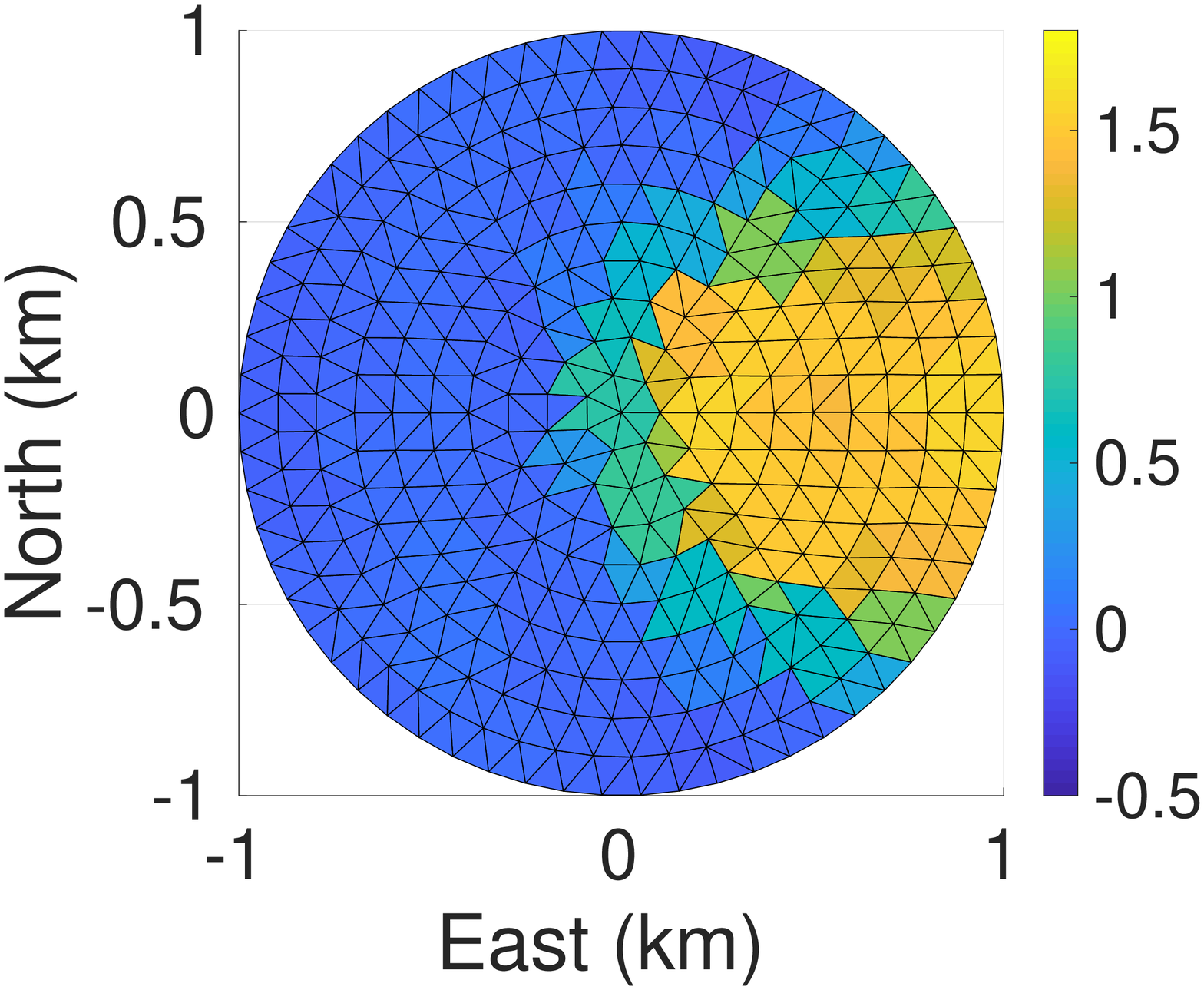}}
\hspace*{-0.2cm}  
\subfigure[Tangential stress]{\includegraphics[scale=0.2]
{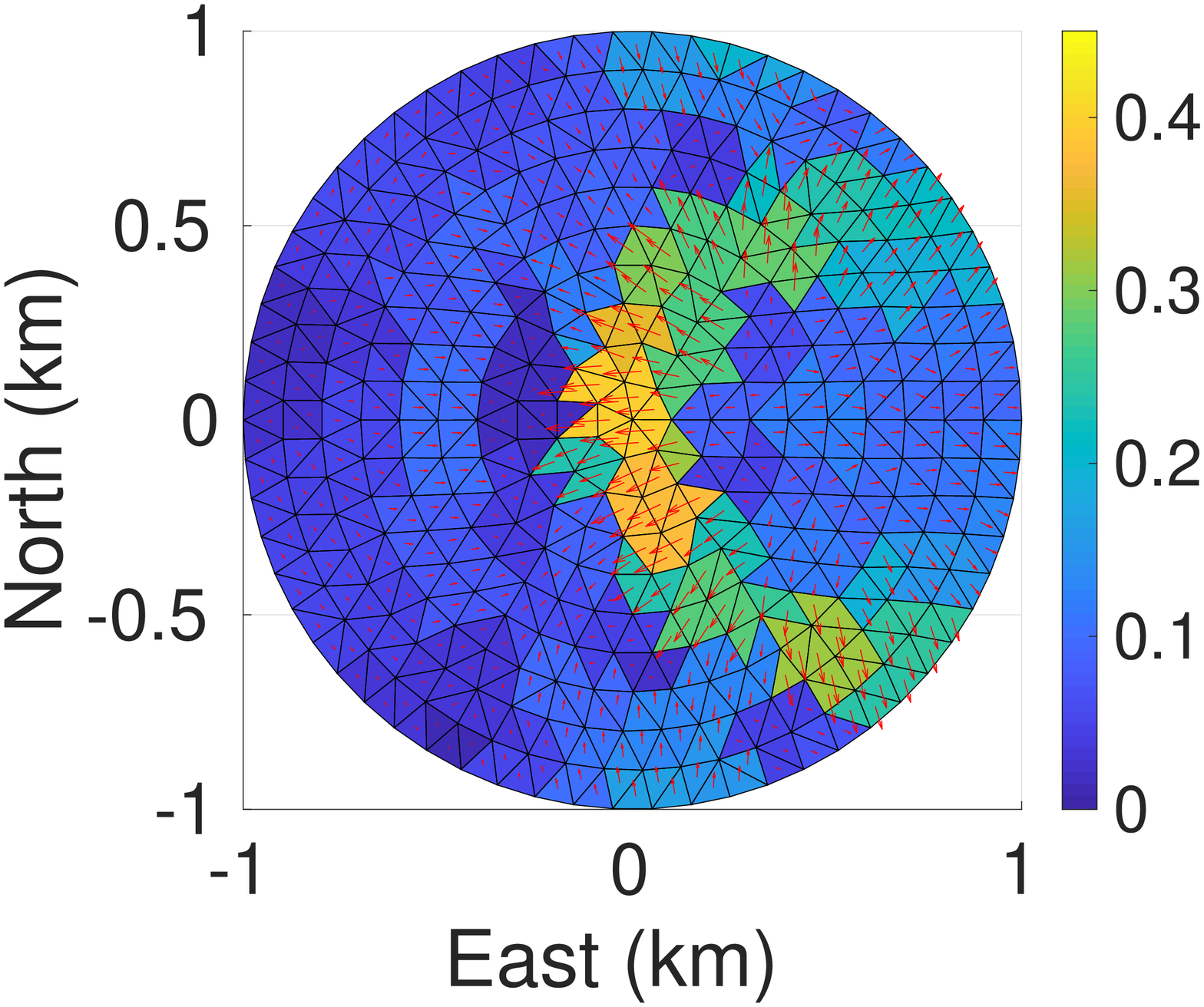}}
\caption{(a) L-curves for a circular fracture located at ${-0.3}$ km, corresponding to $\alpha_1$. (b), (c) fracture stress with reduced identity covariance matrix for $\alpha_1=1.0\ee0$  for S4 radar look.}\label{reduced_cov_sat}
\end{figure}
\end{center}

In a more realistic scenario, the number of observed data are limited. In the previous numerical experiments, we assumed a synthetic $\u_d$ in $\text{P}2$ finite elements.
The degrees of freedom's nodes of a classical $\text{P}2$ Lagrange elements on a 3D tetrahedron, are located on the vertices or nodal mesh points and the midpoints of the edges. In realistic volcano phenomenons (three-dimensional elasticity problem), using the $\text{P}2$ f\/inite element guarantees a good approximated solution. However, we aim to reduce the synthetic $\u_d$
to nodal mesh points on the ground surface as a $\text{P}1$ f\/inite elements. On the other side, 
to conserve the $\text{P}2$ f\/inite elements for the elasticity problem, we should conserve the dimension of the covariance $\CC^{-\frac{1}{2}}$ and the mass matrix $M_G$ in a $\text{P}2$ elements. Therefore, from implementation point of view, the degrees of freedom's related to the midpoints of the edges in $\CC^{-\frac{1}{2}}$ and $M_G$ are set to zero. 
Despite the increase in the iterations number, the numerical experiments presented still show very satisfactory results (see Section Supplementary Material Figure \ref{reduced_cov_sat}).

Another step to approach the reality is using the dense covariance matrices for limited number of data, obtained by InSAR and cGNSS. 

\begin{center}
\begin{figure}\hspace*{-0.8cm}  
\subfigure[The solution for $\alpha_1=1.0\ee1$]{\includegraphics[scale=0.24]
{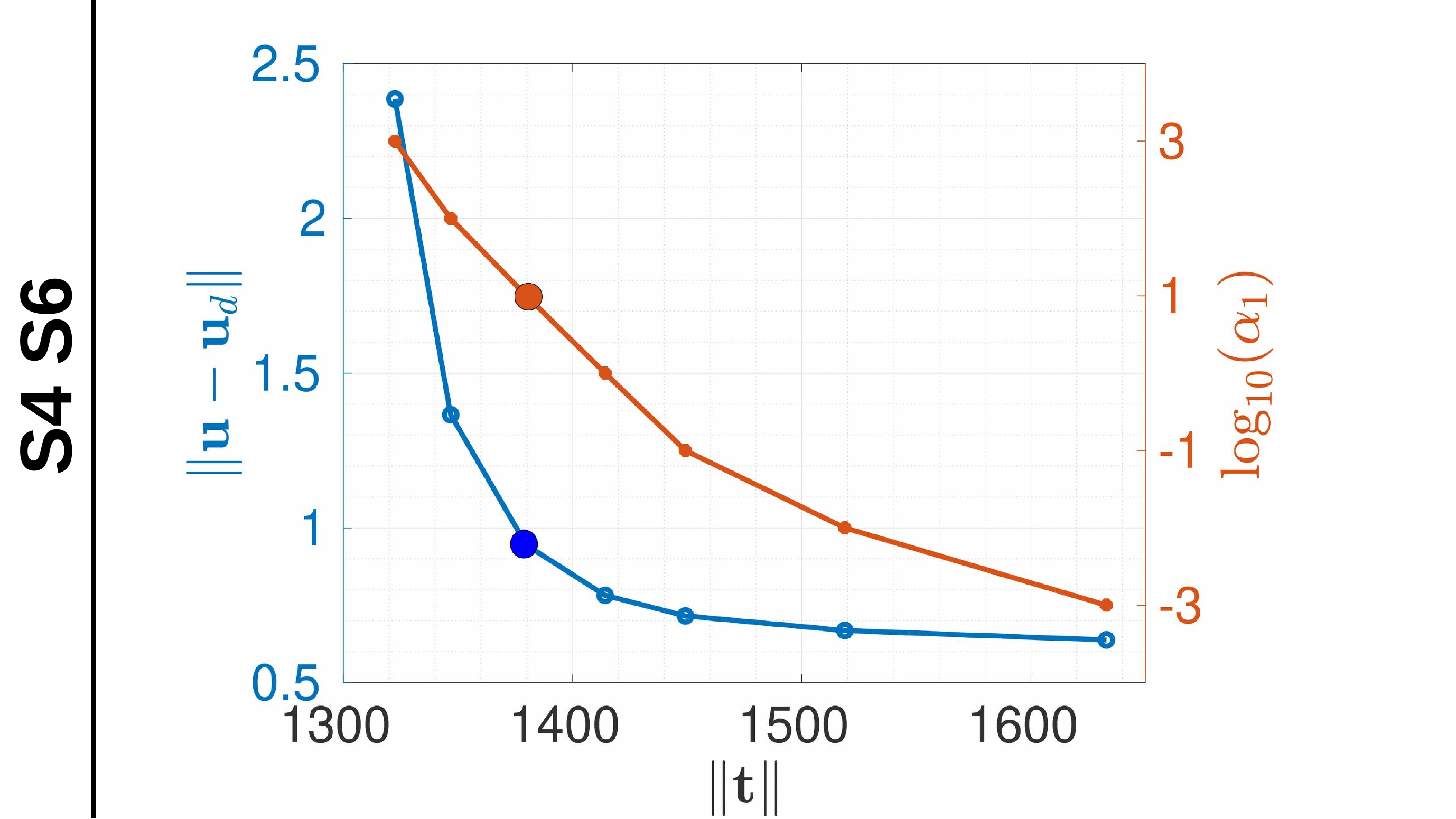}}\hspace*{-0.8cm}  
\subfigure{\includegraphics[scale=0.2]
{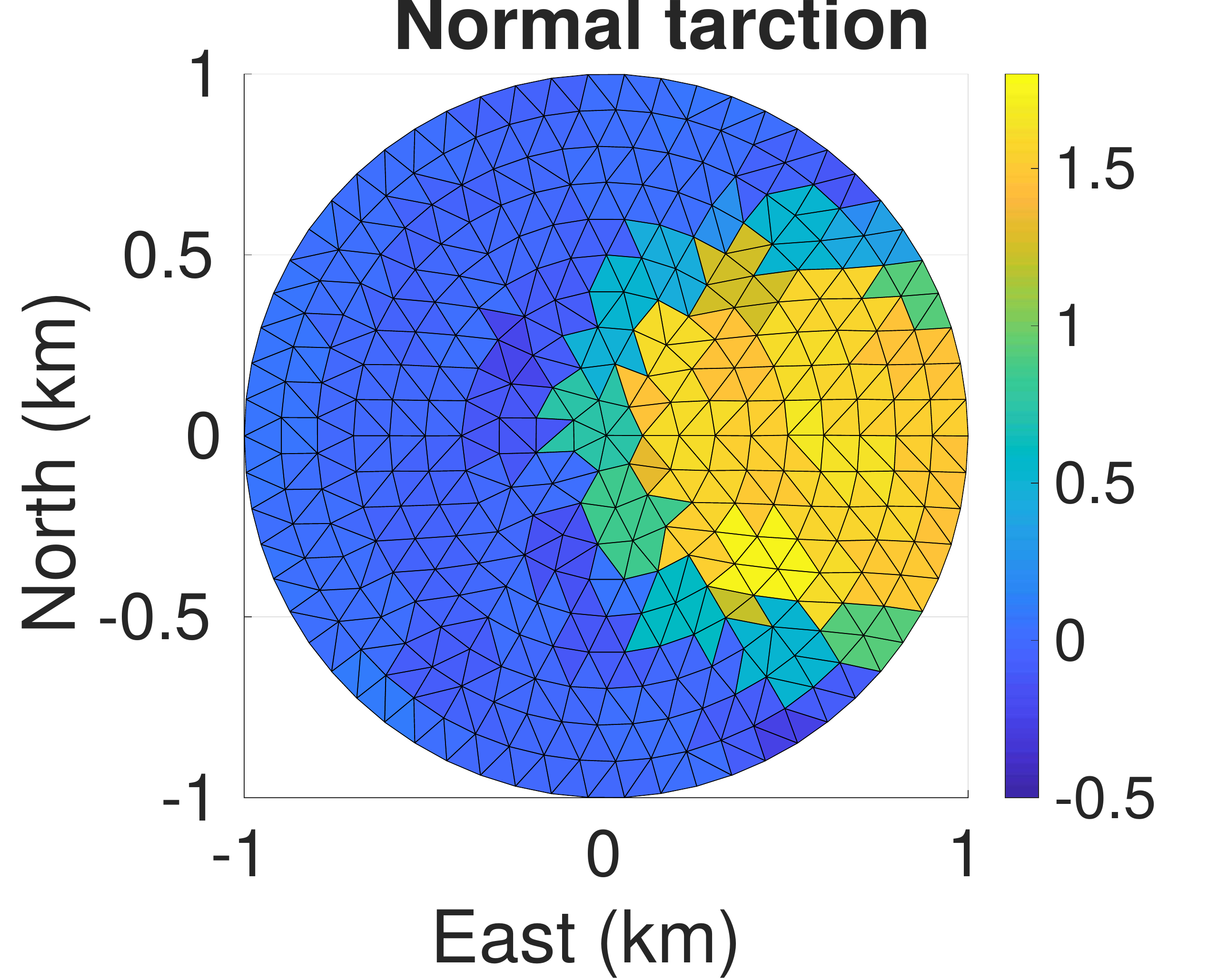}}
\hspace*{-0.4cm}  
\subfigure{\includegraphics[scale=0.18]
{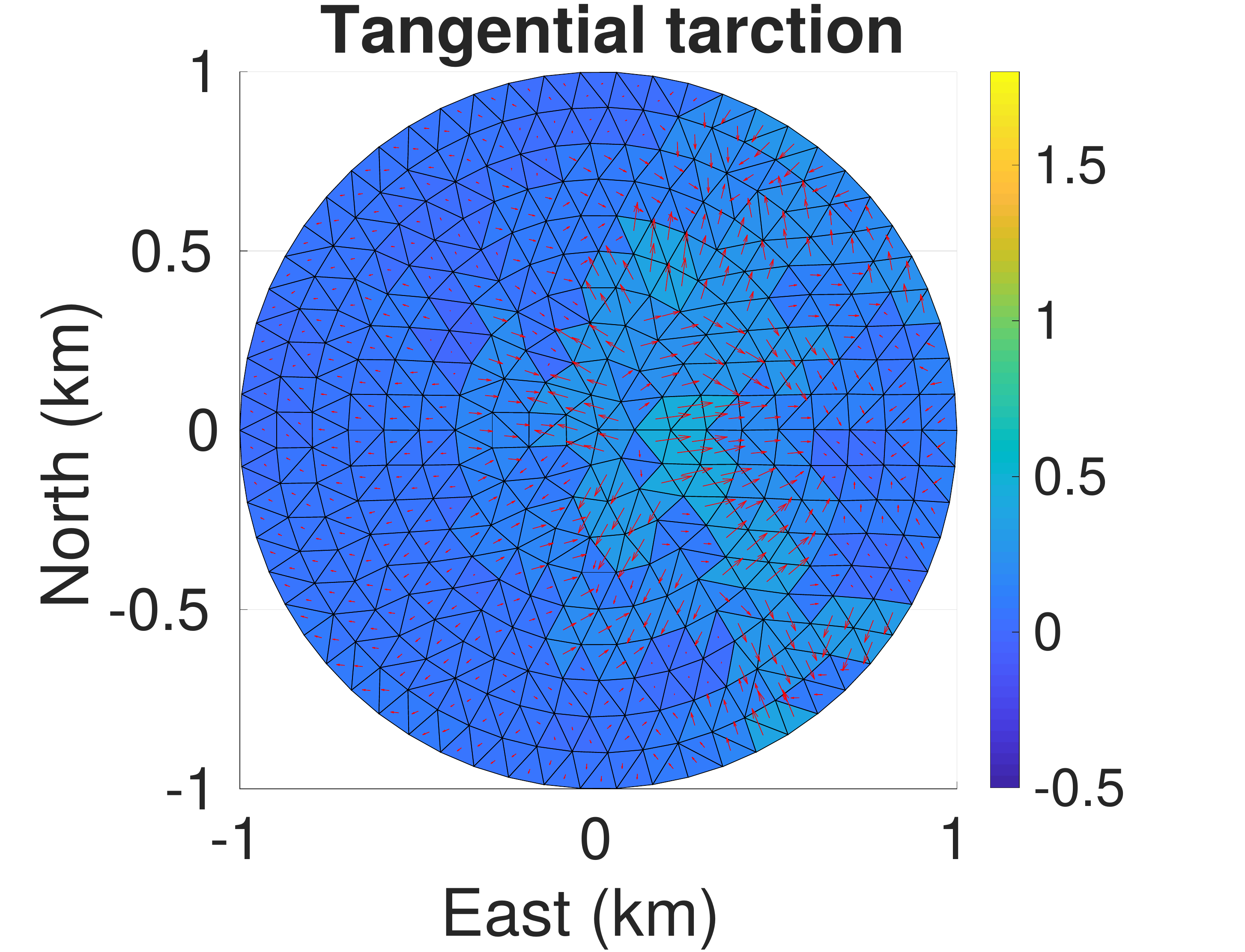}}\\
\hspace*{-0.8cm}  
\subfigure[The solution for $\alpha_1=1.0\ee2$ ]{\includegraphics[scale=0.24]
{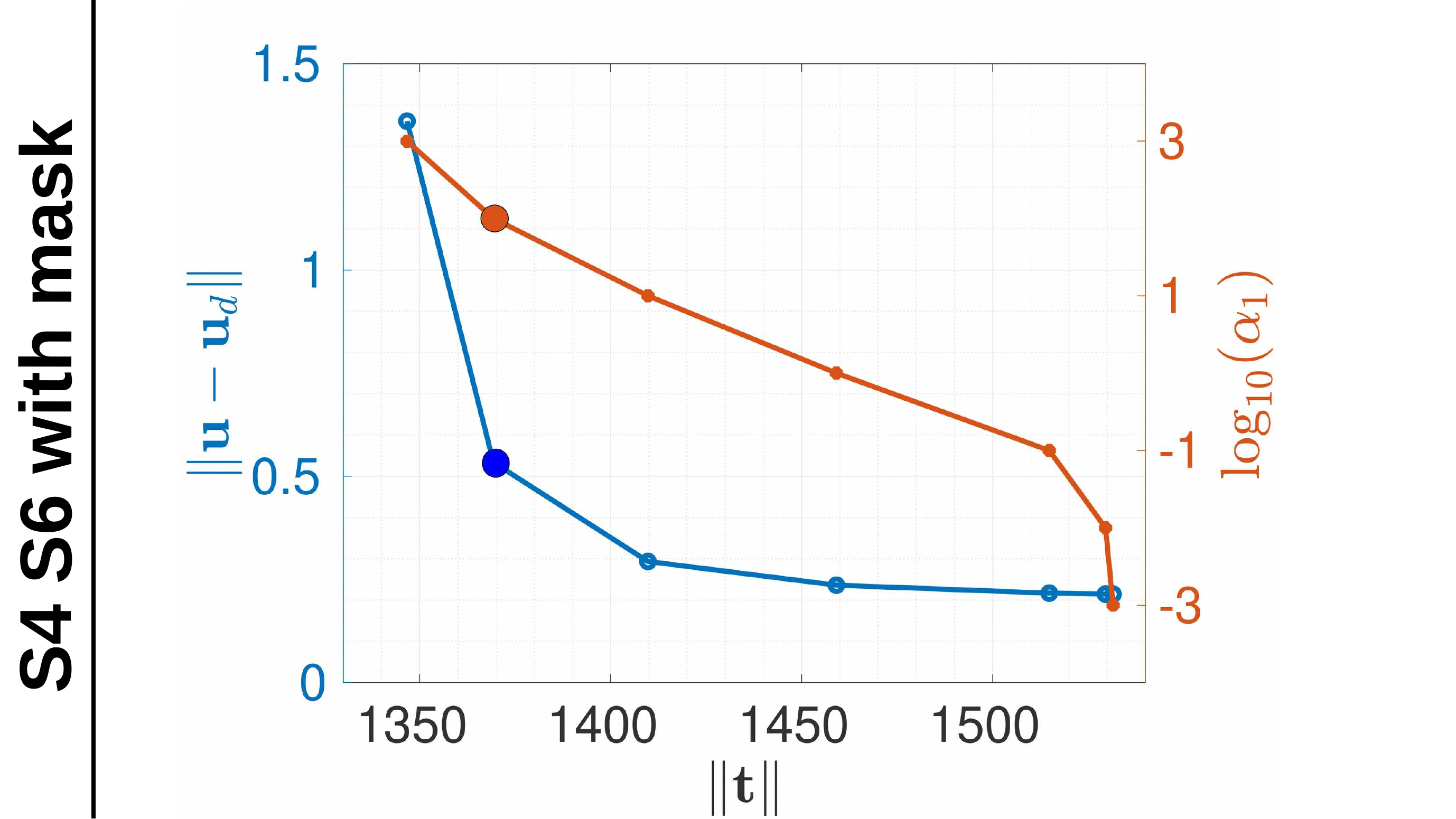}}\hspace*{-0.8cm}  
\subfigure{\includegraphics[scale=0.2]
{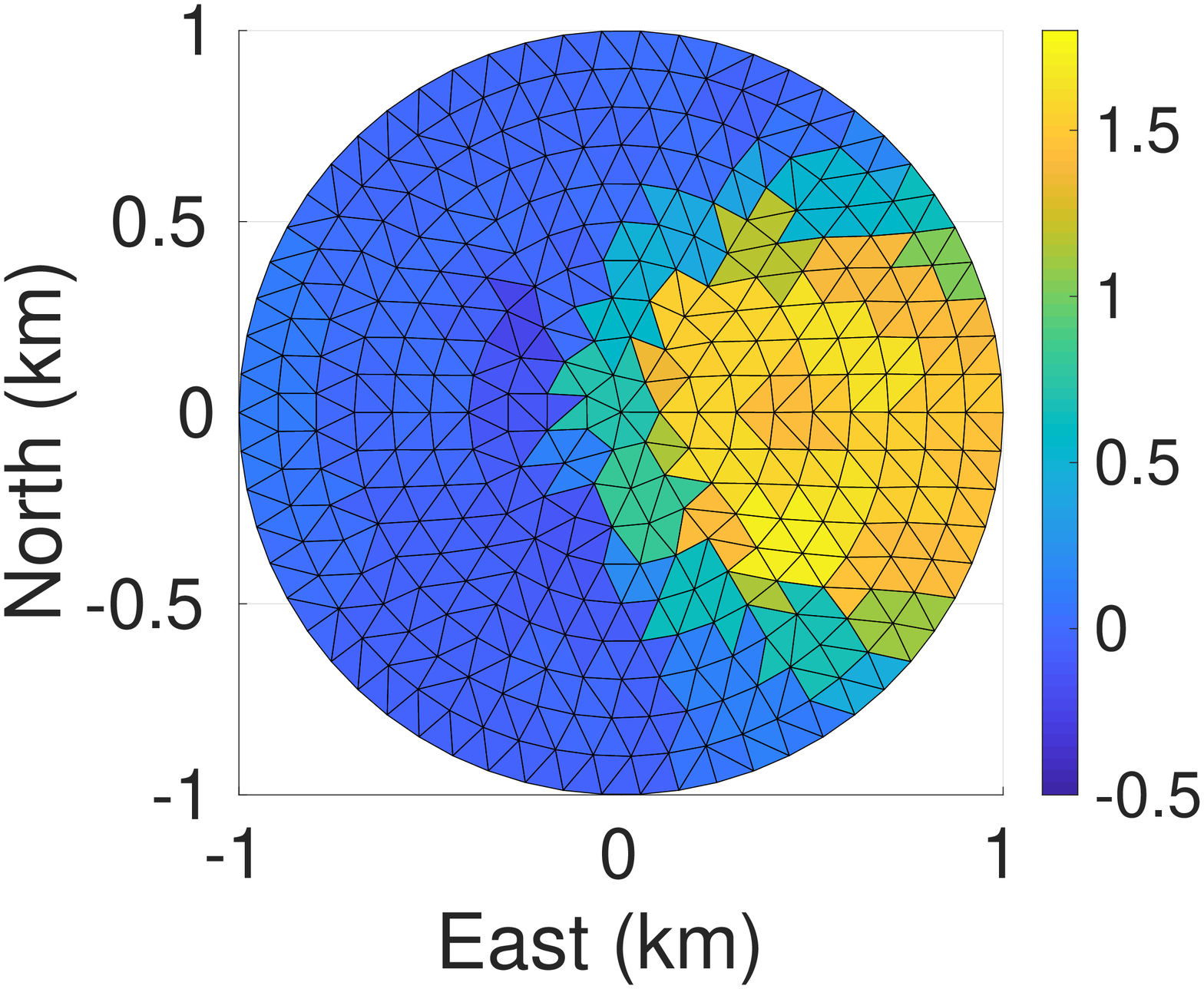}}
\hspace*{-0.4cm}  
\subfigure{\includegraphics[scale=0.18]
{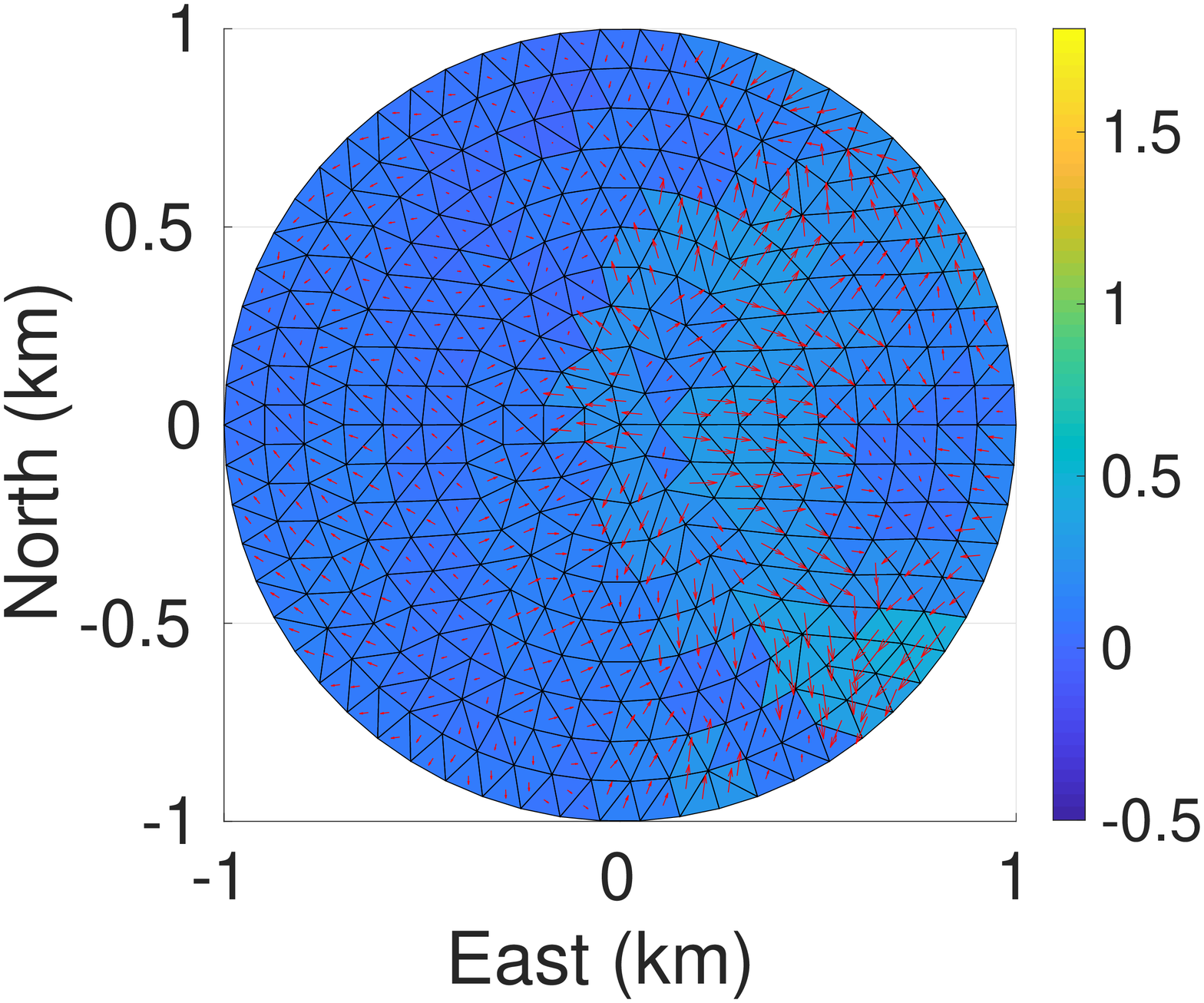}}\\
\hspace*{-0.8cm} 
\subfigure[The solution for $\alpha_1=1.0\ee2.5$]{\includegraphics[scale=0.24]
{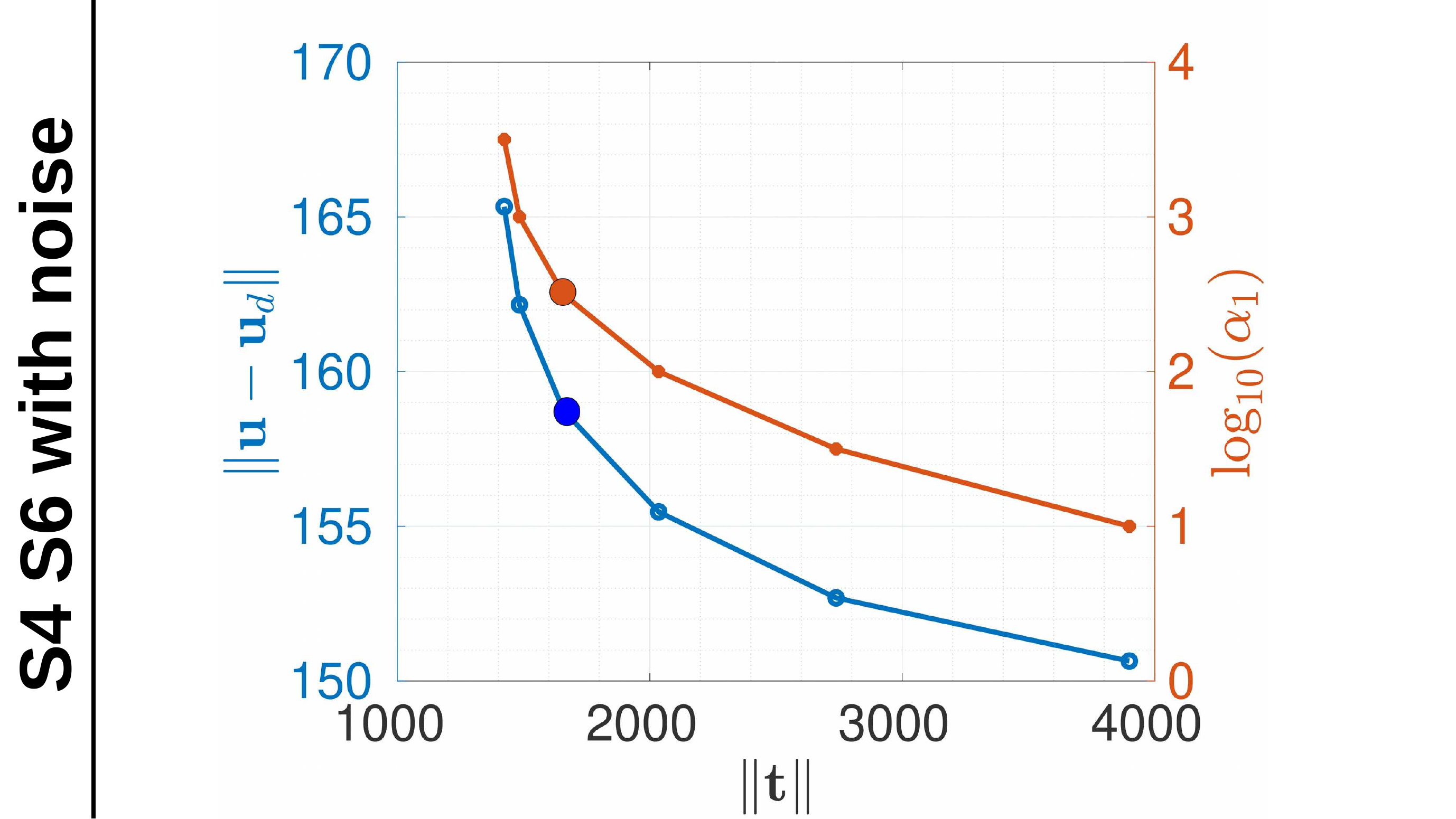}}\hspace*{-0.8cm}  
\subfigure{\includegraphics[scale=0.18]
{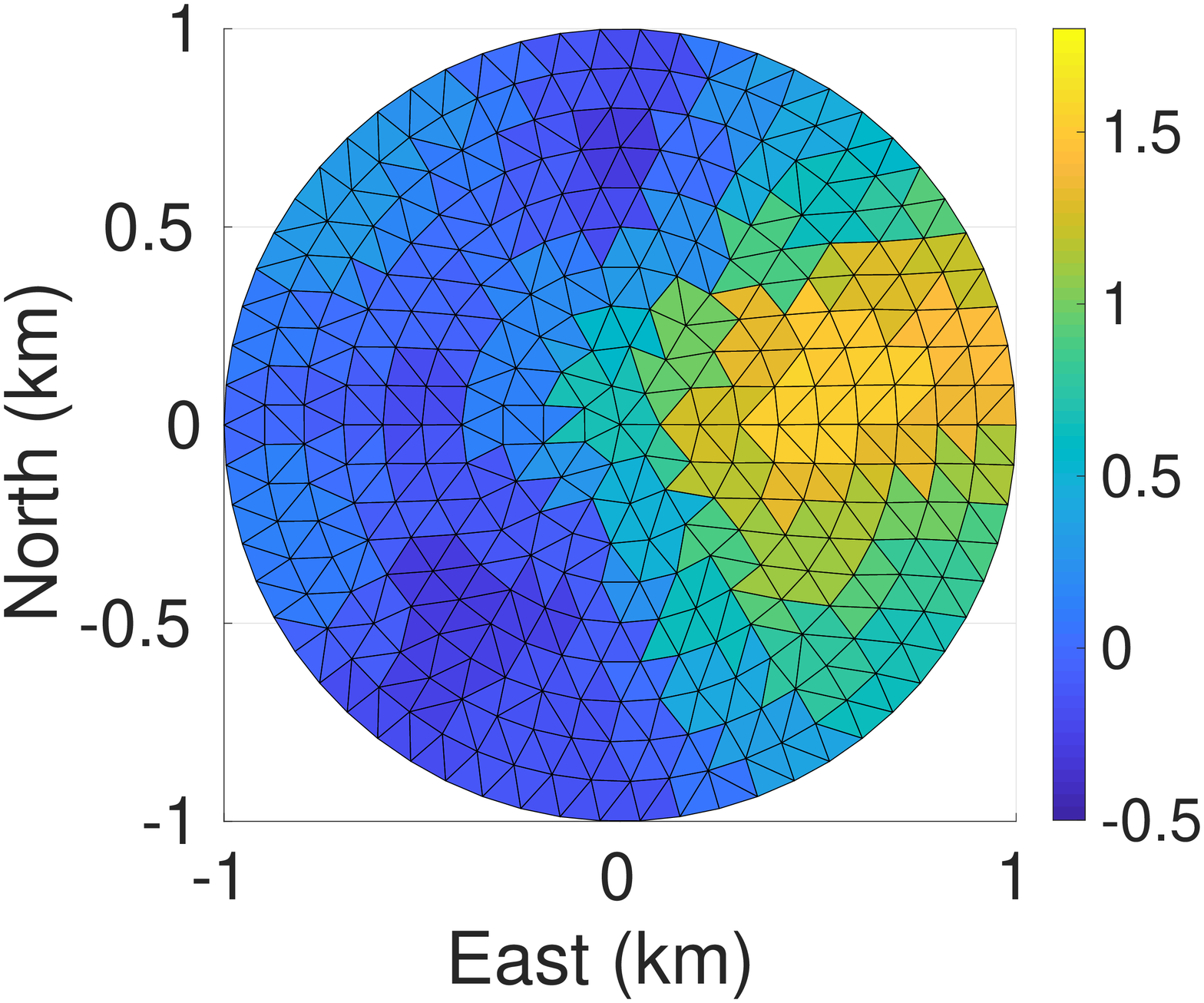}}
\hspace*{-0.2cm}  
\subfigure{\includegraphics[scale=0.17]
{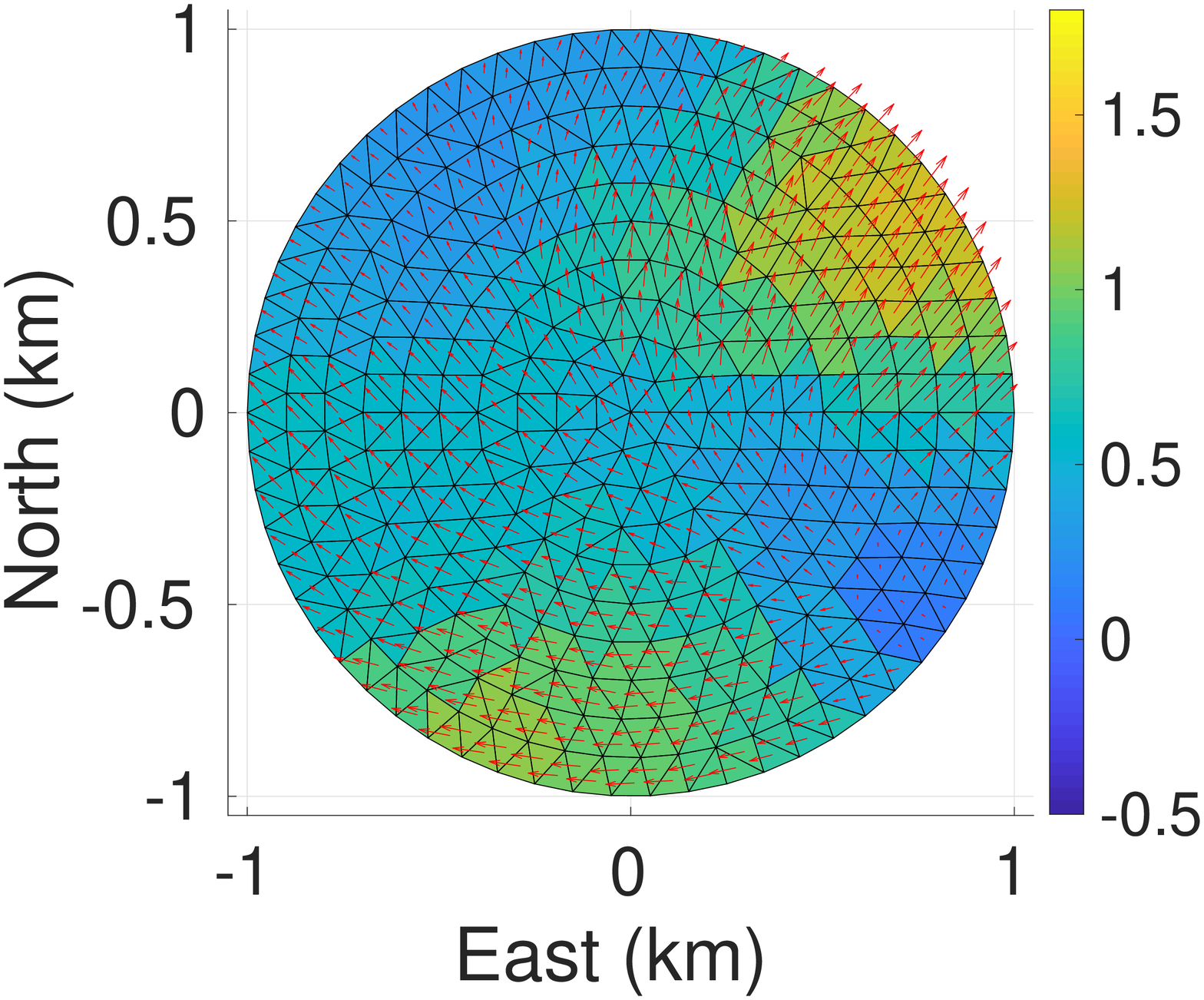}}\\

\hspace*{-0.8cm} 
\subfigure[The solution for $\alpha_1=1.0\ee3$]{\includegraphics[scale=0.24]
{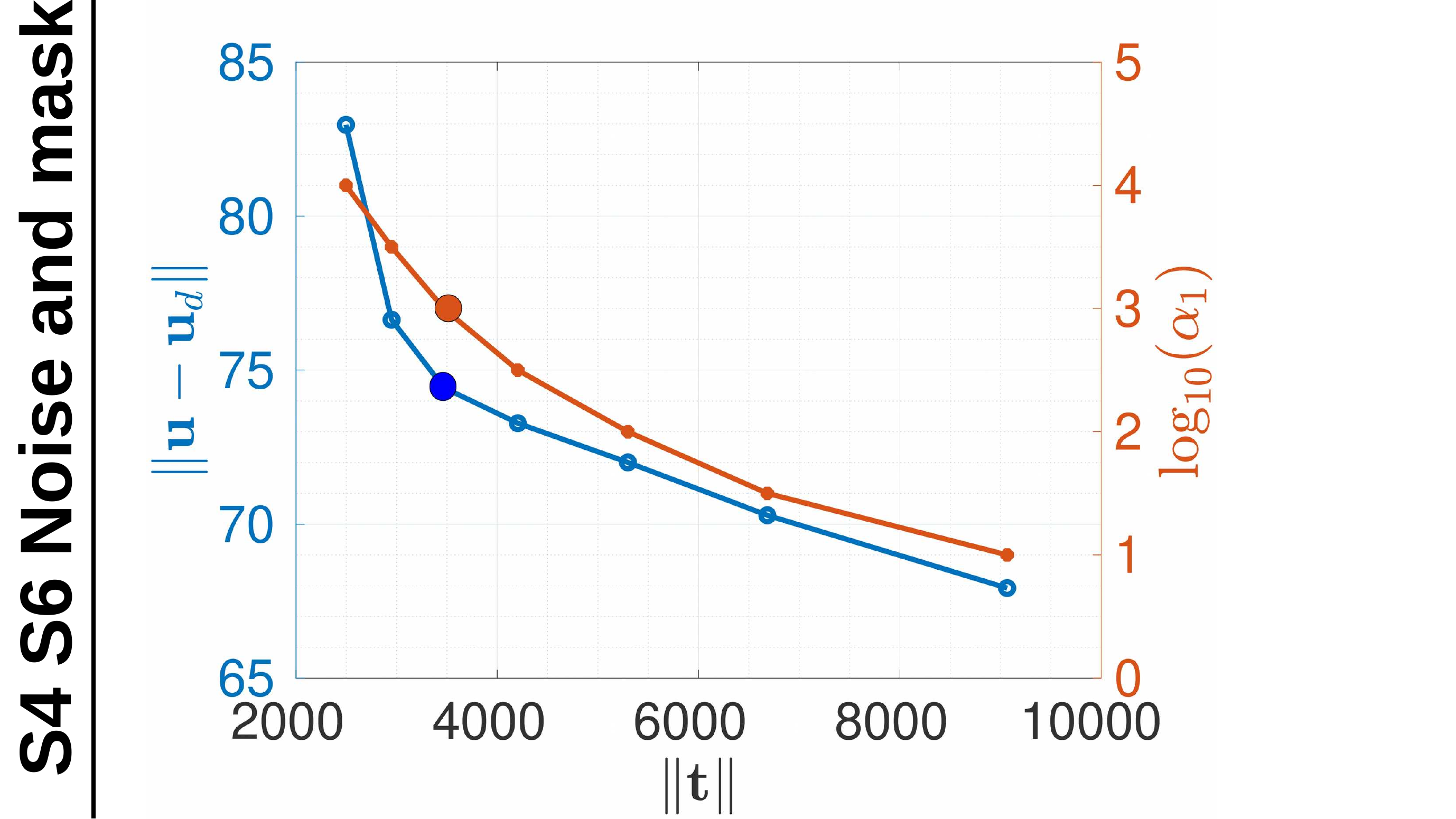}}\hspace*{-0.9cm}  
\subfigure{\includegraphics[scale=0.18]
{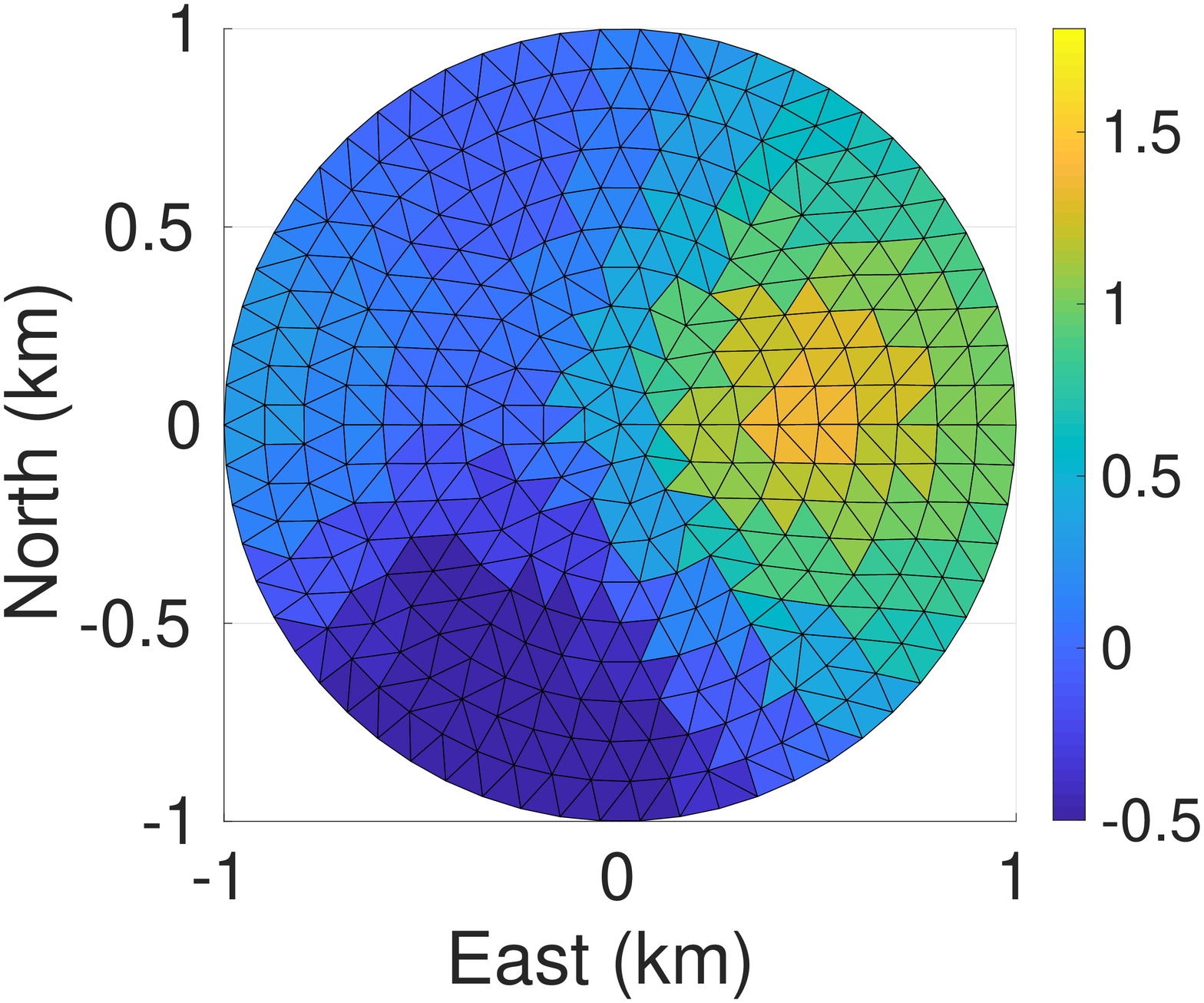}}
\hspace*{-0.2cm}  
\subfigure{\includegraphics[scale=0.17]
{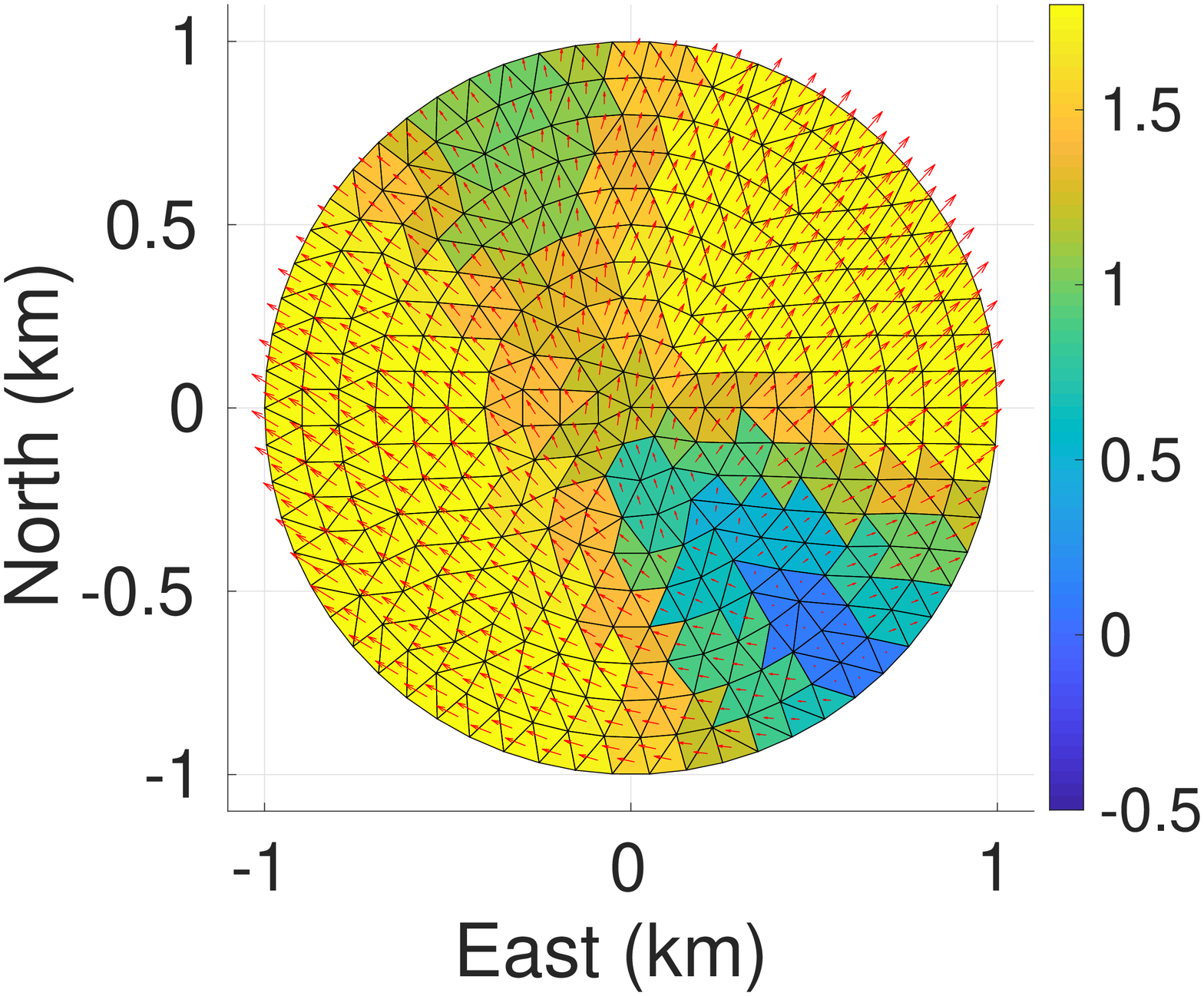}}

\caption{L-curves used to find $\alpha_1$ representing the best compromise between the data fit (equation \eqref{misfit}) and the smoothing (equation \eqref{smoothing}) in the cost function (equation \eqref{min_prob_esd}) taking into account the earth-satellite directions with the dense covariance matrix. The fracture is located at $-0.3$ km, when solving for the traction $\t$. The larger points indicate the best compromise. The normal and tangential traction are presented for the appropriate $\alpha_1$.  The synthetic data $\u_d$ is projected in S4 and S6 radar looks. (a) without any noise and mask, (b) with mask, (c) with noise (d) with noise and mask.
Here, we set $\alpha_0 = 1.0\ee-7 $.}\label{realcov}
\end{figure}
\end{center}

To do that, we are using the DEfvolc interfaces   
The dense covariance matrices are provided by 
DefVolc, pre and post-processor software.
We started by creating the synthetic $\u_d$ and the dense covariance matrix in S4 radar look. 
The results obtained in Figure \ref{realcov} and first row of Table \ref{iterations_comparison_sat_realcov}, confirm the adaptability of our method to the dense covariance matrices. 
In reality, most of the time, the atmospheric contribution caused masked and noisy InSAR data. Therefore, some numerical experiments is performed by creating  masked and noisy synthetic $\u_d$ and covariance matrix. 
The numerical experiments are presented in 
Figure \ref{realcov}. The L-curves are depicted for synthetic $\u_d$ projected to S4 an S6 directions. First without any mask and noise in the data and then different possible cases by adding the mask and noise to the synthetic data.
We summarized appropriate choices for $\alpha_1$
and the number of iterations to achieve the convergence with these $\alpha_1$, for different cases of synthetic $\u_d$, in Table \ref{iterations_comparison_sat_realcov}. 
Despite of choosing a greater $\alpha_1$, for the masked data,  the number of iteration is reduced. However, for the noisy data the convergence of the inversion process is much more harder. Moreover the vector size of the synthetic $\u_d$ for each case are listed in Table \ref{iterations_comparison_sat_realcov}.

\begin{table}[htbp]
\footnotesize
\centering
\begin{tabular}{|c|c|c|c|}
\hline
radar look & $\alpha_1$& Number of iteration& vector size $\u_d$ \\
\hline
S4&$1.0\ee1$ &$98$& 1616\\
 \hline
S4 S6&$1.0\ee1$ &$102$&3232\\
 \hline
 S4 S6 with mask&$1.0\ee2$ &$56$&1087\\
 \hline
  S4 S6 with noise&$1.0\ee2.5$ &$344 $&3232\\
 \hline
   S4 S6 with mask and noise&$1.0\ee3$ &$306 $&1087\\
 \hline
\end{tabular}
\caption{Comparison between the number of iterations for fracture at 0.3 km depth and when the unknowns are pressure $\p$ and traction $\t$
for different radar looks with identity covariance matrix.}\label{iterations_comparison_sat_realcov}
\end{table}


\newpage
\section{Appendix}\label{appendix}
\subsection*{The optimal step size}
This section is dedicated to some mathematical details for interested readers. First part is concerned to compute the optimal step size. To simplify the presentation, let us introduce $c_N$ defined on $\Gc$
$$
c_N(\bw,\bw)=\int_{\Gn} \bw^\top \CC^{-1}\bw \dd \Gn,
$$
so that the cost functional in \eqref{min_prob} becomes
$$
J(\t):=\frac{1}{2}c_N(\u-\u_{d},\u-\u_{d}) 
+\frac{\alpha_0}{2}\ds\int_{\Gc}\lvert \t\rvert^2\dd \Gc+\frac{\alpha_1}{2}\ds\int_{\Gc} \lvert\nabla \t\rvert^2\dd \Gc.
$$
Then the directional derivative of $J$ in the direction of a given $\d$ reads 
$$
\langle\frac{\partial J}{\partial \t}(\t), \d\rangle_{\alpha_0,\alpha_1}= c_N(\u-\u_{d},\bw)+\alpha_0\ds\int_{\Gc}(\t\cdot\d)\dd \Gc+\alpha_1\ds\int_{\Gc} (\nabla \t \cdot\nabla\d) \dd \Gc,
$$
where $\bw$ is the solution to \eqref{forml} with $\t=\d$.
Therefore, we may compute the optimal step size $\rho$,  
with a search direction $d$ by solving 
$$
\langle\frac{\partial J}{\partial \t}(\t +\rho \d), \d\rangle_{\alpha_0,\alpha_1}= 0,
$$
that is 
$$
c_N(\bu+\rho\bw-\bu_d,\bw)+\alpha_0\ds\int_{\Gc}\big((\t +\rho \d)\cdot\d\big)\dd \Gc+\alpha_1\ds\int_{\Gc} \big((\nabla \t+\rho \nabla\d) \cdot\nabla\d \big)\dd \Gc=0,
$$
which gives
\begin{align*}
&\rho\left[c_N(\bw,\bw)+\alpha_0\ds\int_{\Gc}(\d\cdot\d)\dd\Gc+\alpha_1\ds\int_{\Gc}(\nabla \d\cdot \nabla\d)\dd\Gc\right]+c_N(\u-\u_d,\bw)\\[1em]
&+\alpha_0\ds\int_{\Gc}(\t\cdot\d)\dd\Gc+\alpha_1\ds\int_{\Gc}(\nabla \t\cdot \nabla\d)\dd\Gc = 0.
\end{align*}
The optimal step size is therefore
\begin{equation}\label{step_size}
\rho^\ast=-\frac{ c_N(\u-\u_d,\bw)+\alpha_0\ds\int_{\Gc}(\t\cdot\d)\dd\Gc+\alpha_1\ds\int_{\Gc}(\nabla \t\cdot \nabla\d)\dd\Gc  }{c_N(\bw,\bw)+\alpha_0\ds\int_{\Gc}(\d\cdot\d)\dd\Gc+\alpha_1\ds\int_{\Gc}(\nabla \d\cdot \nabla\d)\dd\Gc}.
\end{equation}

\section*{The optimal step size for Earth-Satellite direction}
The optimal step size for Earth-Satellite direction as previous, is still obtained by solving 
$$
\langle\frac{\partial J}{\partial \t}(\t +\rho \d), \d\rangle_{\alpha_0,\alpha_1}=0
$$
that is
$$
c_N(\bP\bu+\rho\bP\bw-\br_d,\bP\bw)
+\alpha_0\ds\int_{\Gc}\big((\t +\rho \d)\cdot\d\big)\dd \Gc+\alpha_1\ds\int_{\Gc}\big( (\nabla \t+\rho \nabla\d) \cdot\nabla\d\big) \dd \Gc
=0,
$$
where $\bw$ is still the solution to \eqref{forml} with $\t=\d$.
This gives the optimal step size
\begin{equation}\label{step_size_esd}
\rho^\ast=-\frac{ c_N(\bP\u-\br_{d},\bP\bw)+\alpha_0\ds\int_{\Gc}(\t\cdot\d)\dd\Gc+\alpha_1\ds\int_{\Gc}(\nabla \t\cdot \nabla\d)\dd\Gc  }{c_N(\bP\bw,\bP\bw)+\alpha_0\ds\int_{\Gc}(\d\cdot\d)\dd\Gc+\alpha_1\ds\int_{\Gc}(\nabla \d\cdot \nabla\d)\dd\Gc}.
\end{equation}

\subsection*{L-BFGS Update}
The second part is concerned to L-BFGS Update in minimization Algorithm \ref{IP_DDM1}
\begin{equation}\label{BFGS_Update}
    H^{k+1}= (I-\theta^k \s^k{\y^k}^\top)H^k(I-\theta^k \y^k{\s^k}^\top)+\theta^k \s^k{\s^k}^\top,
\end{equation}
with
 \begin{equation*}
 \s^k= \t^{k+1}-\t^{k}, \quad \y^k=\g^{k+1}-\g^{k}, \quad \theta^k=\frac{1}{{\y^k}^\top \s^k}.
    \end{equation*}

\section{Supplementary Material}

\newpage
\bibliographystyle{plain}
\bibliography{mybib.bib}
\end{document}